\documentstyle[aps,psfig]{revtex}
\begin{document}
\draft
\renewcommand{\thefootnote}{\fnsymbol{footnote}}
\begin{title}
{\bf Molecular hydrodynamics of the moving contact line in two-phase
immiscible flows}
\end{title}
\author{Tiezheng Qian\footnote{To whom correspondence should be addressed. 
E-mail: maqian@ust.hk} and Xiao-Ping Wang}
\address{Department of Mathematics,
The Hong Kong University of Science and Technology,\\ 
Clear Water Bay, Kowloon, Hong Kong, China} 
\author{Ping Sheng}
\address{Department of Physics,
The Hong Kong University of Science and Technology,\\ 
Clear Water Bay, Kowloon, Hong Kong, China} 
\maketitle
\begin{abstract}

The ``no-slip'' boundary condition, i.e., zero fluid velocity relative to 
the solid at the fluid-solid interface, has been very successful in 
describing many macroscopic flows, yet there is no compelling 
theoretical argument to justify this standard boundary condition
of textbook continuum hydrodynamics. A problem of principle arises 
when the no-slip boundary condition is used to model the hydrodynamics
of immiscible-fluid displacement in the vicinity of the moving contact line, 
where the interface separating two immiscible fluids intersects the solid wall.
Decades ago it was already known that
the moving contact line is incompatible with the no-slip boundary condition, 
since the latter would imply infinite dissipation due to 
a non-integrable singularity in the stress near the contact line. 
While subsequent molecular dynamics (MD) studies have clearly demonstrated 
fluid slipping relative to the wall at the contact line, the exact rule that 
governs this relative slip has eluded numerous prior attempts.
In fact, over the years there have been numerous {\it ad hoc} models
and proposals aiming to resolve the incompatibility of the no-slip
boundary condition with the moving contact line, but none was able to 
quantitatively account for the near-complete slip of the contact line 
observed in MD simulations.

In this paper we first present an introductory review of the problem, 
including (1) the cause of the stress singularity, (2) the ad hoc
introduction of the slip boundary condition, (3) the MD evidence of 
fluid slipping, (4) the gap between the existing MD results and 
a continuum hydrodynamic description, 
and (5) a preliminary account on how to bridge the MD results and
a continuum description. We then present a detailed review of our 
recent results on the contact-line motion in immiscible two-phase flow, 
from MD simulations to continuum hydrodynamics calculations. 
Through extensive MD studies and detailed analysis, we have uncovered 
the slip boundary condition governing the moving contact line, 
denoted the generalized Navier boundary condition.  
We have used this discovery to formulate a continuum
hydrodynamic model whose predictions are in remarkable quantitative
agreement with the MD simulation results at the molecular level. 
These results serve to affirm the validity of 
the generalized Navier boundary condition, as well as to open up 
the possibility of continuum hydrodynamic calculations of
immiscible flows that are physically meaningful at the molecular level.

\end{abstract}

\section{Introduction}

\begin{figure}[h]
\bigskip
\centerline{\psfig{figure=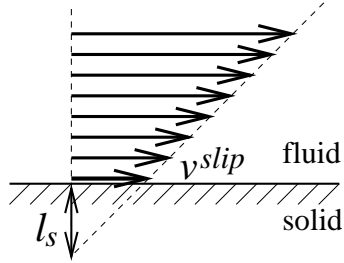,height=3.50cm}}
\caption{ Slip length introduced in the Navier boundary condition,
defined as the distance from the fluid-solid interface to where the linearly
extrapolated tangential velocity vanishes.
}\label{NBC1823}
\end{figure}

The no-slip boundary condition, i.e., zero relative tangential velocity
between the fluid and solid at the interface, serves as a cornerstone
in continuum hydrodynamics \cite{batchelor}. 
Although fluid slipping is generally detected in molecular dynamics (MD) 
simulations for microscopically small systems at high flow rate
\cite{Th-Rob,nbc,barrat,ckb}, the no-slip boundary condition still works well 
for macroscopic flows at low flow rate. 
This is due to the Navier boundary condition which actually accounts for 
the slip observed in MD simulations \cite{Th-Rob,nbc,barrat,ckb}.
This boundary condition, proposed by Navier in 1823 \cite{navier1823}, 
introduces a slip length $l_s$ and assumes that the amount of slip
is proportional to the shear rate in the fluid at the solid surface:
$${\bf v}^{slip}=-l_s{\bf n}\cdot
\left[(\nabla {\bf v})+(\nabla {\bf v})^T\right],$$
where ${\bf v}^{slip}$ is the slip velocity at the surface, measured relative
to the (moving) wall, $\left[(\nabla {\bf v})+(\nabla {\bf v})^T\right]$
is the tensor of the rate of strain, and
$\bf n$ denotes the outward surface normal (directed out of the fluid).
According to the Navier boundary condition, the slip length is defined
as the distance from the fluid-solid interface to where the linearly
extrapolated tangential velocity vanishes (see Fig. \ref{NBC1823}).
Typically, $l_s$ ranges from a few angstroms to a few nanometers 
\cite{Th-Rob,nbc,barrat,ckb}. 
For a flow of characteristic length $R$ and velocity $U$,
the slip velocity is on the order of $Ul_s/R$. This explains why 
the Navier boundary condition is practically indistinguishable from 
the no-slip boundary condition in macroscopic flows where 
$v^{slip}/U\sim l_s/R\rightarrow 0$.

\begin{figure}[h]
\bigskip
\centerline{\psfig{figure=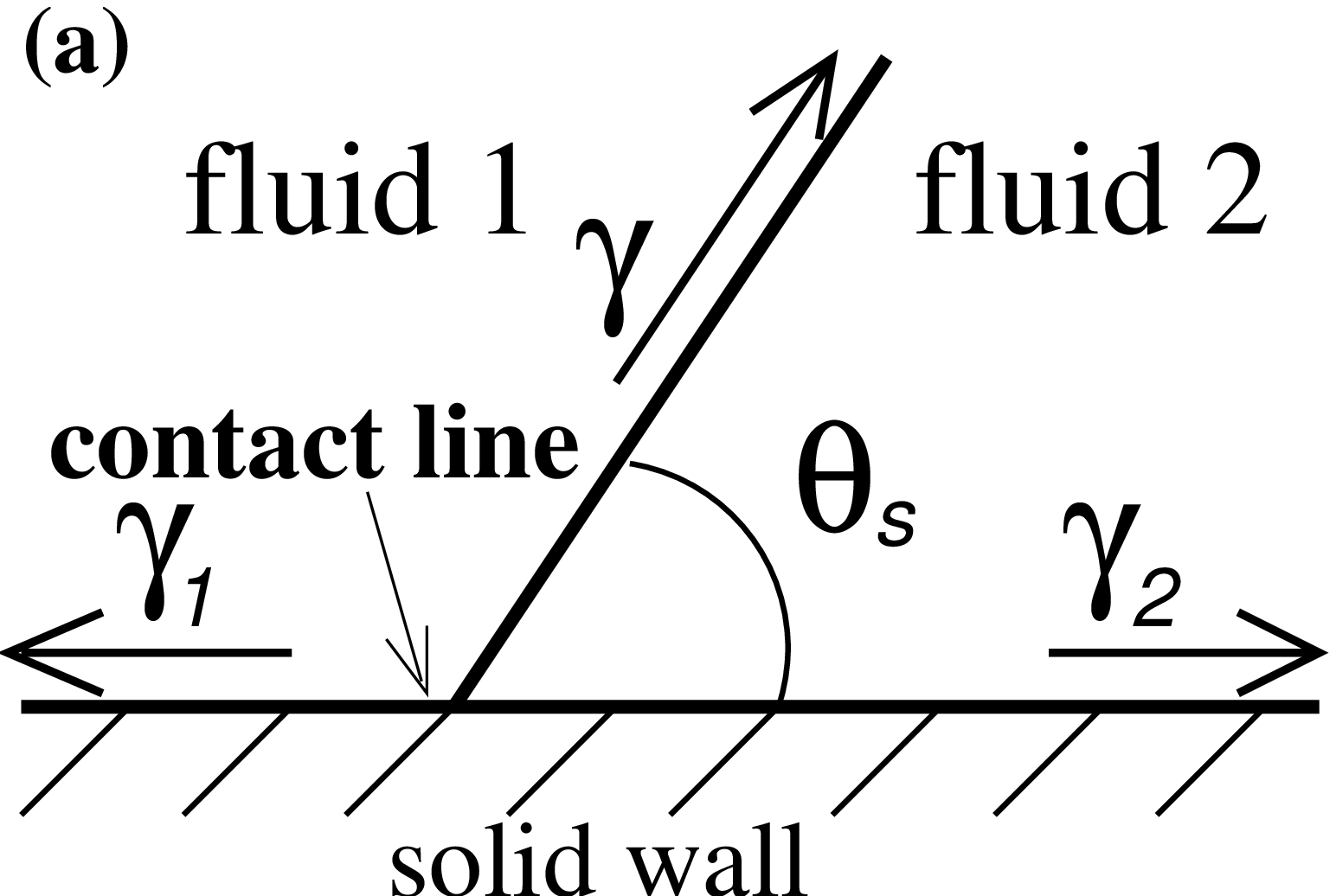,height=3.50cm}}
\bigskip
\centerline{\psfig{figure=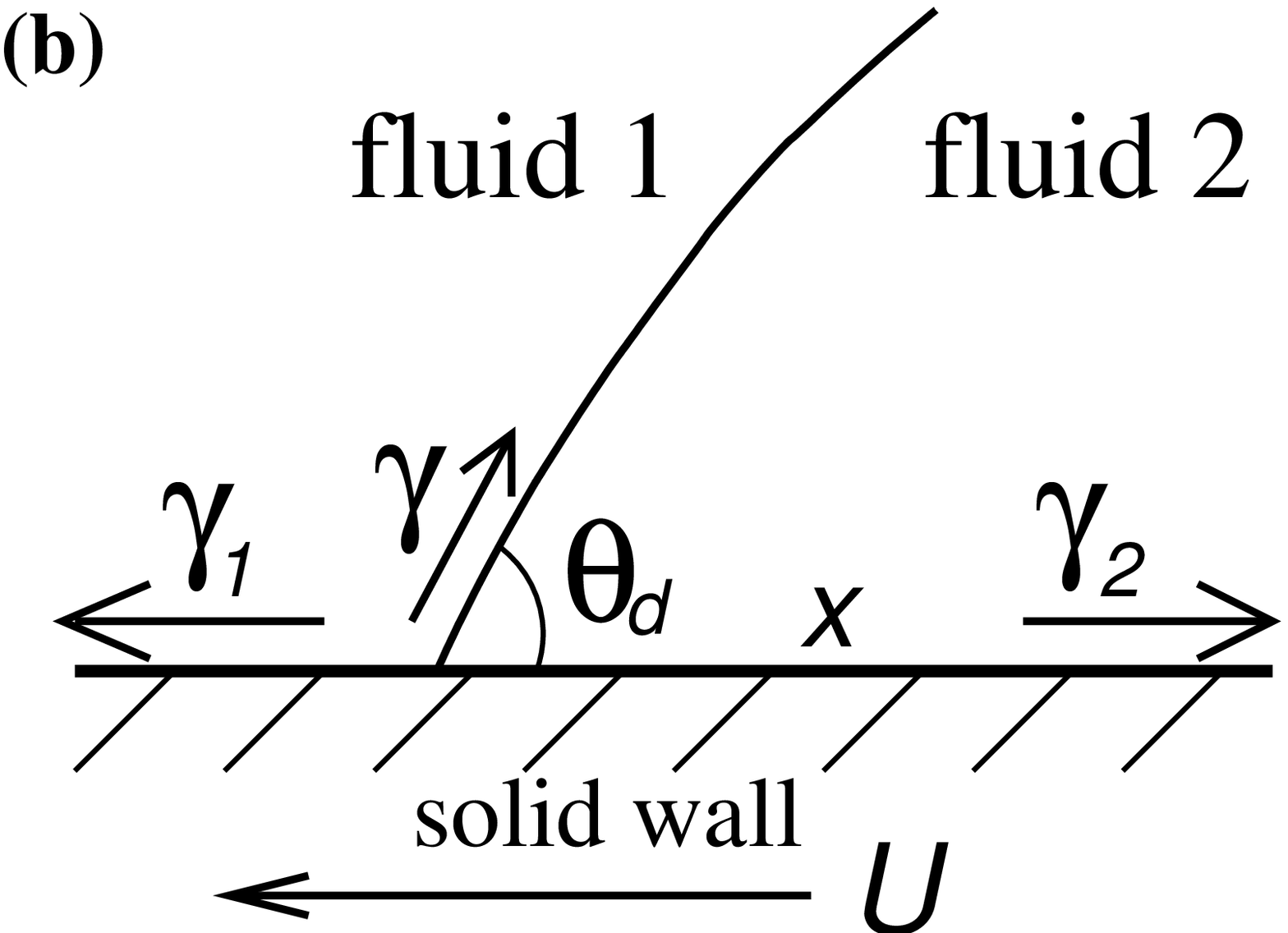,height=3.50cm}}
\caption{
(a) Static contact line. A fluid-fluid interface is formed between
two immiscible fluids and intersects the solid wall at a contact line.
The static contact angle $\theta_s$ is determined by the Young's equation:
$\gamma\cos\theta_s+\gamma_2-\gamma_1=0$, where $\gamma_1$, $\gamma_2$,
and $\gamma$ are the three interfacial tensions at the three interfaces
(two fluid-solid and one fluid-fluid).
(b) Moving contact line. When one fluid displaces another immiscible fluid,
the contact line is moving relative to the solid wall. 
(Here fluid 1 displaces fluid 2.) 
Due to the contact-line motion, the dynamic contact angle $\theta_d$
deviates from the static contact angle $\theta_s$. We will show that
this deviation is primarily responsible for the near-complete slip
at the contact line.
}\label{CL}
\end{figure}

It has been well known that the no-slip boundary condition is not 
applicable to the moving contact line (MCL) where the fluid-fluid interface 
intersects the solid wall \cite{hua-scriven,dussan,degennes} 
(see Fig. \ref{CL} for both the static and moving contact lines). 
The problem may be simply stated as follows.
In the two-phase immiscible flow where one fluid displaces another
fluid, the contact line appears to ``slip'' at the solid surface, in
direct violation of the no-slip boundary condition. Furthermore,
the viscous stress diverges at the contact line if the no-slip 
boundary condition is applied everywhere along the solid wall. 
This stress divergence is best illustrated in the reference frame where
the fluid-fluid interface is time-independent while the wall moves with
velocity $U$ (see Fig. \ref{CL}b). 
As the fluid velocity has to change from $U$ at the wall
(as required by the no-slip boundary condition) to zero at the fluid-fluid
interface (which is static), the viscous stress varies as $\eta U/x$, where
$\eta$ is the viscosity and $x$ is the distance along the wall away from 
the contact line. Obviously, this stress diverges as $x\rightarrow 0$
because the distance over which the fluid velocity changes from $U$ 
to zero tends to vanish as the contact line is approached.
In particular, this stress divergence is non-integrable (the integral
of $1/x$ yields $\ln x$), thus implying infinite viscous dissipation.

Over the years there have been numerous models and proposals aiming to 
resolve the incompatibility of the no-slip boundary condition with the MCL.
For example, there have been the kinetic adsorption/desorption model 
by Blake \cite{blake}, the fluid slip models by Hocking \cite{hocking},
by Huh and Mason \cite{hua-mason}, and by Zhou and Sheng \cite{sheng}, 
and the Cahn-Hilliard-van der Waals models by Seppecher \cite{seppecher},
by Jacqmin \cite{jacqmin}, and by Chen {\it et al.} \cite{vinal}.
In the kinetic adsorption/desorption model by Blake \cite{blake},
the role of molecular processes was investigated. A deviation of 
the dynamic contact angle from the static contact angle was shown to 
be responsible for the fluid/fluid displacement at the MCL.
In the slip model by Hocking \cite{hocking},  the effect of 
a slip coefficient (slip length) on the flow in the neighborhood of 
the contact line was examined.
Two slip models were used by Huh and Mason \cite{hua-mason}: 
Model I for classical slippage assumes the slip velocity is 
proportional to the shear stress exerted on the solid; Model II
for local slippage assumes that over a (small) distance the liquid slips 
freely where fluid stress vanishes, but thereafter the liquid/solid bonding 
has been completed and the no-slip boundary condition is applied.
In the slip model by Zhou and Sheng \cite{sheng}, the incompressible 
Navier-Stokes equation was solved using a prescribed tangential velocity 
profile as the boundary condition, which exponentially interpolates
between the complete-slip at the contact line and the no-slip
far from the contact line.
The Cahn-Hilliard-van der Waals models by Seppecher \cite{seppecher},
by Jacqmin \cite{jacqmin}, and by Chen {\it et al.} \cite{vinal} 
suggested a resolution when perfect no-slip is retained. 
With the fluid-fluid interface modeled to be diffuse, 
the contact line can thus move relative to the solid wall
through diffusion rather than convection.
All the above models are at least mathematically valid because 
the divergence of stress has been avoided, either by introducing 
some molecular process to drive the contact line \cite{blake}, 
or by allowing slip to occur \cite{hocking,hua-mason,sheng}, or 
by modeling a diffuse fluid-fluid interface \cite{seppecher,jacqmin,vinal}.
Pismen and Pomeau have presented a rational analysis of the hydrodynamic 
phase field (diffuse interface) model based on 
the lubrication approximation \cite{pismen-pomeau}.

The most usual (and natural) approach to resolve the stress divergence 
has been to allow slip at the solid wall close to the contact line.
In fact, molecular dynamics (MD) studies have clearly demonstrated 
the existence of fluid slipping in the molecular-scale
vicinity of the MCL \cite{koplik,robbins}.
However, the exact rule that governs this relative slip has eluded 
numerous prior attempts. As a matter of fact, none of the existing models 
has proved successful by 
quantitatively accounting for the contact-line slip velocity profile 
observed in MD simulations.
In a hybrid approach by Hadjiconstantinou \cite{hadji1}, the MD slip
profile was used as the boundary condition for finite-element
continuum calculation. The continuum results so obtained match 
the corresponding MD results, therefore demonstrating the feasibility of 
hybrid algorithm \cite{hadji2,weiqing}. 
But the problem concerning the boundary condition governing the
contact-line motion was still left unsolved.  

In this paper we first present an introductory review of the problem, 
including (1) the origin of the stress singularity, (2) the ad hoc
introduction of the slip boundary condition, (3) the MD evidence of 
fluid slipping, (4) the gap between the existing MD results and 
a continuum hydrodynamic description, 
and (5) a preliminary account on how to bridge the MD results and
a continuum description. We then present a detailed review of our 
recent results on the contact-line motion in immiscible two-phase flow, 
from MD simulations to continuum hydrodynamics calculations \cite{qws}. 
In our MD simulations, 
we consider two immiscible dense fluids of identical density and viscosity,
with the temperature controlled above the gas-liquid critical point.
(Similar results would be obtained if the temperature is below 
the critical point.) For fluid-solid interactions, we choose  
the wall density and interaction parameters to make sure (1) there is 
no epitaxial locking of fluid layer(s) to the solid wall, (2) a finite 
amount of slip is allowed in the single-phase flow for each of 
the two immiscible fluids, (3) the fluid-fluid interface makes 
a finite microscopic contact angle with the solid surface. 
Through extensive MD simulations and detailed analysis, 
we have uncovered the boundary condition governing the fluid slipping 
in the presence of a MCL. With the help of this discovery, we have formulated 
a continuum hydrodynamic model of two-phase immiscible flow. 
Numerical solutions have been obtained through an explicit finite-difference 
scheme. A comparison of the MD and continuum results shows that
velocity and fluid-fluid interface profiles from the MD simulations
can be quantitatively reproduced by the continuum model. 

The paper is organized as follows.
In Sec. \ref{review} we review a few known facts in mathematics and 
physics concerning the contact line motion. Together, they point out
the right direction to approach and elucidate the problem. In fact, they
almost tell us what is expected for a boundary condition governing
the slip at the MCL and in its vicinity.
In Sec. \ref{statement} we outline the main results from MD simulations
to continuum hydrodynamic modeling.
In Sec. \ref{md1} we present the first part of the MD results. From 
the slip velocity and the tangential wall force measured
at the fluid-solid interface, a slip boundary condition is deduced.
In Sec. \ref{continuum-hydrodynamics} we formulate a continuum hydrodynamic 
model of two-phase immiscible flow. The continuum differential 
expression for the tangential stress at the solid surface is derived, 
from which the generalized Navier boundary condition (GNBC) is obtained 
from the slip boundary condition deduced in Sec. \ref{md1}.
In Sec. \ref{comparison} we show a systematic comparison of the MD and 
continuum results. The validity of the GNBC and the continuum model
is demonstrated.
In Sec. \ref{md2} we present the second part of the MD results, concerning
the tangential force balance in a boundary layer at the fluid-solid interface
and the decomposition of the tangential stress in the fluid-fluid interfacial 
region. 
In Sec. \ref{gnbc-derivation} we establish the correspondence between 
the stress components measured in MD and those defined in the continuum 
hydrodynamics.
Based on this correspondence, the continuum GNBC is obtained in 
an integrated form from the MD results in Secs. \ref{md1} and \ref{md2}. 
This may be regarded as a direct MD verification of the continuum GNBC. 
It also justifies the use of the Cahn-Hilliard hydrodynamic formulation of 
two-phase flow, from which the continuum form of GNBC, as verified by 
the MD results, is derived.
The paper is concluded in Sec. \ref{remarks}.

\section{Stress and Slip: A Brief Review}\label{review}

\subsection{Non-integrable stress singularity: the Huh-Scriven model}
\label{HS-model}

\begin{figure}[h]
\bigskip
\centerline{\psfig{figure=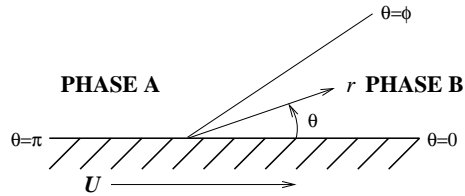,height=2.50cm}}
\caption{ The Hua-Scriven model.
}\label{singularity}
\end{figure}

Hua and Scriven \cite{hua-scriven} considered a simple model
for the immiscible two-phase flow near a MCL.
A flat solid surface is moving with steady velocity $U$ in its
own plane and a flat fluid-fluid interface, formed between
two immiscible phases A and B, intersects the solid surface at
a contact line (see Fig. \ref{singularity}). The contact line
is taken as the origin of a polar coordinate system $(r,\theta)$,
in which the contact angle is $\phi$.

The two-dimensional flow of Newtonian and incompressible fluids
is dominated by viscous stress for $r\ll \eta/\rho U$, 
where $\eta$ is the viscosity and $\rho$ the mass density.
In the viscous flow approximation, the Navier-Stokes equation is linearized
and the steady flow is solved from the biharmonic equation for 
the stream function $\psi(r,\theta)$:
$$\nabla^2(\nabla^2\psi)=0.$$
The boundary conditions to be imposed at $\theta=0,\pi$ (solid surface)
and $\theta=\phi$ (fluid-fluid interface) are:
(i) vanishing normal component of velocity at the solid surface
and fluid-fluid interface,
(ii) continuity of tangential velocity at the fluid-fluid interface,
(iii) continuity of tangential viscous stress at the fluid-fluid interface,
(iv) no slip at the solid surface.
Here conditions (i)--(iii) are well justified while (iv) is usually
taken as a postulate of continuum hydrodynamics.
The no-slip boundary condition can be justified {\it a posteriori}
in macroscopic flow by checking the correctness of its consequences.
In the present model, however, it leads to physically incorrect results
for stress, in the microscopic vicinity of the MCL.

The similarity solution of the biharmonic equation is in the form of
$$\psi(r,\theta)=r(a\sin\theta+b\cos\theta+
c\theta\sin\theta+d\theta\cos\theta),$$
in which the eight constants (4 for phase A and 4 for phase B)
are determined by the eight boundary conditions in (i)--(iv).
What Hua and Scriven found is that the shear stress and pressure fields
vary as $1/r$ and hence increase in magnitude without limit as 
the contact line $r=0$ is approached. As a consequence, 
the total tangential force exerted on the solid surface, which is
an integral of the tangential stress along the surface, is logarithmically
infinite. That indicates a singularity in viscous dissipation, which is
physically unacceptable. 

Obviously, the Hua-Scriven model is defective in the immediate vicinity
of the MCL. This is also seen from the normal stress difference across 
the fluid-fluid interface, which varies as $1/r$ as well. According to
the Laplace's equation, the interface curvature should increase rapidly
as the contact line ($r=0$) is approached, in order to balance 
the normal stress difference by curvature force. This is clearly
inconsistent with the assumption of a flat fluid-fluid interface.
Nevertheless, the flow field solved from the Hua-Scriven model may 
approximately describe the asymptotic region at a large distance 
from the contact line (where the viscous flow approximation is still valid).
In that region, the no-slip boundary condition is considered valid
and the fluid-fluid interface is almost flat, due to the reduced stress.

\subsection{Introducing the slip boundary condition}\label{dussan-sensitivity}

The deficiency of the Hua-Scriven model results from the incompatibility
of the no-slip boundary condition with a MCL: no slip means
$v_r=\pm U$ at the solid surface $(\theta=0,\pi)$ where $r>0$
while at $r=0$ the MCL requires a perfect slip. That is, 
the no-slip boundary condition leads to a velocity discontinuity at the MCL.
In order to remove the stress singularity at the MCL,
while retaining the Newtonian behavior of stress, the continuity of 
velocity field must be restored. For this purpose, a slip profile
can be introduced to continuously interpolate between the complete slip 
at the MCL and the no-slip boundary condition that must hold at regions 
far away. 

\begin{figure}[h]
\bigskip
\centerline{\psfig{figure=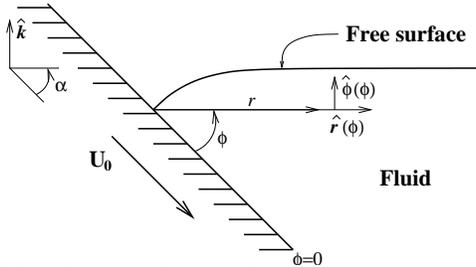,height=3.50cm}}
\caption{ A plate is inserted into a semi-infinite domain of fluid 
at a constant velocity ${\bf U}_0$. The angle between the plate and 
the free surface at $r\rightarrow \infty$ is $\alpha$.
}\label{free-surface}
\end{figure}

Dussan V. \cite{sensitivity} considered a plate of infinite extent
either inserted into or withdrawn from a semi-infinite domain of fluid 
at a constant velocity ${\bf U}_0$ (see Fig. \ref{free-surface}).
The contact line is taken as the origin of a polar coordinate system 
$(r,\phi)$, in which the apparent contact angle is $\alpha$ at
$r\rightarrow \infty$. The equation of motion is 
the Navier-Stokes equation with the incompressibility condition 
$\nabla\cdot{\bf u}=0$. The boundary conditions at the solid surface
$\phi=0$ and the free surface $\{(r,\phi(r)\}|0\le r<\infty\}$ are:
(i) the kinematic boundary conditions
${\bf u}\cdot\hat{\mbox{\boldmath$\phi$}}=0$ at $\phi=0$ and 
${\bf u}\cdot{\bf n}=0$ at $\phi=\phi(r)$, where $\bf n$ is an outward
unit vector normal to the free surface,
(ii) the dynamic boundary condition 
${\bf t}\cdot{\bf T}\cdot{\bf n}=0$ at $\phi=\phi(r)$, where
$\bf T$ is the stress tensor and ${\bf t}$ is a unit vector
tangent to the free surface, i.e., ${\bf t}\cdot{\bf n}=0$,
(iii) the Laplace condition
${\bf n}\cdot{\bf T}\cdot{\bf n}=\sigma\kappa$ at $\phi=\phi(r)$, where
$\sigma$ is the surface tension and $\kappa$ the interface curvature,
(iv) the slip boundary condition
${\bf u}=U(r)\hat{\bf r}$ at $\phi=0$, where $U(r)$ is a {\it prescribed}
function, and (v) the critical static contact angle $\phi(0)=\phi_s$.

The slip boundary condition must continuously interpolate
between the complete slip at the MCL ($r=0$)
and the no-slip boundary condition at $r\rightarrow\infty$:
$$\lim_{r\rightarrow 0} U(r)=0,\;\;\;
\;\;\;\lim_{r\rightarrow \infty} U(r)=U_0.$$
However, the form of $U(r)$ in the intermediate region, where $U$
varies from $0$ to $U_0$, is unknown. Three different slip
boundary conditions were used for $U(r)$, in order to assess
the sensitivity of the overall flow field to the form of 
the slip boundary condition. They are
$$U_1=\displaystyle\frac{r/L_s}{1+r/L_s}U_0,\;\;\;
U_2=\displaystyle\frac{(r/L_s)^2}{1+(r/L_s)^2}U_0,\;\;\;
U_1=\displaystyle\frac{(r/L_s)^{1/2}}{1+(r/L_s)^{1/2}}U_0.$$
where $L_s$ is a length scale called the slip length 
(not the one defined in the Navier boundary condition).
It was found that on the slip length scale the flow fields are quite
different whereas on the meniscus length scale, i.e., the length scale
on which almost all fluid-mechanical measurements are performed, 
all the flow fields are virtually the same. 
That is, identical macroscopic flow behaviors are expected from 
different slipping models.

\subsection{Slip observed in molecular dynamics simulations}
\label{slip-md}

According to the conclusion in Ref. \cite{sensitivity}, only in 
the microscopic slip region (of length scale $\sim L_s$) can
different slip models be distinguished. This makes the experimental
verification of a particular slip model very difficult because
experiments usually probe distances ($\sim 1\mu{\rm m}$) much larger
than $L_s$. Naturally, computer experiments become very useful
in elucidating the problem \cite{md-book}. 

\begin{figure}[h]
\bigskip
\centerline{\psfig{figure=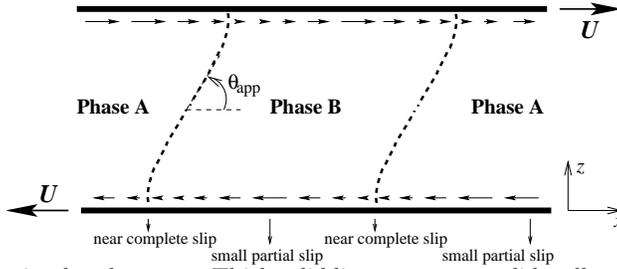,height=3.50cm}}
\caption{ The $xz$ projection of the simulated system.
Thick solid lines represent solid walls and dashed lines represent
fluid-fluid interfaces. The arrows indicate fluid velocity close to 
the solid walls, from which the variation of the amount of slip
is clearly seen.
}\label{couette-robbins}
\end{figure}

Non-equilibrium MD simulations were carried out to
investigate the fluid motion in the vicinity of the MCL,
in both the Poiseuille-flow \cite{koplik} and Couette-flow
\cite{robbins} geometries. In these MD simulations, interactions between
fluid molecules were modeled with the Lennard-Jones potential, modified
to segregate immiscible fluids. The confining walls were constructed
with a molecular structure. Wall-fluid interactions were also
modeled with the Lennard-Jones potential, with energy and length scales 
different from those of the fluid-fluid interactions.
In the simulations performed in the Couette geometry \cite{robbins},
two immiscible fluids were confined between two planar walls
parallel to the $xy$ plane and a shear flow was induced by moving 
the two solid walls in $\pm x$ directions at constant speed $U$
(see Fig. \ref{couette-robbins}). Steady-state velocity fields 
were obtained from the time average of fluid molecular velocities
in small bins. 

There was clean evidence for a slip region 
in the vicinity of the MCL: appreciable slip occurs within 
a length scale $\sim 10\sigma$, where $\sigma$ is the length scale in 
the fluid-fluid Lennard-Jones potential, and at the MCL the slip is
near-complete. Far from the interfaces, viscous damping makes the flow
insensitive to the fast variations near the interfaces, and hence
uniform shear flow was observed, with a negligible amount of slip.
Therefore, MD simulations provide evidence for a cutoff, below which
the no-slip boundary condition breaks down, introduced phenomenologically  
in various slip models to remove the stress singularity.
However, the exact boundary condition that governs the observed slip
was left unresolved. In particular, a breakdown of local hydrodynamics
in the molecular scale slip region was suggested \cite{robbins},
considering the extreme velocity variations there.

The Navier slip model assumes that the amount of slip is proportional
to the tangential component of the stress tensor, $P_{xz}$, in the fluid
at the solid surface.
In Ref. \cite{robbins}, the microscopic value of $P_{xz}$ was directly 
measured. A comparison to the slip profile is roughly consistent
with the Navier slip model. However, a large discrepancy was found 
between the microscopic value of $P_{xz}$ and the shear rate
$\partial v_x/\partial z$. According to the authors,
this discrepancy suggests a breakdown of local hydrodynamics.
On the contrary, we will show in this paper: \\
(i) The Navier slip model is the correct model describing 
the fluid slipping near the MCL.\\
(ii) The tangential stress in the Navier model is not merely viscous.\\
(iii) The interfacial tension plays an important role in 
governing the contact-line slip in immiscible two-phase flows.\\
(iv) There is no breakdown of local hydrodynamics.

\subsection{Fluid/fluid displacement driven by unbalanced Young force}
\label{displacement}
\subsubsection{Unbalanced Young force}\label{unYf}

For a cylindrical capillary of radius $r$, if the steady displacement
is sufficiently slow, then the pressure drop across the moving fluid-fluid
interface is given by $\Delta p=2\gamma_{12}\cos\theta/r$, 
where $\gamma_{12}$ is the interfacial tension, and $\theta$ is 
an appropriate dynamic contact angle. At equilibrium, the pressure drop 
is given by $\Delta p^0=2\gamma_{12}\cos\theta^0/r$, where $\theta^0$
is the equilibrium contact angle. In general, $\theta$ and $\theta_0$
differ. Physically, $\Delta p^0$ measures the change of the interfacial
free energy at the fluid-solid interface when the fluid-fluid interface
moves relative to the wall, and $\Delta p$ measures the external work
done to the system. Therefore, the difference $\Delta p-\Delta p^0$
is a measure of the dissipation due to the presence of the MCL.

\begin{figure}[h]
\bigskip
\centerline{\psfig{figure=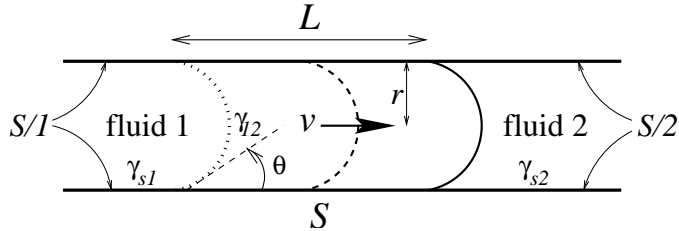,height=3.0cm}}
\caption{A displacement of fluid 2 by fluid 1 over distance $L$
in a cylindrical capillary of radius $r$.
}\label{displace}
\end{figure}

Let's consider a displacement of fluid 2 by fluid 1 over distance $L$
(see Fig. \ref{displace}). 
According to the Young's equation for the equilibrium contact angle, 
the force $\pi r^2\Delta p^0$ equals $2\pi r\gamma_{12}\cos\theta^0
=2\pi r(\gamma_{S1}-\gamma_{S2})$, where $\gamma_{S1}$ and $\gamma_{S2}$
are the interfacial tensions for the $S/1$ and $S/2$ interfaces, respectively.
Thus the change of the interfacial free energy at the fluid-solid interface
is given by $\pi r^2\Delta p^0L=2\pi rL(\gamma_{S1}-\gamma_{S2})$. 
The external work done to the system is simply $\pi r^2\Delta pL$. 
It follows that the dissipation due to the presence of the MCL is given by
$\pi r^2(\Delta p-\Delta p^0)L=2\pi rL\gamma_{12}
(\cos\theta-\cos\theta^0)$, where $\gamma_{12}(\cos\theta-\cos\theta^0)=
\gamma_{12}\cos\theta+\gamma_{S2}-\gamma_{S1}$ is the {\it unbalanced
Young force} \cite{degennes}.

In order to find a relation between the displacement velocity
$v$ and the unbalanced Young force $F_Y$, two classes of models have
been proposed to describe the contact line motion: a) An Eyring approach, 
based on the molecular adsorption/desorption processes at the contact line 
\cite{blake}. 
b) A hydrodynamic approach, assuming that dissipation is dominated by 
viscous shear flow inside the wedge \cite{degennes}. Viscous flows 
in wedges were calculated by Hua and Scriven \cite{hua-scriven}. 
For wedges of small (apparent) contact angle, a lubrication approximation
can be used to simplify the calculations \cite{degennes}. As discussed in 
Secs. \ref{HS-model} and \ref{dussan-sensitivity}, a (molecular scale) 
cutoff has to be introduced to remove the logarithmic singularity in 
viscous dissipation. 
On the contrary, the Eyring approach assumes that the molecular dissipation
at the tip is dominant. A quantitative theory is briefly reviewed below.

\subsubsection{Blake's kinetic model}\label{kinetic-model}

\begin{figure}[h]
\bigskip
\centerline{\psfig{figure=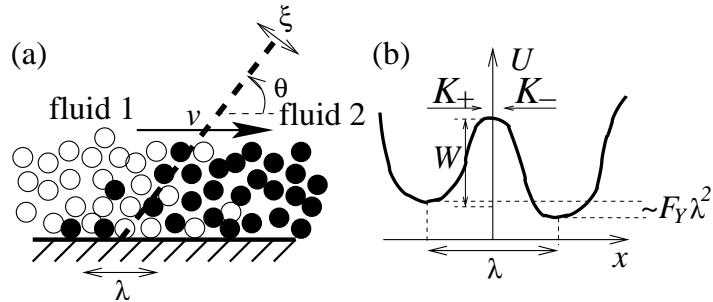,height=4.0cm}}
\caption{(a) A molecular picture of the three-phase zone.
(b) Shifted potential profile.
}\label{kinetic}
\end{figure}

The role of interfacial tension was investigated in a kinetic model
by Blake and Haynes \cite{blake}. Consider a fluid-fluid interface
in contact with a flat solid surface at a line of three-phase contact
(see Fig. \ref{kinetic}a). Viewed on a molecular scale, 
the three-phase line is actually a fluctuating three-phase zone, 
where adsorbed molecules of one fluid (at the solid surface) interchange 
with those of the other, either by migration at the solid surface or 
through the contiguous bulk phases. 
In equilibrium the net rate of exchange will be zero.

For a three-phase zone moving relative to the solid wall 
(Fig. \ref{kinetic}a),
the net displacement, driven by the unbalanced Young force 
$F_Y=\gamma_{12}\cos\theta+\gamma_{S2}-\gamma_{S1}$, is due to
a nonzero net rate of exchange.
Let $\xi$ be the interfacial thickness, $\sigma$ be the area of 
an adsorption site, and $\lambda$ be the hopping distance of molecules.
The force per adsorption site is approximately $\sigma F_Y/\xi$ and 
the energy shift over distance $\lambda$ is approximately
$\lambda\sigma F_Y/\xi\sim F_Y\lambda^2$ (see Fig. \ref{kinetic}b). 
This energy shift leads to two different rates $K_+$ and $K_-$:
$$K_{\pm}=k\exp\left[-\displaystyle\frac{1}{k_BT}\left(W\mp
\displaystyle\frac{1}{2}F_Y\lambda^2\right)\right],$$
for forward and backward hopping events, respectively.
Here $W$ is an activation energy for hopping. It follows that
the velocity of the MCL is $v=\lambda(K_+-K_-)\propto
\sinh\left({F_Y\lambda^2}/{2k_BT}\right)$. For very small $F_Y$,
$v\propto F_Y$.

Blake's kinetic model shows that fluid slipping can be induced
by the unbalanced Young force at the contact line. Therefore,
it emphasizes the role of interfacial tension, 
though not in a hydrodynamic formulation. On the contrary, 
the authors of Ref. \cite{robbins} considered the viscous shear stress
as the only driving force. In fact a large discrepancy was found 
between the shear rate and the microscopic value of the tangential stress 
(the driving force according to the Navier slip model).
In this paper we will show that in the two-phase interfacial region,
such a discrepancy is exactly caused by the neglect of 
the non-viscous contribution from interfacial tension.

\subsection{From the Navier boundary condition to the generalized
Navier boundary condition: a preliminary discussion}\label{primer}

Here we give a preliminary account on the main finding reported in 
Ref. \cite{qws}, the GNBC. 
Based on the results reviewed in Secs. \ref{HS-model}, 
\ref{dussan-sensitivity}, \ref{slip-md}, and \ref{displacement}, 
we try to show: a) The Navier boundary condition may actually work
for the single-phase slip region near the MCL.
b) In the two-phase contact-line region, 
the GNBC is a natural extension of the Navier boundary condition, 
with the fluid-fluid interfacial tension taken into account. 

\subsubsection{Navier boundary condition and slip length}\label{micro-nbc}

The validity of the Navier boundary condition
${\bf v}^{slip}=-l_s{\bf n}\cdot
\left[(\nabla {\bf v})+(\nabla {\bf v})^T\right]$ 
has been well established by many MD studies for single-phase fluids
\cite{Th-Rob,nbc,barrat,ckb}.
This boundary condition is a constitutive equation
that {\it locally} relates the amount of slip to the shear rate 
at the solid surface, though in most of the reported simulations, 
the hydrodynamics involves no spatial variation along the solid surface.
Physically, the Navier boundary condition is local in nature simply 
because intermolecular wall-fluid interactions are short-ranged.

\begin{figure}[h]
\bigskip
\centerline{\psfig{figure=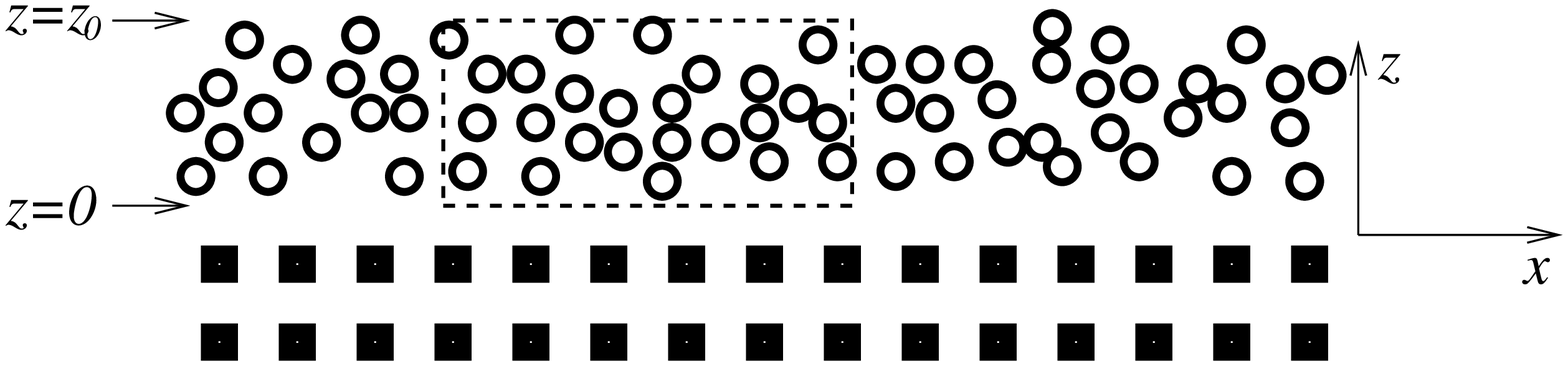,height=2.0cm}}
\caption{A boundary layer of fluid at the fluid-solid interface,
responsible for viscous momentum transport between fluid and solid.
Circles represent fluid molecules and solid squares represent wall molecules.
}\label{boundary}
\end{figure}

The slip length $l_s$ is a phenomenological parameter that measures
the local viscous coupling between fluid and solid.
Figure \ref{boundary} illustrates the viscous momentum transport 
between fluid and solid through a boundary layer of fluid.
The thickness of this boundary layer, $z_0$, must be of molecular scale, 
within the range of wall-fluid interactions. Now we show that 
the slip length $l_s$ is defined based on a linear law for 
tangential wall force and the Newton's law for shear stress.

Hydrodynamic motion of fluid at the solid surface is measured by 
the slip velocity $v_x^{slip}$ in the $x$ direction, 
defined relative to the (moving) wall. When $v_x^{slip}$ is present, 
a tangential wall force $G_x^w$ is exerted on the boundary-layer fluid, 
defined as the rate of tangential ($x$) momentum 
transport per unit wall area, from the wall to the (boundary-layer) fluid.
Physically, this force represents a time-averaged effect of
wall-fluid interactions, to be incorporated into a hydrodynamic
slip boundary condition. The linear law for $G_x^w$ is expressed by
$G_x^w=-\beta v^{slip}$, where $\beta$ is called the slip coefficient
and the minus sign means the fluid-solid coupling is viscous (frictional).
For the boundary layer of molecular thickness, the tangential wall force
$G_x^w$ must be balanced by the tangential fluid force 
$G_x^f=\int_0^{z_0}dz
\left(\partial_x\sigma_{xx}+\partial_z\sigma_{zx}\right)$, where
the integral is over the $z$-span of the boundary layer 
(see Fig. \ref{boundary}), $\sigma_{xx(zx)}$ are
the $xx(zx)$ components of the fluid stress tensor, and 
$\partial_x\sigma_{xx}+\partial_z\sigma_{zx}$ is the fluid force density
in the $x$ direction. From the equation of tangential force balance
$G_x^w+G_x^f=0$, we obtain 
$$\int_0^{z_0}dz\left(\partial_x\sigma_{xx}+\partial_z\sigma_{zx}\right)
=\partial_x\int_0^{z_0}dz\sigma_{xx}(x,z)+\sigma_{zx}(x,z_0)=
\beta v^{slip}(x).$$
This equation should be regarded as a microscopic expression for
the Navier slip model: the amount of slip is proportional to the tangential
fluid force at the fluid-solid interface.
When the normal stress $\sigma_{xx}$ varies slowly in the $x$ direction
and the tangential stress $\sigma_{zx}$ is caused by shear viscosity only,
the above equation becomes 
$\eta\partial_z v_x(x,z_0)=\beta v^{slip}(x)$,
where $\eta$ is the viscosity and $\partial_z v_x(x,z_0)$ is the shear rate
``at the solid surface''. For flow fields whose characteristic length scale
is much larger than $z_0$, the boundary layer thickness, we replace
$z_0$ by $z=0$ and recover the Navier boundary condition 
for continuum hydrodynamics, in which the slip length is given by 
$l_s=\eta/\beta$. 
To summarize, the hydrodynamic viscous coupling between fluid and solid 
is actually measured by the slip coefficient $\beta$. 
The Navier boundary condition, in which the slip length is introduced, 
is due to the Newton's law for viscous shear stress. 
For very weak viscous coupling between fluid and solid, 
$\beta\rightarrow 0$, and thus $l_s\rightarrow\infty$, making
$\partial_z v_x\rightarrow 0$: the fluid-solid interface becomes
a free surface.

\subsubsection{Single-phase region}\label{gnbc-I}

A number of MD studies have shown that for single-phase flows,
the Navier boundary condition is valid in describing the fluid slipping 
at solid surface \cite{Th-Rob,nbc,barrat,ckb}.
Therefore, we expect that it can also describe the partial slip
observed in the single-phase region near the MCL \cite{robbins}.
However, according to the authors of Ref. \cite{robbins},
the Navier boundary condition failed even in the single-phase region:
the velocity gradient $\partial v_x/\partial z$ was not proportional
to the amount of slip. This discrepancy was attributed to 
a breakdown of local hydrodynamics, considering the very large velocity
variations observed near the MCL. Here we present a heuristic discussion,
to explain why such a discrepancy is expected even if 
the Navier boundary condition actually works for the single-phase region.

First, the success of the hybrid approaches outlined below strongly
indicates that local hydrodynamics doesn't break down in the slip region.
In Ref. \cite{robbins}, an apparent contact angle $\theta_{app}$ was 
defined at half the distance between two solid surfaces 
(see Fig. \ref{couette-robbins}).
For $\theta_{app}<135^\circ$, fluid-fluid interfaces could be approximated 
by planes. The Navier-Stokes equation was solved for the simplified geometry.
For this purpose, the tangential velocity along fluid-fluid interface 
was set to be zero (according to the simulation results) and 
the slip velocity $\Delta V(r)$ (measured relative to the moving wall) 
was specified as a function of distance $r$ from the MCL: 
$\Delta V(r)=\pm U\exp[-r\ln 2/S]$, with $+$ and $-$ for the lower and upper
walls, respectively.
This $\Delta V(r)$, proposed according to the MD slip profile, 
continuously interpolates between the complete slip at the MCL 
($\Delta V(r)=\pm U$ at $r=0$) and the no-slip boundary condition
($\Delta V(r)\rightarrow 0$ at $r\rightarrow\infty$). 
With a proper value chosen for $S$ ($\approx 1.8\sigma$), 
the MD flow fields were reproduced by solving the Navier-Stokes equation 
with the above boundary conditions. Recently, an improved hybrid approach 
has been used to reproduce the MD simulation results for contact-line motion 
in a Poiseuille geometry \cite{hadji1}.

\begin{figure}[h]
\bigskip
\centerline{\psfig{figure=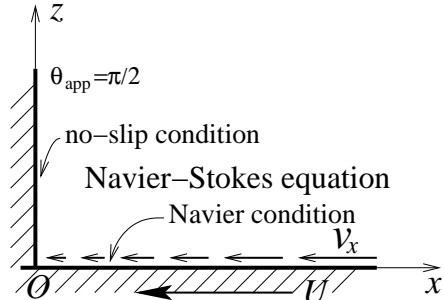,height=4.0cm}}
\caption{Two-dimensional corner flow caused by one rigid plane 
sliding over another, with constant inclination $\theta_{app}=90^\circ$. 
In the reference frame of the vertical plane,
the horizontal plane is moving to the left with constant speed $U$.
}\label{corner-flow}
\end{figure}

To further test the validity of continuum modeling,
we have solved the Navier-Stokes equation for a corner flow \cite{qian-wang}.
Consider one rigid plane sliding steadily over another, with 
constant inclination $\theta_{app}$ (see Fig. \ref{corner-flow}). 
The fluid is Newtonian and incompressible. The no-slip boundary condition 
is applied at the vertical plane and the Navier boundary condition
applied at the horizontal plane. The kinematic condition of vanishing
normal component of velocity at the solid surface requires
$v_x=0$ at the vertical plane and $v_z=0$ at the horizontal plane, and
hence ${\bf v}=0$ at the intersection $O$, which is taken as the origin of
the coordinate system $(x,z)$. This corner-flow model resembles
the continuum model used in the hybrid approach above, and produces similar
flow fields for quantitative comparison with MD results. In particular,
the corner flow exhibits a slip profile very close to $\Delta V(r)$:
the slip velocity $v_x+U$ shows a linear decrease over a length scale
$\sim l_s$, the slip length in the Navier condition, followed by 
a more gradual decrease. (Note that $v_x+U=U$ at $O$ implies complete slip.)

From the hybrid approach with the prescribed slip velocity $\Delta V(r)$
to the corner-flow model with the Navier boundary condition, 
they indicate that the single-phase region near the MCL can be modeled 
by the Navier-Stokes equation for an incompressible Newtonian fluid,
supplemented by appropriate boundary conditions. 
Then a simple question arises: Given a continuum hydrodynamic model
that uses the Navier boundary condition and approximately reproduces 
the MD flow fields in the single-phase region near the MCL, why do the MD
simulation results show a large discrepancy between the velocity gradient 
$\partial v_x/\partial z$ and the amount of slip?

The answer lies in the fast velocity variation in the vicinity of the MCL,
where the flow is dominated by viscous stress. With the characteristic 
velocity scale set by $U$ and the characteristic length scale set by $l_s$, 
the normal stress $\sigma_{xx}$ must show as well a fast variation along 
the solid surface: $\partial_x\sigma_{xx}\sim \eta U/l_s^2$. 
According to Sec. \ref{micro-nbc}, the microscopic value of 
the tangential fluid force $G_x^f$ is given by 
$\partial_x\int_0^{z_0}dz\sigma_{xx}(x,z)+\sigma_{zx}(x,z_0)$.
This expression is necessary because $G_x^f$ is distributed in 
a boundary layer of molecular thickness $z_0$. Obviously,
to represent $G_x^f$ by $\sigma_{zx}\approx\eta\partial v_x/\partial z$ 
at $z=z_0$ alone, the normal stress $\sigma_{xx}$ near the solid surface
has to vary slowly in the $x$ direction. This is not the case 
in the vicinity of the MCL if the Navier slip model works there,
as implied by the continuum corner-flow calculation which yields
$\partial_x\sigma_{xx}\sim \eta U/l_s^2$ near the intersection.
(It is reasonable to expect that if the Navier slip model is valid,
then the normal stress measured in MD simulations should in general
agree with that from continuum model calculation with the Navier condition.)
Given $\partial_x\int_0^{z_0}dz\sigma_{xx}\sim \eta Uz_0/l_s^2$ according
to the corner-flow model, and that $z_0$ and $l_s$ are both $\sim 1\sigma$,
it is obvious that considering only $\eta\partial v_x/\partial z$ at $z=z_0$ 
would lead to an appreciable underestimate of $G_x^f\sim \eta U/l_s$ 
in the slip region. In short, the large discrepancy between 
the velocity gradient $\partial v_x/\partial z$ and the amount of slip,
observed in the single-phase slip region near the MCL, cannot be used
to exclude the microscopic Navier slip model and the hydrodynamic 
Navier slip boundary condition, for if the Navier model is valid, then
the tangential viscous stress $\eta\partial v_x/\partial z$, measured 
at some level away from the solid surface, is not enough for a complete evaluation of the tangential fluid force $G_x^f$.

\subsubsection{Two-phase region}\label{gnbc-II}

The linear law for tangential wall force, $G_x^w=-\beta v_x^{slip}$,
describes the hydrodynamic viscous coupling between fluid and solid.
Assume that in the two-phase region, the two fluids interact with 
the wall independently. Then the tangential wall force becomes
$G_x^w=-\beta v_x^{slip}$ at the contact line, with the slip coefficient
given by $\beta=(\beta_1\rho_1+\beta_2\rho_2)/(\rho_1+\rho_2)$,
where $\beta_1$ and $\beta_2$ are the slip coefficients for the two
single-phase regions separated by the fluid-fluid interface, and
$\rho_1$ and $\rho_2$ are the local densities of the two fluids in 
the contact-line region. Obviously, $\beta$ varies between $\beta_1$
and $\beta_2$ across the fluid-fluid interface.

The equation of tangential force balance $G_x^w+G_x^f=0$ must hold
as well in the two-phase region. Therefore the Navier slip model is
still of the form
$$\partial_x\int_0^{z_0}dz\sigma_{xx}(x,z)+\sigma_{zx}(x,z_0)=
\beta v^{slip}(x).$$
Nevertheless, this does not lead to the Navier boundary condition
$\eta\partial_zv_x=\beta v^{slip}$ anymore because in the two-phase
region, the tangential stress is not contributed by shear viscosity only:
there is a non-viscous component in $\sigma_{zx}$, caused by 
the fluid-fluid interfacial tension. Put in an ideal form of decomposition,
the tangential stress $\sigma_{zx}(x,z)$ at level $z$ can be
expressed as
$$\sigma_{zx}(x,z)=\sigma_{zx}^v(x,z)+\sigma_{zx}^Y(x,z),$$
where $\sigma_{zx}^v$ is the viscous component due to shear viscosity
and $\sigma_{zx}^Y$ (the tangential Young stress) is the non-viscous 
component, which is narrowly distributed in the two-phase interfacial region
and related to the interfacial tension through 
$\int_{int}dx\sigma_{zx}^Y(x,z)=\gamma\cos\theta(z)$. 
Here $\int_{int}dx$ denotes the integration across
the fluid-fluid interface along the $x$ direction, $\gamma$ is 
the interfacial tension, and $\theta(z)$ is the angle between 
the interface and the $x$ direction at level $z$. Physically,
the existence of a fluid-fluid interface causes an anisotropy
in the stress tensor in the two-phase interfacial region.
The interfacial tension is an integrated measure of that stress anisotropy.
When expressed in a coordinate system different from the principal system,
the stress tensor is not diagonalized and a nonzero $\sigma_{zx}^Y$ appears.
In the presence of shear flow, while the shear viscosity leads to
the viscous component $\sigma_{zx}^v$ in $\sigma_{zx}$,
the non-viscous component $\sigma_{zx}^Y$ is still present.
A detailed discussion on the stress decomposition will be given in 
Sec. \ref{stress-decomposition}.

Consider an equilibrium state in which the tangential stress 
$\sigma_{zx}^0$ is balanced by the gradient of 
the normal stress $\sigma_{xx}^0$:
$$\partial_x\int_0^{z_0}dz\sigma_{xx}^0(x,z)+\sigma_{zx}^0(x,z_0)=0,$$
where the superscript $0$ denotes equilibrium quantities.
Here the equilibrium tangential stress $\sigma_{zx}^0$
is narrowly distributed in the two-phase interfacial region
and related to the interfacial tension through 
$\int_{int}dx\sigma_{zx}^0(x,z)=\gamma\cos\theta^0(z)$.
(There is no viscous stress in equilibrium, and hence $\sigma_{zx}^0$
is caused by the interfacial tension only.)
Subtracting the equation of equilibrium force balance from 
the expression of Navier slip model, we obtain
$$\begin{array}{ll}
& \partial_x\int_0^{z_0}dz[\sigma_{xx}(x,z)-\sigma_{xx}^0(x,z)]+
[\sigma_{zx}(x,z_0)-\sigma_{zx}^0(x,z_0)]\\
=& \partial_x\int_0^{z_0}dz[\sigma_{xx}(x,z)-\sigma_{xx}^0(x,z)]+
[\sigma_{zx}^v(x,z_0)+\sigma_{zx}^Y(x,z_0)-\sigma_{zx}^0(x,z_0)]\\
=&\beta v^{slip}(x).\end{array}$$
This equation will be the focus of our continuum deduction from 
molecular hydrodynamics. In fact it leads to the GNBC which governs
the fluid slipping everywhere, from the two-phase contact-line region to 
the single-phase regions away from the MCL. 
This will be elaborated in Secs. \ref{md1}, \ref{continuum-hydrodynamics},
\ref{md2}, and \ref{gnbc-derivation}. 

To summarize, our preliminary analysis shows that compared to 
the single-phase region where the tangential stress is 
due to shear viscosity only, the two-phase region has 
the tangential Young stress as the additional component.
Naturally, the Navier boundary condition, which considers
the tangential viscous stress only, needs to be generalized to include 
the tangential Young stress.

\section{Statement of Results}\label{statement}

\begin{figure}[h]
\bigskip
\centerline{\psfig{figure=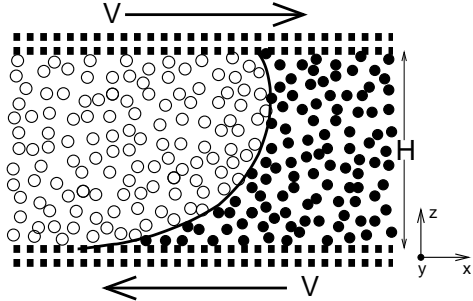,height=4.0cm}}
\caption{Schematic of the immiscible Couette flow. 
Two immiscible fluids (empty and solid circles) are confined between 
two solid walls (solid squares) that are parallel to the $xy$ plane
and separated by a distance $H$. The Couette flow is generated by moving 
the top and bottom walls at a speed $V$ along $\pm x$, respectively. 
}\label{couette}
\end{figure}

We have carried out MD simulations for immiscible two-phase flows
in a Couette geometry (see Fig. \ref{couette}) \cite{qws}. The two 
immiscible fluids were modeled by using the Lennard-Jones (LJ) potentials
for the interactions between fluid molecules. The solid walls were
modeled by crystalline plates. More technical details will be presented 
in Secs. \ref{md1} and \ref{md2}. 
The purpose of carrying out MD simulations is threefold:
(1) To uncover the boundary condition governing the MCL, denoted the GNBC;
(2) To fix the material parameters (e.g. viscosity, interfacial tension, etc)
in our hydrodynamic model;
(3) To produce velocity and interface profiles for comparison with the
continuum hydrodynamic solutions.

Our main finding is the GNBC:
$$\beta v^{slip}_x=\tilde{\sigma}_{zx}=\sigma_{zx}^v+\tilde{\sigma}_{zx}^Y.$$
Here $\beta$ is the slip coefficient, 
$v^{slip}_x$ is the tangential slip velocity 
measured relative to the (moving) wall, and $\tilde{\sigma}_{zx}$ 
is the hydrodynamic tangential stress, given by the sum of 
the viscous stress $\sigma_{zx}^v$ and 
the uncompensated Young stress $\tilde{\sigma}_{zx}^Y$.
The validity of the GNBC has been verified by a detailed analysis of
MD data (Secs. \ref{md1}, \ref{md2}, and \ref{gnbc-derivation}) 
plus a comparison of the MD and continuum results (Sec. \ref{comparison}). 
Compared to the conventional Navier boundary condition which
includes the tangential viscous stress only, the GNBC captures
the uncompensated Young stress as the additional component.
Together, the tangential viscous stress and the uncompensated Young stress
give rise to the near-complete slip at the MCL.
The uncompensated Young stress arises from 
the deviation of the fluid-fluid interface from its static configuration
and is narrowly distributed in the fluid-fluid interfacial region.
Obviously, far away from the MCL, the uncompensated Young stress vanishes
and the GNBC becomes the usual Navier Boundary condition 
$\beta v^{slip}_x=\sigma_{zx}^v$. 

We have incorporated the GNBC into the Cahn-Hilliard (CH) hydrodynamic 
formulation of two-phase flow \cite{jacqmin,vinal} to  
obtain a continuum hydrodynamic model \cite{qws}.
The continuum model may be briefly described as follows.
The CH free energy functional \cite{free-energy} is of the form
$F[\phi]=\int d{\bf r}
\left[K\left(\nabla\phi\right)^2/2+f(\phi)\right]$,
where $\phi({\bf r})$ is the composition field defined by
$\phi({\bf r})=(\rho_2-\rho_1)/(\rho_2+\rho_1)$, with 
$\rho_1$ and $\rho_2$ being the local number densities 
of the two fluid species,
$f(\phi)=-r\phi^2/2+u\phi^4/4$, and $K$, $r$, $u$
are parameters which can be determined in MD simulations 
by measuring the interface width $\xi=\sqrt{K/r}$,
the interfacial tension $\gamma=2\sqrt{2}r^2\xi/3u$, and the two
homogeneous equilibrium phases $\phi_\pm=\pm\sqrt{r/u}$ ($=\pm 1$ in our case). 
The two coupled equations of motion are the CH convection-diffusion 
equation for $\phi$ and the Navier-Stokes equation (with the addition of 
the capillary force density):
\begin{equation}\label{he2}
{\partial\phi\over\partial t}+{\bf v}\cdot\nabla\phi=M\nabla^2\mu,
\end{equation}
\begin{equation}\label{he1}
\rho_m\left[{\partial{\bf v}\over
\partial t}+ \left({\bf v}\cdot\nabla\right){\bf v} \right]=
-\nabla p +\nabla\cdot{\mbox{\boldmath$\sigma$}}^v
+\mu\nabla\phi+m\rho{\bf g}_{ext},
\end{equation}
together with the incompressibility condition 
$\nabla\cdot{\bf v}=0$. 
Here $M$ is the phenomenological mobility coefficient, 
$\rho_m$ is the fluid mass density, $p$ is the pressure, 
${\mbox{\boldmath$\sigma$}}^v$ is the Newtonian viscous stress tensor, 
$\mu\nabla\phi$ is the capillary force density with 
$\mu=\delta F/\delta \phi$ being the chemical potential, and
$m\rho{\bf g}_{ext}$ is the external body force density (for Poiseuille flows).
The boundary conditions at the solid surface are
$v_n=0$, $\partial_n \mu=0$ ($n$ denotes the outward surface normal),
a relaxational equation for $\phi$ at the solid surface:
\begin{equation}\label{he4} 
{\partial\phi\over\partial t}+{\bf v}\cdot\nabla\phi=-\Gamma L(\phi), 
\end{equation}
and the continuum form of the GNBC: 
\begin{equation}\label{he3} 
\beta v_x^{slip}=-\eta\partial_n v_x+L(\phi)\partial_x\phi,
\end{equation}
Here $\Gamma$ is a (positive) phenomenological parameter,
$L(\phi)=K\partial_n\phi+\partial\gamma_{wf}(\phi)/\partial\phi$
with $\gamma_{wf}(\phi)$ being the fluid-solid interfacial free energy
density, $\beta$ is the slip coefficient, $v_x^{slip}$ is the (tangential)
slip velocity relative to the wall, $\eta$ is the viscosity,
and $L(\phi)\partial_x\phi$ is the uncompensated Young stress.

Compared to the Navier boundary condition, the additional component
captured by the GNBC in Eq. (\ref{he3}) is the uncompensated Young
stress $\tilde{\sigma}_{zx}^Y$. Its differential expression is given by 
\begin{equation}\label{unYS0}
\tilde{\sigma}_{zx}^Y=L(\phi)\partial_x\phi=
[K\partial_n\phi+\partial\gamma_{wf}(\phi)/\partial\phi]\partial_x\phi
\end{equation}
at $z=0$. Here the $z$ coordinate is for the lower fluid-solid interface
where $\partial_n=-\partial_z$ (same below), 
with the understanding that the same physics holds at the upper interface. 
It can be shown that the integral of this uncompensated Young stress along 
$x$ across the fluid-fluid interface yields
\begin{equation}\label{unYS1} 
\int_{int}dx[L(\phi)\partial_x\phi](x,0)=\gamma\cos\theta_d^{surf}
+\Delta\gamma_{wf},
\end{equation}
where $\int_{int}dx$ denotes the integration along $x$ across the
fluid-fluid interface, $\gamma$ is the fluid-fluid interfacial tension,
$\theta_d^{surf}$ is the dynamic contact angle at the solid surface,
and $\Delta\gamma_{wf}$ is the change of $\gamma_{wf}(\phi)$ across
the fluid-fluid interface, i.e., $\Delta\gamma_{wf}\equiv\int_{int}dx
\partial_x\gamma_{wf}(\phi)$.
The Young's equation relates $\Delta\gamma_{wf}$ to the static contact angle
$\theta_s^{surf}$ at the solid surface:
\begin{equation}\label{YE} 
\gamma\cos\theta_s^{surf}+\Delta\gamma_{wf}=0,
\end{equation}
which is obtained as a phenomenological expression for 
the tangential force balance at the contact line in equilibrium.
Substituting Eq. (\ref{YE}) into Eq. (\ref{unYS1}) yields
\begin{equation}\label{unYS2} 
\int_{int}dx[L(\phi)\partial_x\phi](x,0)=\gamma(\cos\theta_d^{surf}
-\cos\theta_s^{surf}),
\end{equation}
implying that the uncompensated Young stress arises from the deviation
of the fluid-fluid interface from its static configuration.
Equations (\ref{unYS0}), (\ref{unYS1}), (\ref{YE}), and (\ref{unYS2}) 
will be derived in Sec. \ref{continuum-hydrodynamics}.
In essence, our results show that in the vicinity of the MCL,
the tangential viscous stress $-\eta\partial_n v_x$ as postulated by 
the usual Navier boundary condition can not account for the contact-line
slip profile without taking into account the uncompensated Young stress.
This is seen from both the MD data and the predictions of the continuum model.

Besides the external conditions such as the shear speed $V$ and the wall
separation $H$, there are nine parameters in our continuum model, including
$\rho_m$, $\eta$, $\beta$, $\xi$, $\gamma$, $|\phi_\pm|$, $M$, $\Gamma$, 
and $\theta_s^{surf}$. The values of $\rho_m$, $\eta$, $\beta$, $\xi$, 
$\gamma$, $|\phi_\pm|$, and $\theta_s^{surf}$ were directly obtained from 
MD simulations. The values of the two phenomenological parameters 
$M$ and $\Gamma$ were fixed by an optimized comparison with one MD flow field.
The same set of parameters (corresponding to the same local properties 
in a series of MD simulations) has been used to produce velocity fields 
and fluid-fluid interface shapes for comparison with the MD results obtained 
for different external conditions ($V$, $H$, and flow geometry).
The overall agreement is excellent in all cases, demonstrating the validity 
of the GNBC and the hydrodynamic model.

The CH hydrodynamic formulation of immiscible two-phase flow has been
successfully used to construct a continuum model. We would like to emphasize
that while the phase-field formulation does provide a convenient way of modeling
that is familiar to us, it should not be conceived as the unique way.
After all, what we need is to incorporate our key finding, the GNBC,
into a continuum formulation of immiscible two-phase flow.
The GNBC itself is simply a fact found in MD simulations, independent of
any continuum formulation.

\section{Molecular Dynamics I}\label{md1}

\subsection{Geometry and interactions}

\begin{figure}[h]
\bigskip
\centerline{\psfig{figure=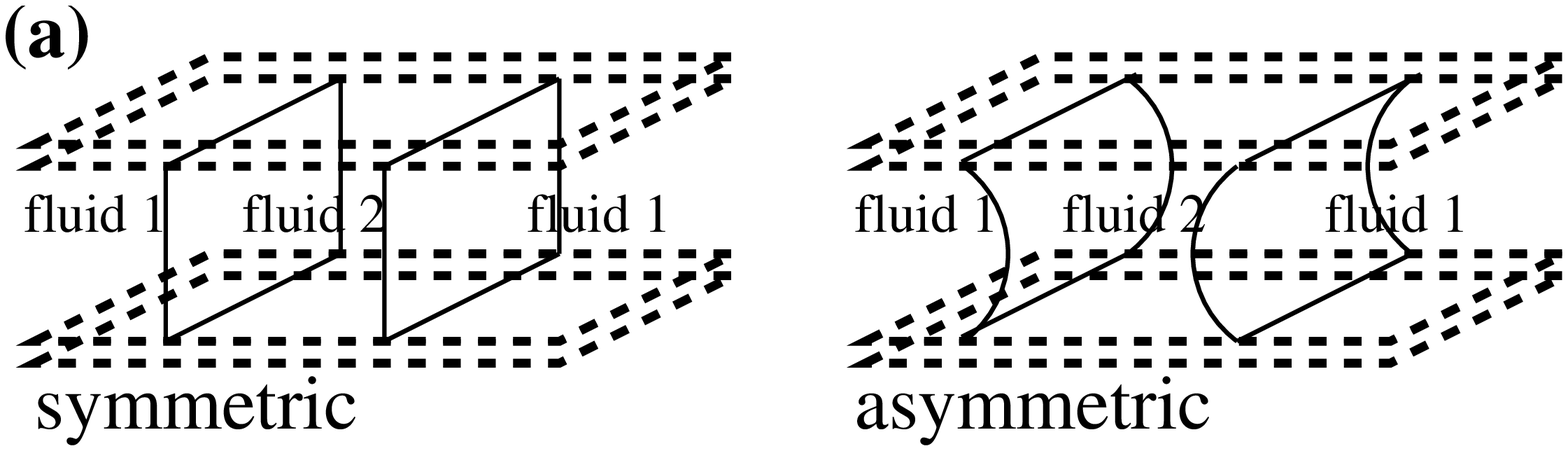,height=2.cm}}
\bigskip
\centerline{\psfig{figure=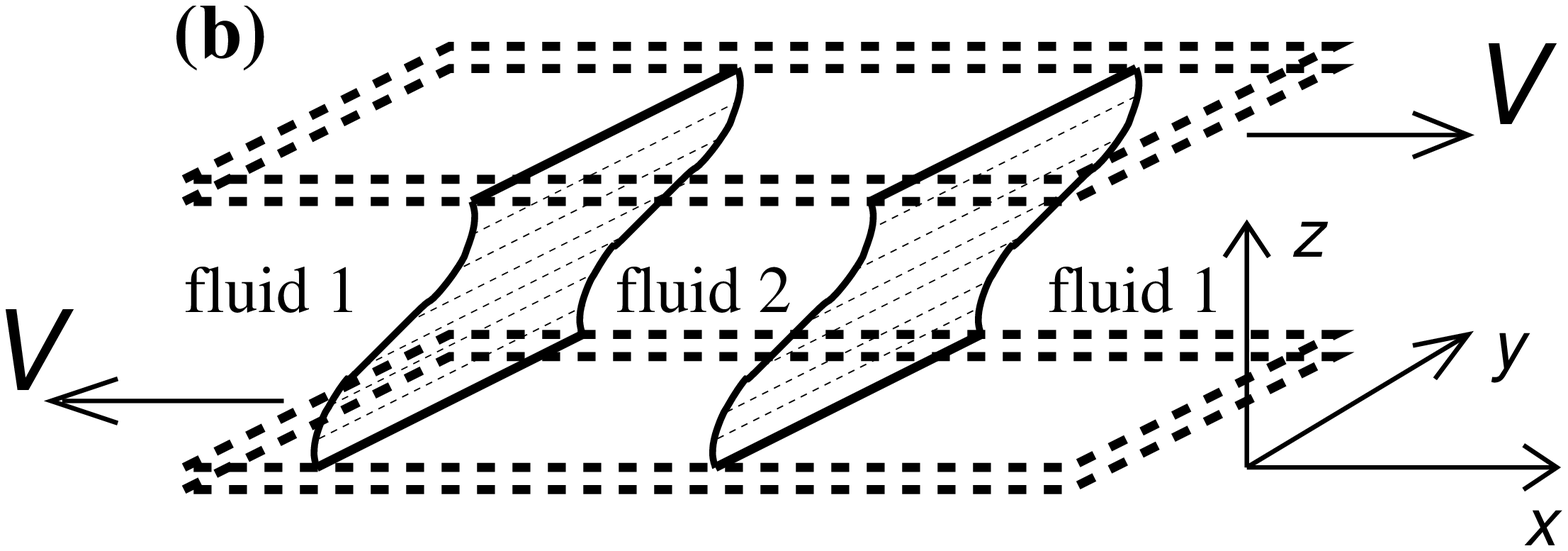,height=2.5cm}}
\caption{
Schematic of simulation geometry.
(a) Static configurations in the symmetric and asymmetric cases. Here 
fluid 2 is sandwiched between fluid 1 due to the periodic boundary condition
along the $x$ direction.
In the symmetric case, the static contact angle is $90^\circ$
and the fluid-fluid interface is flat, parallel to the $yz$ plane.
In the asymmetric case, the static contact angle is not $90^\circ$
and the fluid-fluid interface is curved in the $xz$ plane.
(b) Dynamic configuration in the symmetric case.
}\label{geo12}
\end{figure}

MD simulations have been carried out for two-phase immiscible flows in
Couette geometry (see Figs. \ref{couette} and \ref{geo12}) \cite{qws}. 
Two immiscible fluids were confined between two planar solid walls parallel to
the $xy$ plane, with the fluid-solid boundaries defined by $z=0$, $H$. 
The Couette flow was generated by moving the top and bottom walls 
at a constant speed $V$ in the $\pm x$ directions, respectively. 
Periodic boundary conditions were imposed along the $x$ and $y$ directions.
Interaction between fluid molecules separated by a distance $r$ 
was modeled by a modified LJ potential 
$$U_{ff}=4\epsilon\left[\left(\sigma/r\right)^{12}-
\delta_{ff}\left(\sigma/r\right)^6\right],$$ 
where $\epsilon$ is the energy scale, $\sigma$ is the range scale, with
$\delta_{ff}=1$ for like molecules and $\delta_{ff}=-1$ for
molecules of different species.
(The negative $\delta_{ff}$ was used to ensure immiscibility.)
The average number density for the fluids was set at $\rho=0.81\sigma^{-3}$. 
The temperature was controlled at $2.8\epsilon/k_B$. (This high temperature 
was used to reduce the near-surface layering induced by the solid wall.) 
Each wall was constructed by two [001] planes of an fcc lattice, 
with each wall molecule attached to a lattice site by a harmonic spring. 
The mass of the wall molecule was set equal to that of the fluid molecule $m$. 
The number density of the wall was set at $\rho_w=1.86\sigma^{-3}$.
The wall-fluid interaction was modeled by another LJ potential 
$$U_{wf}=4\epsilon_{wf}\left[\left(\sigma_{wf}/r\right)^{12}-
\delta_{wf}\left(\sigma_{wf}/r\right)^6\right],$$
with the energy and range parameters given by 
$\epsilon_{wf}=1.16\epsilon$ and $\sigma_{wf}=1.04\sigma$, and $\delta_{wf}$ 
for specifying the wetting property of the fluid.
There is no locked layer of fluid molecules at the solid surface.
We have considered two cases. In the symmetric case, the two fluids
have the identical wall-fluid interactions with $\delta_{wf}=1$.
Consequently, the static contact angle is $90^\circ$
and the fluid-fluid interface is flat, parallel to the $yz$ plane.
In the asymmetric case, the two fluids have different wall-fluid
interactions, with $\delta_{wf}=1$ for one and $\delta_{wf}=0.7$ for the other.
As a consequence, the static contact angle is $64^\circ$ and 
the fluid-fluid interface is curved in the $xz$ plane.
In most of our simulations, the shearing speed $V$ was 
on the order of $0.1\sqrt{\epsilon/m}$, the sample dimension along $y$ 
was $6.8\sigma$, the wall separation along $z$ varied from $H=6.8\sigma$ 
to $68\sigma$, and the sample dimension along $x$ was set to be long 
enough so that the uniform single-fluid shear flow was recovered 
far away from the MCL. Steady-state quantities were obtained from 
time average over $10^5\tau$ or longer where $\tau$ is the atomic time scale 
$\sqrt{m\sigma^2/\epsilon}$. Throughout the remainder of this paper, 
all physical quantities are given in terms of the LJ reduced units 
(defined in terms of $\epsilon$, $\sigma$, and $m$).

\subsection{Boundary-layer tangential wall force}\label{tangential-wall-force}

\begin{figure}[h]
\bigskip
\centerline{\psfig{figure=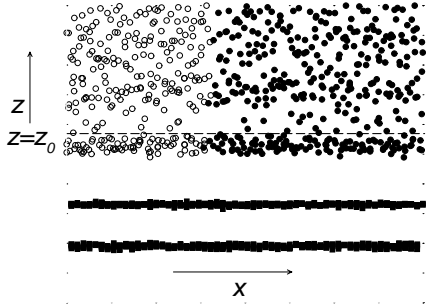,height=4.0cm}}
\caption{Boundary layer at the lower fluid-solid interface. 
The empty and solid circles indicate the instantaneous molecular positions 
of the two fluids projected onto the $xz$ plane. 
The solid squares denote the wall molecules. 
The dashed line indicates the level of $z=z_0$.
}\label{BL}
\end{figure}

\begin{figure}[h]
\bigskip
\centerline{\psfig{figure=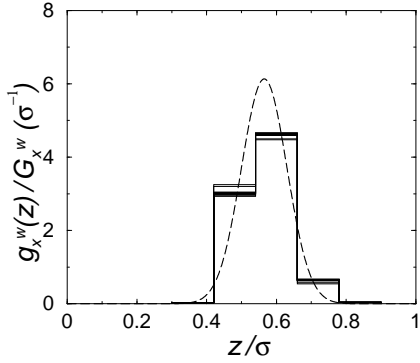,height=5.0cm}}
\caption{
Subdividing the BL into thin sections, we plot the reduced 
tangential wall force density as a function of distance $z$ 
away from the boundary. The solid lines are averaged ${g}_x^w(z)$ 
in thin sections at different $x$, normalized by the corresponding 
total wall force per unit area. The dashed line is a smooth Gaussian fit.
It is seen that ${g}_x^w(z)$ is a function sharply peaked at
$z\approx z_0/2$. Note that $\int_0^{z_0}dz [{g}_x^w(x,z)/G_x^w(x)]=1$.
}\label{peak}
\end{figure}

We denote the region within $z_0=0.85\sigma$ from the wall 
the boundary layer (BL) (see Fig. \ref{BL}). 
It must be thin enough to ensure sufficient precision for 
measuring the slip velocity at the solid surface,
but also thick enough to fully account for the tangential 
wall-fluid interaction force, which is of a finite range.
The wall force can be singled out by separating the force on 
each fluid molecule into wall-fluid and fluid-fluid components. 
The fluid molecules in the BL, being close to the solid wall, 
can experience a strong periodic modulation in interaction with the wall.
This lateral inhomogeneity is generally referred to as the ``roughness''
of the wall potential \cite{nbc}. When coupled with kinetic collisions 
with the wall molecules, there arises a nonzero tangential wall force 
density $g_x^w$ that is sharply peaked at $z\approx z_0/2$ 
and vanishes beyond $z\approx z_0$ (see Fig. \ref{peak}).
From the force density $g_x^w$, we define the tangential wall force 
per unit area as $G_x^w(x)=\int_0^{z_0}dz {g}_x^w(x,z)$, 
which is the total tangential wall force accumulated across the BL.

The short saturation range of the tangential wall force may be understood 
as follows. Out of the BL, each fluid molecule can interact with many 
wall molecules on a nearly equal basis. 
Thus the modulation amplitude (the roughness) of the wall potential 
would clearly decrease with increasing distance from the wall. 
That's why the tangential wall force tends to saturate at $z\approx z_0$,
which is still within the cutoff distance of wall-fluid interaction.
On the contrary, the normal wall force arises from the direct
wall-fluid interaction, independent of whether the wall potential is 
rough or not. Consequently, the normal wall force saturates much slower 
than the tangential component.

\begin{figure}[h]
\bigskip
\centerline{\psfig{figure=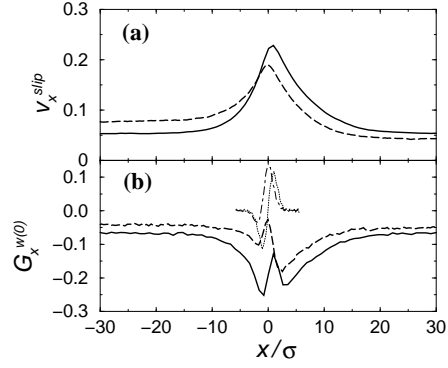,height=5.0cm}}
\caption{Slip velocity and tangential wall force (in reduced units)
measured in the BL at the lower fluid-solid interface. 
(a) The slip velocity $v_x^{slip}=v_x+V$ is plotted as a function of $x$.
The solid line denotes the dynamic symmetric case with $V=0.25\sqrt{\epsilon/m}$
and $H=13.6\sigma$; the dashed line denotes the dynamic asymmetric case with 
$V=0.2\sqrt{\epsilon/m}$ and $H=13.6\sigma$. The slip at the contact line
($x\approx 0$) is near-complete, i.e., $v_x^{slip}\approx V$.
(b) The tangential wall force is plotted as a function of $x$.
The solid line denotes $G_x^w$ in the dynamic symmetric case; 
the dashed line denotes $G_x^w$ in the dynamic asymmetric case. 
The dotted line denotes $G_x^{w0}$ in the static symmetric case; 
the dot-dashed line denotes $G_x^{w0}$ in the static asymmetric case. 
Note that $G_x^{w0}$ vanishes out of the contact-line region.
}\label{wallforce}
\end{figure}

We have measured the slip velocity and the tangential wall force in the BL. 
Spatial resolution along the $x$ direction was achieved by evenly 
dividing the BL into bins, each $\Delta x=0.85\sigma$ or $0.425\sigma$. 
The slip velocity $v_x^{slip}$ was obtained as the time average of 
fluid molecules' velocities inside the BL, 
measured relative to the moving wall; 
the tangential wall force $G_x^w$ was obtained from the time average of 
the total tangential wall force experienced by the fluid molecules 
in the BL, divided by the bin area in the $xy$ plane.
As reference quantities, we also measured $G_x^{w0}$
in the static ($V=0$) configuration. Figure \ref{wallforce} shows
$v_x^{slip}$ and $G_x^w$ measured in the dynamic configuration and
$G_x^{w0}$ measured in the static configuration. It is seen that
in the absence of hydrodynamic motion ($V=0$), the static 
tangential wall force $G_x^{w0}$ is not identically zero everywhere.
Instead, it has some fine features in the contact-line region 
(a few $\sigma$'s) (see Fig. \ref{wallforce}b). 
This nonzero static component in the tangential wall force
arises from the microscopic organization of fluid molecules 
in the contact-line region. 

\begin{figure}[h]
\bigskip
\centerline{\psfig{figure=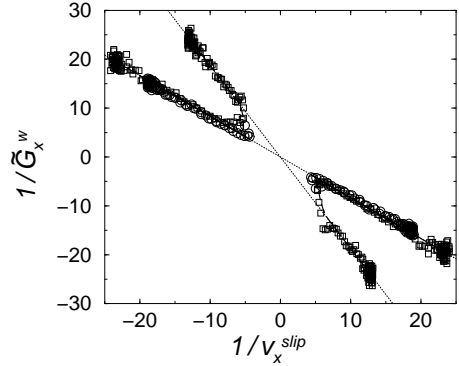,height=5.0cm}}
\caption{
$1/\tilde{G}_x^w$ plotted as a function of $1/v_x^{slip}$ (in reduced units). 
Symbols are MD data measured in the BL at different $x$ locations. 
The circles denote the symmetric case with 
$V=0.25\sqrt{\epsilon/m}$ and $H=13.6\sigma$;
the squares denote the asymmetric case with 
$V=0.2\sqrt{\epsilon/m}$ and $H=13.6\sigma$. 
The solid and dashed lines are calculated from Eq. (\ref{gwslip}) for 
the symmetric and asymmetric cases, respectively, as described in the text. 
The lower-right data segment corresponds to the lower BL,
whereas the upper-left segment corresponds to the upper BL.
The slopes of the two dotted lines are given by $\beta_{1,2}^{-1}$,
which are proportional to the slip length.
}\label{nbc_wf}
\end{figure}

The static component is also present in $G_x^w$ measured 
in the dynamic configuration, as shown by Fig. \ref{wallforce}b.
To see the effects arising purely from the hydrodynamic motion of the fluids,
we subtract $G_x^{w0}$ from $G_x^w$ through the relation
$$\tilde{G}_x^w=G_x^w-G_x^{w0},$$
where $\tilde{G}_x^w$ is the hydrodynamic part in $G_x^w$.
In the notations below, the over tilde will denote 
the difference between that quantity and its static part.
We find the hydrodynamic tangential wall force per unit area, 
$\tilde{G}_x^w$, is proportional to the local slip velocity $v_x^{slip}$:
\begin{equation}\label{gwslip}
\tilde{G}_x^w(x)=-\beta v_x^{slip}(x),
\end{equation}
where the proportionality constant $\beta$ is the slip coefficient.
In Fig. \ref{nbc_wf}, $\tilde{G}_x^w$ is plotted as a function of $v_x^{slip}$.
The symbols represent the values of $\tilde{G}_x^w$ and $v_x^{slip}$
measured in the BL. The lines represent the values of $\tilde{G}_x^w$
calculated from $-\beta v_x^{slip}$ using $v_x^{slip}$ measured in the BL 
and $\beta=(\beta_1\rho_1+\beta_2\rho_2)/(\rho_1+\rho_2)$,
with $\beta_{1,2}$ the slip coefficients for the two fluid species and 
$\rho_{1,2}$ the molecular densities of the two fluid species measured 
in the BL. Independent measurements determined
$\beta_1=\beta_2=1.2\sqrt{\epsilon m}/\sigma^3$ for the symmetric case, 
$\beta_1=1.2\sqrt{\epsilon m}/\sigma^3$ and 
$\beta_2=0.532\sqrt{\epsilon m}/\sigma^3$ for the asymmetric case.
(The dependence of $\beta$ on $\beta_{1,2}$ and $\rho_{1,2}$ assumes
the two fluids interact with the wall independently.
The desired expression is obtained by expressing $\tilde{G}_x^w$ 
as the weighted average of $\tilde{G}_x^{w1}=-\beta_1 v_x^{slip1}$ 
and $\tilde{G}_x^{w2}=-\beta_2 v_x^{slip2}$ and noting
$v_x^{slip1}\approx v_x^{slip2}$ to within $10\%$).

\subsection{Tangential fluid force}\label{tangential-fluid-force}

In either the static equilibrium state (where $\tilde{G}_x^w=0$) or
the dynamic steady state (where $\tilde{G}_x^w\ne 0$), local force balance
necessarily requires the stress tangential to the fluid-solid interface
to be the same on the two sides. Therefore, the hydrodynamic tangential 
fluid force per unit area, $\tilde{G}_x^f$, must be proportional to 
the slip velocity $v_x^{slip}$:
\begin{equation}\label{gfslip}
\tilde{G}_x^f(x)=\beta v_x^{slip}(x),
\end{equation}
such that $\tilde{G}_x^f(x)+\tilde{G}_x^w(x)=0$ according to Eq. (\ref{gwslip}).
(The MD evidence for this force balance will be presented in Sec. \ref{md2}.)
Physically, $\tilde{G}_x^f$ is the hydrodynamic force along $x$ 
exerted on a BL fluid element by the surrounding fluids, 
and may be expressed as
\begin{equation}\label{gf-expression}
\begin{array}{ll}
\tilde{G}_x^f(x) & =\int_0^{z_0}dz
[\partial_x\tilde{\sigma}_{xx}(x,z)+\partial_z\tilde{\sigma}_{zx}(x,z)]\\
& =\tilde{\sigma}_{zx}(x,z_0)+
\partial_x\int_0^{z_0}dz\tilde{\sigma}_{xx}(x,z),
\end{array}
\end{equation}
using the fact that $\tilde{\sigma}_{zx}(x,0)=0$.
(More strictly, $\tilde{\sigma}_{zx}(x,0^-)=0$ because there is no fluid
below $z=0$, hence no momentum transport across $z=0$.)
Here $\tilde{\sigma}_{xx(zx)}=\sigma_{xx(zx)}-\sigma_{xx(zx)}^0$, with
$\sigma_{xx}^{(0)}$ being the normal component and 
$\sigma_{zx}^{(0)}$ the tangential component of the fluid stress tensor 
in the dynamic (static) configuration.

\subsection{Sharp boundary limit}\label{sharp-boundary-limit}

The form of $\tilde{G}_x^f$ in Eq. (\ref{gf-expression}) is derived from 
the fact that the tangential wall force is distributed in a BL
of finite thickness (Figs. \ref{BL} and \ref{peak}).
Now we take the sharp boundary limit by letting the tangential wall force
strictly concentrate at $z=0$: $\tilde{g}_x^w(x,z)=\tilde{G}_x^w(x)
\delta(z)$ with the same $\tilde{G}_x^w(x)$ per unit area. 
As shown in Fig. \ref{peak}, the tangential wall force density
is a sharply peaked function. By taking the sharp boundary limit 
the normalized peaked function is replaced by $\delta(z)$. 
Rewriting $\tilde{G}_x^f$ in Eq. (\ref{gf-expression}) as
$$\begin{array}{ll}
\tilde{G}_x^f(x) & =\int_{0^-}^{z_0}dz
[\partial_x\tilde{\sigma}_{xx}(x,z)+\partial_z\tilde{\sigma}_{zx}(x,z)]\\
& =\tilde{\sigma}_{zx}(x,0^+)+\int_{0^+}^{z_0}dz
[\partial_x\tilde{\sigma}_{xx}(x,z)+\partial_z\tilde{\sigma}_{zx}(x,z)],
\end{array}$$
we obtain 
\begin{equation}\label{gfslipsharp}
\tilde{G}_x^f(x)=\tilde{\sigma}_{zx}(x,0^+)=\beta v_x^{slip}(x),
\end{equation}
because local force balance requires $\partial_x\tilde{\sigma}_{xx}+
\partial_z\tilde{\sigma}_{zx}=0$ above $z=0^+$.
Therefore, in the sharp boundary limit $\tilde{\sigma}_{zx}$ varies from
$\tilde{\sigma}_{zx}(x,0^-)=0$ to $\tilde{\sigma}_{zx}(x,0^+)=
\tilde{G}_x^f(x)$ at $z=0$ such that
$$(\nabla\cdot\tilde{\mbox{\boldmath$\sigma$}})\cdot\hat{\bf x}
=\tilde{G}_x^f(x)\delta(z),$$
in balance with the tangential wall force density 
$\tilde{g}_x^w(x,z)=\tilde{G}_x^w(x)\delta(z)$. 
Equation (\ref{gfslipsharp}) may serve as a boundary condition 
in hydrodynamic calculation if a continuum (differential) form of 
$\tilde{\sigma}_{zx}(x,0^+)$ is given. This will be accomplished in
Sec. \ref{boundary-condition}.

\section{Continuum Hydrodynamic Model}\label{continuum-hydrodynamics}

For decades numerous models have been proposed to resolve 
the boundary condition problem for the contact-line motion 
\cite{blake,hocking,hua-mason,sheng,seppecher,jacqmin,vinal}, 
but so far none has proved successful by reproducing 
the slip velocity profiles observed in MD simulations \cite{koplik,robbins}. 
In particular, based on the extreme (velocity) variations in the slip region, 
a breakdown of local hydrodynamic description in the immediate vicinity of 
the MCL has been suggested \cite{robbins}.

The main purpose of this paper is to present a continuum hydrodynamic model
that is capable of reproducing MD results in the molecular-scale vicinity
of the MCL \cite{qws}. For this purpose, we have derived a differential form 
for Eq. (\ref{gfslipsharp}) (the continuum GNBC, Eq. (\ref{he3})) using 
the Cahn-Hilliard hydrodynamic formulation of two-phase flow 
\cite{jacqmin,vinal}. Our model consists of the convection-diffusion 
equation in the fluid-fluid interfacial region (Eq. (\ref{he2})),
the Navier-Stokes equation for momentum transport (Eq. (\ref{he1})),
the relaxational equation for the composition at the solid surface 
(Eq. (\ref{he4})), and the GNBC (Eq. (\ref{he3})).

\subsection{Cahn-Hilliard free energy functional}

The CH free energy was proposed to phenomenologically describe 
an interface between two coexisting phases \cite{free-energy}.
In terms of the composition order parameter 
$\phi=(\rho_2-\rho_1)/(\rho_2+\rho_1)$, the CH free energy functional reads
$$
F=\int{\rm d}{\bf r}\left[ 
\displaystyle\frac{1}{2}K\left(\nabla\phi\right)^2+f(\phi) \right],
$$
with 
$f(\phi)=-\frac{1}{2}r\phi^2+\frac{1}{4}u\phi^4$.
Two thermally stable phases are given by 
$\phi_{\pm}=\pm \sqrt{{r}/{u}}$ at which ${\partial f}/{\partial\phi}=0$.
An interface can be formed between the phases of $\phi_+$ and $\phi_-$ 
in coexistence. 

\subsubsection{Chemical potential}

The chemical potential $\mu$ is defined by
$$
\mu=\displaystyle\frac{\delta F}{\delta\phi}=-K\nabla^2\phi-r\phi+u\phi^3,
$$
from which the diffusion current ${\bf J}=-M\nabla\mu$ is obtained
with $M$ being the mobility coefficient.
The convection-diffusion equation (Eq. (\ref{he2})) comes from 
the continuity equation
$$
\displaystyle\frac{D\phi}{Dt}=
\displaystyle\frac{\partial\phi}{\partial t}+{\bf v}\cdot\nabla\phi
=-\nabla\cdot {\bf J}.
$$

\subsubsection{Interfacial tension}

A few important physical quantities can be derived from the CH free energy. 
We first derive the interfacial tension $\gamma$ for 
the interface formed between $\phi_+$ and $\phi_-$.
In equilibrium, the spatial variation of $\phi$ is determined by 
the condition that $\mu({\bf r})$ is constant, i.e., 
$$-K\partial_z^2\phi-r\phi+u\phi^3={\rm constant}.$$
Here the interface is assumed to be in the $xy$ plane with the interface
normal along the $z$ direction and 
the constant equals to zero because $\lim_{z\rightarrow\pm\infty}\phi
=\phi_{\pm}$ and $\lim_{z\rightarrow\pm\infty}\mu=0$. 
The interfacial profile is solved to be
$$\phi_0(z)=\phi_{+}\tanh\displaystyle\frac{z}{\sqrt{2}\xi},$$ 
with $\xi=\sqrt{{K}/{r}}$ being a characteristic length along 
the interface normal.
The first integral is
$$
-\displaystyle\frac{1}{2}K\left(\partial_z\phi\right)^2+f(\phi)=C,
$$
where the integral constant $C$ equals $f(\phi_{\pm})$.
It follows the interfacial free energy per unit area, i.e., 
the interfacial tension, is given by 
$$
\gamma=\int dz\left[ 
\displaystyle\frac{1}{2}K\left(\partial_z\phi\right)^2+
f(\phi)-f(\phi_{\pm})\right]=\int dzK\left(\partial_z\phi\right)^2.
$$
Using the interfacial profile $\phi_0(z)$, we obtain
$$\gamma=\displaystyle\frac{K\phi_{\pm}^2}{\sqrt{2}\xi}
\int d\bar{z}\cosh^{-4}\bar{z}=
\displaystyle\frac{2\sqrt{2}K\phi_{\pm}^2}{3\xi}=
\displaystyle\frac{2\sqrt{2}r^2\xi}{3u}.$$

\subsubsection{Capillary force and Young stress}\label{two-phase-force}

We now turn to the forces arising from the interface. Consider 
a virtual displacement ${\bf u}({\bf r})$ and the corresponding variation
in $\phi$, $\delta\phi({\bf r})=-{\bf u}({\bf r})\cdot\nabla\phi$.
The change of the free energy due to this $\delta\phi$ is 
$$
\begin{array}{ll}
\delta F & =\int d{\bf r}\left [\displaystyle\frac
{\partial f(\phi)}{\partial\phi}\delta\phi\right ]+
\int d{\bf r}\left\{\displaystyle\frac
{\partial\left[\frac{1}{2}K\left(\nabla\phi\right)^2\right]}
{\partial(\partial_j\phi)}\delta\left(\partial_j\phi\right)\right\}
\\ & =\int d{\bf r}\left[\mu\delta\phi\right] +
\int ds\left[K\partial_n\phi\delta\phi\right]\\
& =-\int d{\bf r}\left[{\bf g}\cdot{\bf u}\right] +
\int ds\left[\sigma_{ni}^Yu_i\right],
\end{array}
$$
where ${\bf g}=\mu\nabla\phi$ is the capillary force density in the
Navier-Stokes equation (Eq. (\ref{he1})), and $\sigma_{ni}^Y=
-K\partial_n\phi\partial_i\phi$ is the tangential Young stress 
(the $i$ direction is along the fluid-solid interface, ${\bf i}\perp {\bf n}$).

The body force ${\bf g}({\bf r})=\mu\nabla\phi$ can be reduced to 
the familiar curvature force in the sharp interface limit
\cite{sharp-interface-limit}.
The unit vector normal to the level sets of constant $\phi$ is given by
${\bf m}={\nabla\phi}/{|\nabla\phi|}$ and 
$$
\begin{array}{ll}
\mu\nabla\phi & =[-K\nabla^2\phi-r\phi+u\phi^3]|\nabla\phi|{\bf m}\\
& =-K\nabla_t^2\phi |\nabla\phi|{\bf m}+
[-K\partial_m^2\phi-r\phi+u\phi^3]|\nabla\phi|{\bf m}
\end{array}
$$
where $\nabla_t$ and $\partial_m$ denote the differentiations 
tangential and normal to the interface respectively.
For gently curved interfaces, the order parameter $\phi$ along 
the interface normal can be approximated by the one-dimensional stationary
solution $\phi_0$, i.e., $-K\partial_m^2\phi-r\phi+u\phi^3\approx 0$. Hence,
$\mu\nabla\phi\approx -K\nabla_t^2\phi |\nabla\phi|{\bf m}$, from which 
we obtain the desired relation
$$
\mu\nabla\phi\approx K|\nabla\phi|^2\kappa {\bf m}\approx 
\gamma\kappa\delta(l_m){\bf m},
$$
where $\kappa=-{\nabla_t^2\phi}/{|\nabla\phi|}$ is the curvature 
and $\gamma\approx\int dl_mK\left(\nabla\phi\right)^2
\approx\int dl_mK\left(\partial_m\phi\right)^2$
is the interfacial tension, with $l_m$ being the coordinate along the
interface normal and the interface located at $l_m=0$. 

For gently curved interfaces, $K\partial_n\phi\approx 
K\partial_m\phi\cos\theta^{surf}$, where $\bf n$ is the outward (solid) 
surface normal, $\bf m$ the (fluid-fluid) interface normal, and 
$\theta^{surf}$ the angle at which the interface intersects the solid surface
(${\bf n}\cdot{\bf m}=\cos\theta^{surf}$).
For the tangential Young stress $\sigma_{zx}^Y=
K\partial_n\phi\partial_x\phi$ at $z=0$ where ${\bf n}=-{\bf z}$ and 
${\bf i}={\bf x}$, the integral $\int_{int} dx \sigma_{zx}^Y$ along $x$ 
across the interface equals to $\int_{int} dx K\partial_n\phi\partial_x\phi=
(\int_{int} d\phi K\partial_m\phi)\cos\theta^{surf}$, where
$\int_{int} d\phi K\partial_m\phi=
\int_{int} dl_mK\left(\partial_m\phi\right)^2=\gamma$.
Hence, 
\begin{equation}\label{integral-YS}
\int_{int}dx\sigma_{zx}^Y=\gamma\cos\theta^{surf},
\end{equation}
where $\theta^{surf}$ may be the dynamic contact angle at the solid surface
$\theta_d^{surf}$ or the static contact angle $\theta_s^{surf}$.
This $\int_{int}dx\sigma_{zx}^Y$ is the tangential force per unit length 
at the contact line (aligned along $y$), exerted by the fluid-fluid interface
of tension $\gamma$, which intersects the solid wall at the contact angle 
$\theta^{surf}$. So it equals to $\gamma\cos\theta^{surf}$.

\subsubsection{Young's equation}

The Young's equation for the static contact angle $\theta_s^{surf}$
can be derived as well.
Consider the interfacial free energy at the fluid-solid interface, 
$F_{wf}=\int ds\gamma_{wf}(\phi)$. 
Minimizing the total free energy $F+F_{wf}$ with respect to 
$\phi$ at the solid surface yields 
\begin{equation}\label{YE1}
\left[K\partial_n\phi+\displaystyle\frac
{\partial\gamma_{wf}(\phi)}{\partial\phi}\right]_{\phi_{eq}}=0,
\end{equation}
from which an equation of local tangential force balance
\begin{equation}\label{YE2}
\left[\tilde{\sigma}_{zx}^Y\right]_{\phi_{eq}}=
\left[\sigma_{zx}^Y+\partial_x\gamma_{wf}(\phi)\right]_{\phi_{eq}}=
\sigma_{zx}^{0}+\partial_x\gamma_{wf}(\phi_{eq})=0,
\end{equation}
is obtained at $z=0$. Here $\tilde{\sigma}_{zx}^Y=
\sigma_{zx}^Y+\partial_x\gamma_{wf}$ is the uncompensated Young stress
(first introduced in Eq. (\ref{unYS0})), 
$\phi_{eq}$ is the equilibrium composition field, 
and $\sigma_{zx}^{0}$ denotes the static Young stress 
$\sigma_{zx}^Y({\phi_{eq}})$. Integrating Eq. (\ref{YE2}) along $x$ 
across the interface leads to the Young's equation 
$\gamma\cos\theta_s^{surf}+\Delta\gamma_{wf}=0$ (Eq. (\ref{YE})), where 
$\gamma\cos\theta_s^{surf}=\int_{int}dx\sigma_{zx}^{0}$
and $\Delta\gamma_{wf}\equiv\int_{int} dx\partial_x\gamma_{wf}(\phi)$ 
is the change of fluid-solid interfacial 
free energy per unit area across the fluid-fluid interface.
A microscopic picture for the Young's equation as an (integrated) equation 
of tangential force balance will be elaborated in Sec. \ref{molecular-YE}.

\subsection{Two boundary conditions}\label{boundary-condition}

Equations (\ref{YE1}) and (\ref{YE2}) are boundary conditions
for the equilibrium state. In the dynamic steady state, however,
neither $K\partial_n\phi+{\partial\gamma_{wf}(\phi)}/{\partial\phi}=
L(\phi)$ nor $\sigma_{zx}^Y+\partial_x\gamma_{wf}(\phi)=
L(\phi)\partial_x\phi$ vanishes. In fact, the nonzero $L(\phi)$ 
is responsible for the relaxation of $\phi$ at the solid surface 
while the nonzero $L(\phi)\partial_x\phi$ is necessary to a slip boundary 
condition that is able to account for the near-complete slip at the MCL.

The convection-diffusion equation (Eq. (\ref{he2})) is fourth-order in space.
Consequently, besides the usual impermeability condition $\partial_n\mu=0$,
one more boundary condition is needed. 
The dynamics of $\phi$ at the solid surface is plausibly assumed to be
relaxational, governed by the first-order extension of Eq. (\ref{YE1}).
More explicitly, when the system is driven away from the equilibrium,
both ${\partial\phi}/{\partial t}+{\bf v}\cdot\nabla\phi$
and $L(\phi)$ become nonzero, and they are related to each other by 
a linear relation
$${\partial\phi\over\partial t}+{\bf v}\cdot\nabla\phi\propto L(\phi).$$
This leads to Eq. (\ref{he4}) with $\Gamma$ introduced as a phenomenological
parameter.

The GNBC (Eq. (\ref{he3})) is obtained by substituting 
\begin{equation}\label{continuum-hydro-tangential-stress}
\begin{array}{lll}
\tilde{\sigma}_{zx}(x,0^+)=\sigma_{zx}(x,0)-\sigma_{zx}^0(x,0)
& =\sigma_{zx}^v(x,0)+\sigma_{zx}^Y(x,0)-\sigma_{zx}^0(x,0)\\
& = \sigma_{zx}^v(x,0)+\sigma_{zx}^Y(x,0)+\partial_x\gamma_{wf}\\
& = \sigma_{zx}^v(x,0)+\tilde{\sigma}_{zx}^Y.
\end{array}
\end{equation}
into Eq. (\ref{gfslipsharp}). Here the hydrodynamic tangential stress
$\tilde{\sigma}_{zx}$ is decomposed into a viscous component 
${\sigma}_{zx}^v$ and a non-viscous component $\tilde{\sigma}_{zx}^Y$.
The viscous component is simply given by ${\sigma}_{zx}^v=\eta\partial_zv_x$;
the non-viscous component is the uncompensated Young stress
$\tilde{\sigma}_{zx}^Y$, given by $\tilde{\sigma}_{zx}^Y=
\sigma_{zx}^Y+\partial_x\gamma_{wf}(\phi)$ (Eq. (\ref{unYS0})). 
According to Eq. (\ref{YE2}), this uncompensated Young stress vanishes
in the equilibrium state. But in a dynamic configuration, from the integral 
of $\tilde{\sigma}_{zx}^Y$ along $x$ across the fluid-fluid interface
(Eqs. (\ref{unYS1}), (\ref{YE}), and (\ref{unYS2}))
$$\int_{int} dx \tilde{\sigma}_{zx}^Y=\gamma\cos\theta_d^{surf}+
\Delta\gamma_{wf}=\gamma(\cos\theta_d^{surf}-\cos\theta_s^{surf}),$$
there is always a non-viscous contribution to the total tangential stress 
$\tilde{\sigma}_{zx}$ as long as the fluid-fluid interface deviates from 
its static configuration.

In Sec. \ref{comparison} we will show that the GNBC, 
with the uncompensated Young stress included, 
can account for the slip velocity profiles in the vicinity of the MCL,
especially the near-complete slip at the contact line. 
In Secs. \ref{tangential-force-balance} and \ref{stress-decomposition} 
we will present more MD evidence supporting the GNBC. 
A ``derivation'' of the GNBC, based on the tangential force balance
(Sec. \ref{tangential-force-balance}) and the tangential stress decomposition
(Sec. \ref{stress-decomposition}), will be given in Sec. \ref{gnbc-derivation}.

\subsection{Dimensionless equations}

Dimensionless equations suitable for numerical computation
are obtained as follows. We scale $\phi$ by $|\phi_\pm|=\sqrt{{r}/{u}}$, 
length by $\xi=\sqrt{{K}/{r}}$, velocity by the wall speed $V$, 
time by $\xi/V$, and pressure/stress by $\eta V/\xi$.
In dimensionless forms, the convection-diffusion equation is
\begin{equation}\label{dmlcde}
\displaystyle\frac{\partial\phi}{\partial t}+
{\bf v}\cdot\nabla\phi={\cal L}_d\nabla^2(-\nabla^2\phi-\phi+\phi^3),
\end{equation}
the Navier-Stokes equation is
\begin{equation}\label{dmlnse}
{\cal R}\left[\displaystyle\frac{\partial{\bf v}}{\partial t}+
\left({\bf v}\cdot\nabla\right){\bf v}\right]=-\nabla p+
\nabla^2{\bf v}+{\cal B}(-\nabla^2\phi-\phi+\phi^3 )\nabla\phi,
\end{equation}
the relaxational equation for $\phi$ at the solid surface is 
\begin{equation}\label{dmlrelax}
\displaystyle\frac{\partial\phi}{\partial t}+v_x\partial_x\phi=
-{\cal V}_s\left[\partial_n\phi-
\displaystyle\frac{\sqrt{2}}{3}\cos\theta_s^{surf} s_\gamma(\phi)\right],
\end{equation}
and the GNBC is
\begin{equation}\label{dmlgnbc}
\left[{\cal L}_s(\phi)\right]^{-1} v^{slip}_x={\cal B}
\left[\partial_n\phi-\displaystyle\frac{\sqrt{2}}{3}\cos\theta_s^{surf} 
s_\gamma(\phi)\right]\partial_x\phi-\partial_nv_x.
\end{equation}
Here $s_\gamma(\phi)=(\pi/2)\cos(\pi\phi/2)$ is from the fluid-solid
interfacial free energy $$\gamma_{wf}(\phi)=(\Delta\gamma_{wf}/2)
\sin(\pi\phi/2),$$ which denotes a smooth interpolation between 
$\pm\Delta\gamma_{wf}/2$.
Five dimensionless parameters appear in the above equations. 
They are (1) ${\cal L}_d={Mr}/{V\xi}$, 
which is the ratio of a diffusion length $Mr/V$ to $\xi$, 
(2) ${\cal R}={\rho V\xi}/{\eta}$,
(3) ${\cal B}={r^2\xi}/{u\eta V}={3\gamma}/{2\sqrt{2}\eta V}$,
which is inversely proportional to the capillary number $Ca=\eta V/\gamma$, 
(4) ${\cal V}_s={K\Gamma}/{V}$, 
and (5) ${\cal L}_s(\phi)={\eta}/{\beta(\phi)\xi}$, which is 
the ratio of the slip length $l_s(\phi)={\eta}/{\beta(\phi)}$ to $\xi$,
where $\beta(\phi)=(1-\phi)\beta_1/2+(1+\phi)\beta_2/2$.
A numerical algorithm based on a fixed uniform mesh has been presented 
in Ref. \cite{qws}.

\section{Comparison of MD and Continuum Results}\label{comparison} 

To demonstrate the validity of our continuum model, we have obtained 
numerical solutions that can be directly compared to the MD results 
for flow field and fluid-fluid interface shape.
We have carried out the MD-continuum comparison in such a way that 
virtually {\it no} adjustable parameter is involved in the continuum
calculations. This is achieved as follows.
 
There are totally nine material parameters in our continuum model.
They are $\rho_m$, $\eta$, $\beta$, $\xi$, $\gamma$, $|\phi_\pm|$, 
$M$, $\Gamma$, and $\theta_s^{surf}$. (Note (1) For the asymmetric case,
two unequal slip coefficients $\beta_1$ and $\beta_2$ are involved in $\beta$;
(2) The three parameters $\xi$, $\gamma$, and $|\phi_\pm|$ 
are equivalent to the three parameters $K$, $r$, and $u$ in 
the CH free energy density; (3) $\theta_s^{surf}$
is for $\Delta\gamma_{wf}=-\gamma\cos\theta_s^{surf}$.)
Among the nine parameters, seven are directly obtainable (measurable)
in MD simulations. They are $\rho_m$, $\eta$, $\beta_{1,2}$, 
$\xi$, $\gamma$, $|\phi_\pm|$, and $\theta_s^{surf}$.
(The fluid mass density $\rho_m$ is set in MD simulations,
the viscosity $\eta$ and the slip coefficients $\beta_{1,2}$
can be measured in suitable single-fluid MD simulations, 
the interfacial thickness $\xi$ can be obtained by measuring the interfacial
profile $\phi=(\rho_2-\rho_1)/(\rho_2+\rho_1)$ in MD simulations,
the interfacial tension $\gamma$ can be obtained by measuring an integral
of the pressure/stress anisotropy in the interfacial region 
\cite{interfacial-tension}, $|\phi_\pm|=1$ means the total immiscibility
of the two fluids, and the static contact angle $\theta_s^{surf}$
is directly measurable.) 
The two phenomenological parameters $M$ and $\Gamma$ have been introduced 
to describe the composition dynamics in the interfacial region.
Their values are fixed by an optimized MD-continuum comparison.
That is, one MD flow field is best matched by varying the continuum flow field 
with respect to the values of $M$ and $\Gamma$.
Once all the parameter values are obtained ($7$ measured in MD simulations and
$2$ fixed by one MD-continuum comparison), our continuum hydrodynamic model
can yield predictions that can be readily compared to the results from 
a series of MD simulations with different external conditions 
($V$, $H$, and flow geometry). The overall agreement is excellent in 
all cases, thus demonstrating the validity of the GNBC and the hydrodynamic
model. We emphasize that the MD-continuum agreement has been achieved
both in the molecular-scale vicinity of the contact line and far way
from the contact line. This opens up the possibility of not only 
continuum simulations of nano- and microfluidics involving
immiscible components, but also macroscopic immiscible flow
calculations that are physically meaningful at the molecular level.
(Molecular-scale details may be resolved through the iterative grid 
redistribution method without significantly compromising computation
efficiency, see \cite{ren-wang,power-law}).

\subsection{Immiscible Couette flow}

\subsubsection{Two symmetric cases}

In Figs. \ref{8symflow} and \ref{16symflow} we show the MD and continuum 
velocity fields for two symmetric cases of immiscible Couette flow.
In MD simulations, these two cases have the same local properties 
(fluid density, temperature, fluid-fluid interaction, wall-fluid interaction,
etc) but different external conditions ($H$ and $V$). Correspondingly, 
the continuum results are obtained using the same set of nine
material parameters $\rho_m$, $\eta$, $\beta$ ($=\beta_1=\beta_2$), 
$\xi$, $\gamma$, $|\phi_\pm|$, $M$, $\Gamma$, and $\theta_s^{surf}$.

\subsubsection{Two asymmetric cases}

In Figs. \ref{8asymflow} and \ref{16asymflow} we show the MD and continuum 
velocity fields for two asymmetric cases of immiscible Couette flow.
In MD simulations, these two cases have the same local properties 
(fluid density, temperature, fluid-fluid interaction, wall-fluid interaction,
etc) but different external conditions ($H$ and $V$). Correspondingly, 
the continuum results are obtained using the same set of ten
material parameters $\rho_m$, $\eta$, $\beta_1$, $\beta_2$,
$\xi$, $\gamma$, $|\phi_\pm|$, $M$, $\Gamma$, and $\theta_s^{surf}$.
In particular, among these parameters, $\beta_2$ and $\theta_s^{surf}$
are measured in MD simulations while all the others directly come from 
the symmetric cases. Therefore, the comparison here is
{\it without adjustable parameters}.

\subsubsection{From near-complete slip to uniform shear flow}

\begin{figure}
\bigskip
\centerline{\psfig{figure=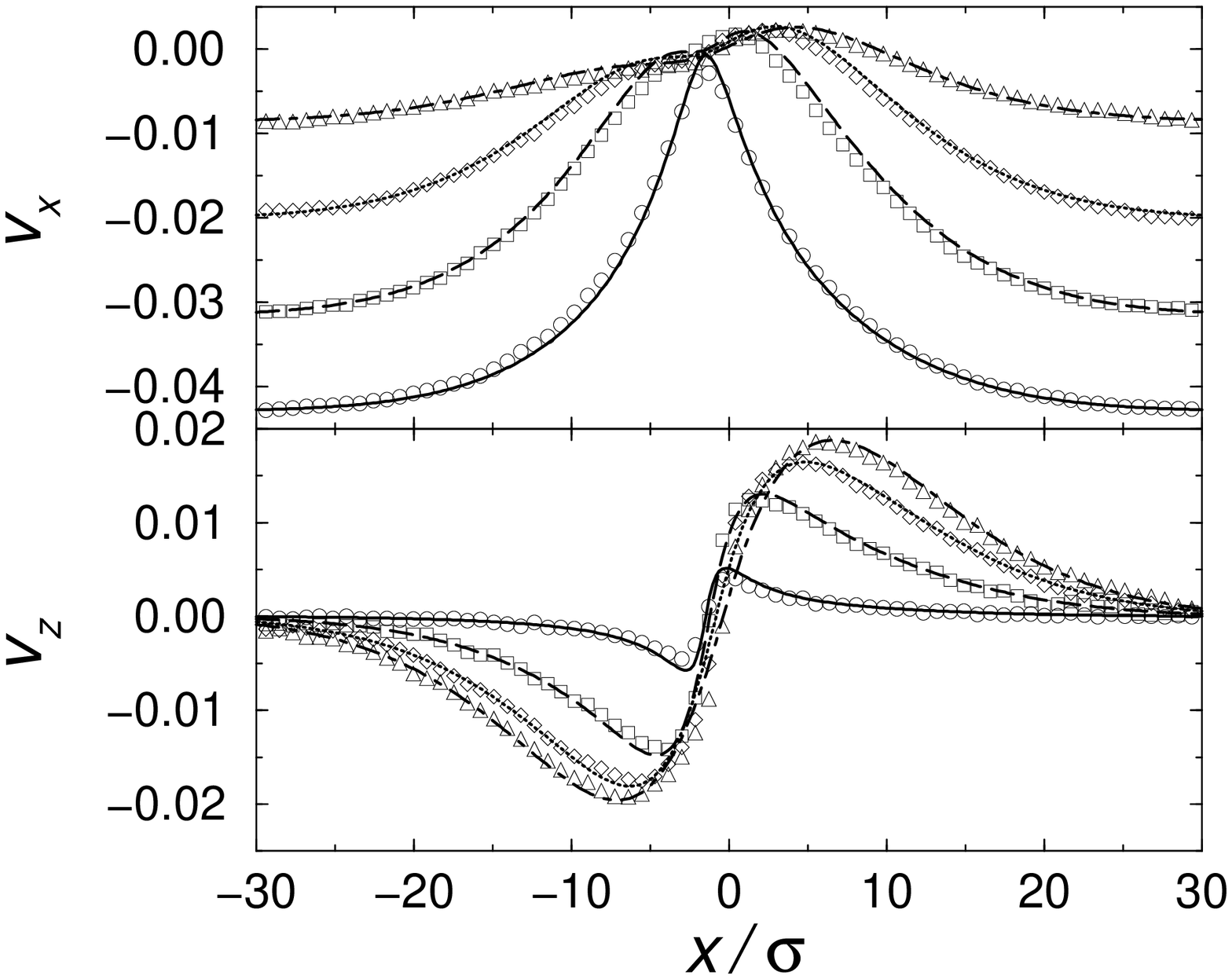,height=7.0cm}}
\caption{Comparison of the MD (symbols) and continuum (lines) 
velocity profiles ($v_x(x)$ and $v_z(x)$ at different $z$ levels)
for a symmetric case of immiscible Couette flow
($V=0.05(\epsilon/m)^{1/2}$ and $H=27.2\sigma$).
The profiles are symmetric about the center plane $z=H/2$,
hence only the lower half is shown at
$z=0.85\sigma$ (circles and solid lines), 
$4.25\sigma$ (squares and dashed lines),
$7.65\sigma$ (diamonds and dotted line), 
and $11.05\sigma$ (triangles and dot-dashed lines).
The MD velocity profiles were measured by dividing the fluid space 
into $16$ layers along $z$, each of thickness $H/16=1.7\sigma$.
}\label{16symflow}
\end{figure}

\begin{figure}
\bigskip
\centerline{\psfig{figure=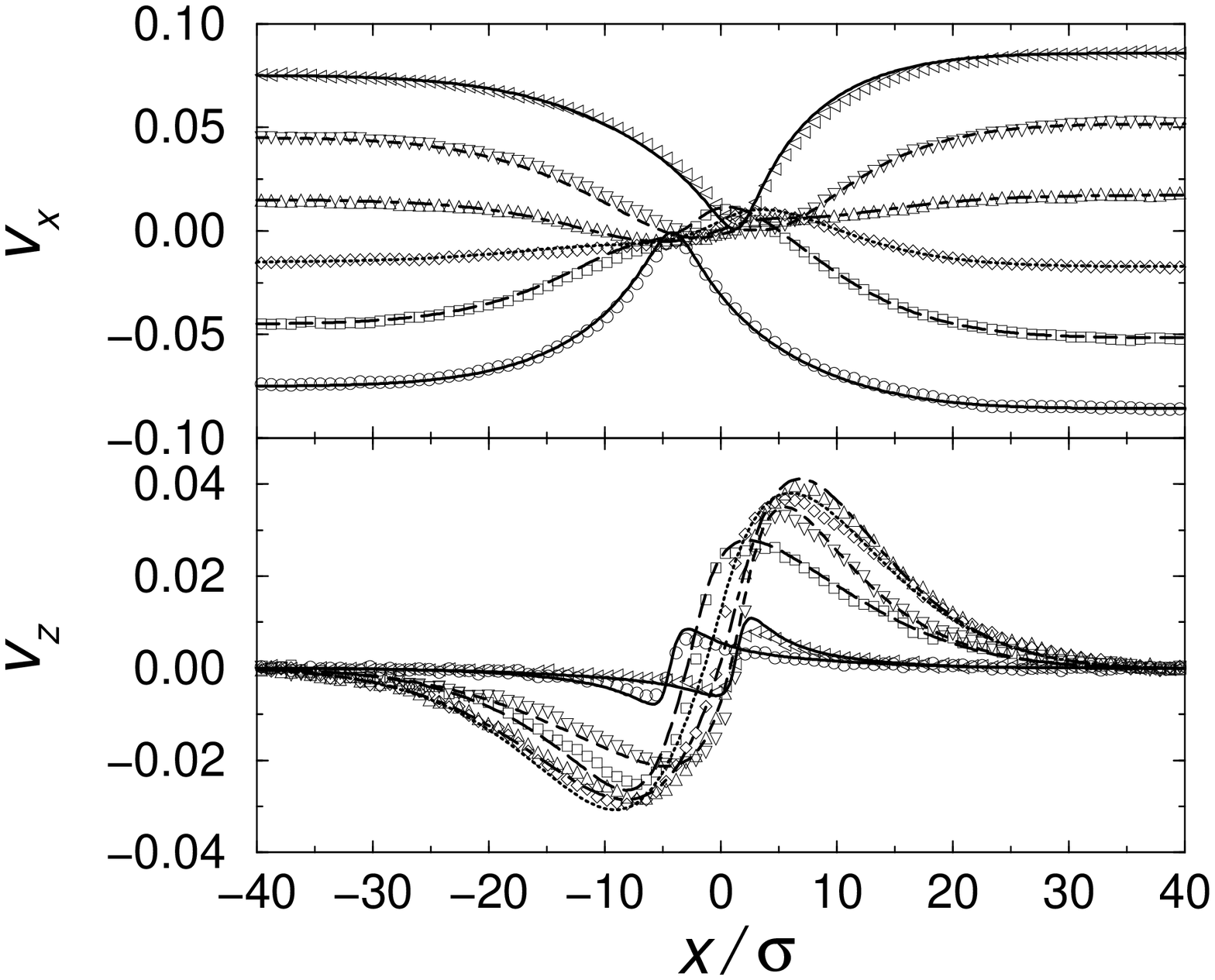,height=7.0cm}}
\caption{Comparison of the MD (symbols) and continuum (lines) 
velocity profiles ($v_x(x)$ and $v_z(x)$ at different $z$ levels)
for an asymmetric case of immiscible Couette flow
($V=0.1(\epsilon/m)^{1/2}$ and $H=27.2\sigma$), shown at
$z=0.85\sigma$ (circles and solid lines), 
$5.95\sigma$ (squares and long-dashed lines),
$11.05\sigma$ (diamonds and dotted line), 
$16.15\sigma$ (up-triangles and dot-dashed lines),
$21.25\sigma$ (down-triangles and dashed lines), 
$26.35\sigma$ (left-triangles and solid lines).
The MD velocity profiles were measured by dividing the fluid space 
into $16$ layers along $z$, each of thickness $H/16=1.7\sigma$.
}\label{16asymflow}
\end{figure}

From Figs. \ref{8symflow}, \ref{16symflow}, \ref{8asymflow}, and 
\ref{16asymflow}, we see that at the MCL, the slip is near-complete, 
i.e., $v_x\approx 0$ and $|v_x^{slip}|\approx V$, while far away from 
the contact line, the flow field is not perturbed by the fluid-fluid
interface and the single-fluid uniform shear flow is recovered. 
The slip amount in the uniform shear flow is $2l_sV/(H+2l_s)$, vanishing 
in the limit of $H\gg l_s$. Here we encounter an intriguing question: 
In a mesoscopic or macroscopic system, what is the slip profile 
which consistently interpolates between the near-complete slip at the MCL 
and the no-slip boundary condition that must hold at regions far away?
Large-scale MD and continuum simulations have been carried out to answer 
this question \cite{power-law}.

\subsubsection{Steady-state fluid-fluid interface}

\begin{figure}
\bigskip
\centerline{\psfig{figure=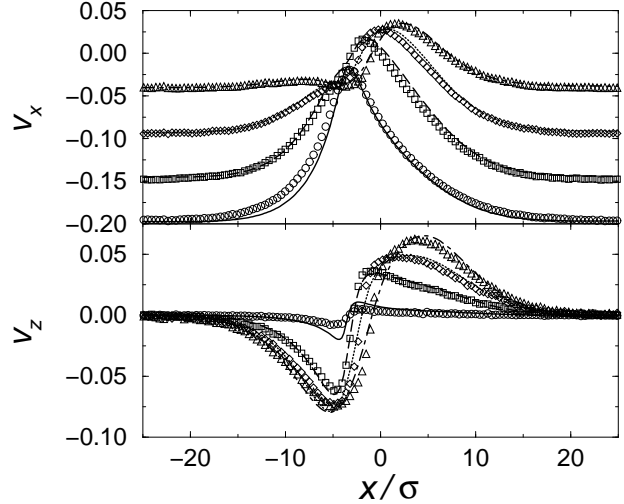,height=7.0cm}}
\caption{Comparison of the MD (symbols) and continuum (lines) 
velocity profiles ($v_x(x)$ and $v_z(x)$ at different $z$ levels)
for a symmetric case of immiscible Couette flow
($V=0.25(\epsilon/m)^{1/2}$ and $H=13.6\sigma$).
The profiles are symmetric about the center plane $z=H/2$,
hence only the lower half is shown at
$z=0.425\sigma$ (circles and solid lines), 
$2.125\sigma$ (squares and dashed lines),
$3.825\sigma$ (diamonds and dotted line), 
and $5.525\sigma$ (triangles and dot-dashed lines).
The MD velocity profiles were measured by dividing the fluid space 
into $16$ layers along $z$, each of thickness $H/16=0.85\sigma$.
}\label{8symflow}
\end{figure}

\begin{figure}
\bigskip
\centerline{\psfig{figure=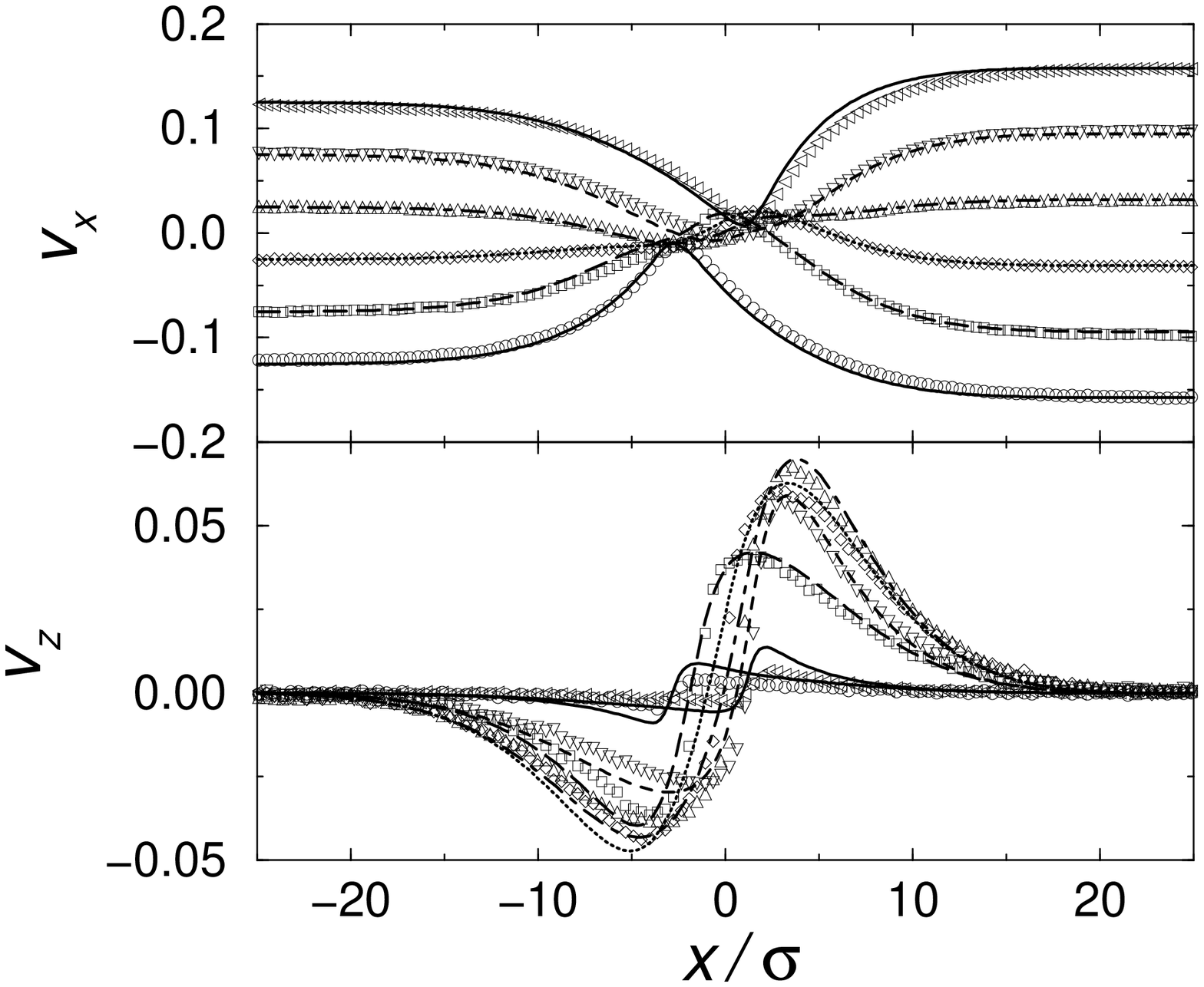,height=7.0cm}}
\caption{Comparison of the MD (symbols) and continuum (lines) 
velocity profiles ($v_x(x)$ and $v_z(x)$ at different $z$ levels)
for an asymmetric case of immiscible Couette flow
($V=0.2(\epsilon/m)^{1/2}$ and $H=13.6\sigma$), shown at
$z=0.425\sigma$ (circles and solid lines), 
$2.975\sigma$ (squares and long-dashed lines),
$5.525\sigma$ (diamonds and dotted line), 
$8.075\sigma$ (up-triangles and dot-dashed lines),
$10.625\sigma$ (down-triangles and dashed lines), 
$13.175\sigma$ (left-triangles and solid lines).
Although the solid lines are used to denote two different $z$ levels,
for each solid line, whether it should be compared to circles or
left-triangles is self-evident (same for the next figure).
The MD velocity profiles were measured by dividing the fluid space 
into $16$ layers along $z$, each of thickness $H/16=0.85\sigma$.
}\label{8asymflow}
\end{figure}

\begin{figure}
\bigskip
\centerline{\psfig{figure=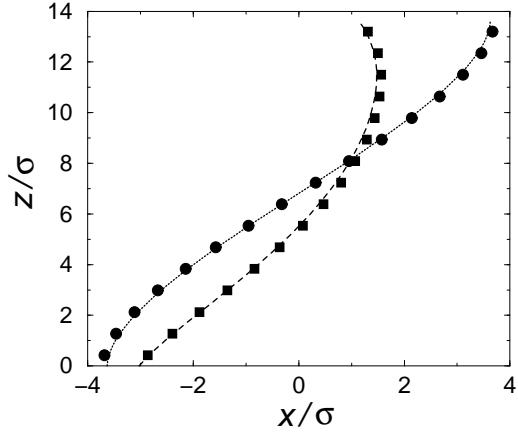,height=6.0cm}}
\caption{Comparison of the MD (symbols) and continuum (lines)
fluid-fluid interface profiles, defined by $\rho_1=\rho_2$ ($\phi=0$). 
The circles and dotted line denote the symmetric immiscible Couette flow 
with $V=0.25(\epsilon/m)^{1/2}$ and $H=13.6\sigma$; 
the squares and dashed line denote the asymmetric immiscible Couette flow 
with $V=0.2(\epsilon/m)^{1/2}$ and $H=13.6\sigma$.
The MD profiles were measured by dividing the fluid space 
into $16$ layers along $z$, each of thickness $H/16=0.85\sigma$.
}\label{interface8}
\end{figure}

In Fig. \ref{interface8} we show the MD and continuum fluid-fluid 
interface profiles for one symmetric and one asymmetric cases 
whose velocity fields are shown in Figs. \ref{8symflow} and \ref{8asymflow}.

\subsection{Immiscible Poiseuille flow}

\begin{figure}
\bigskip
\centerline{\psfig{figure=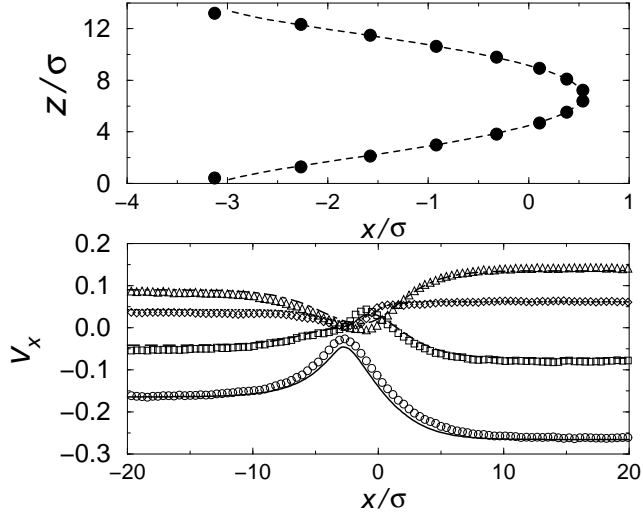,height=7.0cm}}
\caption{Comparison of the MD (symbols) and continuum (lines)
results for an asymmetric case of immiscible Poiseuille flow.
An external force $mg_{ext}=0.05\epsilon/\sigma$ is
applied on each fluid molecule in the $x$ direction,
and the two walls, separated by $H=13.6\sigma$, move at a constant speed 
$V=0.51(\epsilon/m)^{1/2}$ in the $-x$ direction in order to maintain 
a time-independent steady-state interface.
(a) Fluid-fluid interface profiles, defined by $\rho_1=\rho_2$ ($\phi=0$). 
(b) $v_x(x)$ at different $z$ levels.
The profiles are symmetric about the center plane $z=H/2$,
hence only the lower half is shown at
$z=0.425\sigma$ (circles and solid line), 
$2.125\sigma$ (squares and dashed line),
$3.825\sigma$ (diamonds and dotted line), 
and $5.525\sigma$ (triangles and dot-dashed line).
The MD profiles were measured by dividing the fluid space 
into $16$ layers along $z$, each of thickness $H/16=0.85\sigma$.
}\label{psf}
\end{figure}

In order to further verify that the continuum model is local 
and the parameter values are local properties, hence applicable to  
different flow geometries, we have carried out MD simulations and
continuum calculations for immiscible Poiseuille flows. We find that
the continuum model with the same set of parameters is capable of 
reproducing the MD results for velocity field and fluid-fluid
interface profile, shown in Fig. \ref{psf}.
Similar to what we have observed in Couette flows, here at the MCL
the slip is near-complete, i.e., $v_x\approx 0$ and $|v_x^{slip}|\approx V$,
while far away from the contact line, the flow field is not perturbed 
by the fluid-fluid interface and the single-fluid unidirectional 
Poiseuille flow is recovered. In particular, the slip amount in 
the unidirectional Poiseuille flow vanishes in the limit of $H\gg l_s$.

We emphasize that the overall agreement is excellent in all cases 
(from Fig. \ref{8symflow} to \ref{psf}), therefore the validity of the GNBC 
and the hydrodynamic model is well affirmed.

\subsection{Flow in narrow channels}

\begin{figure}
\bigskip
\centerline{\psfig{figure=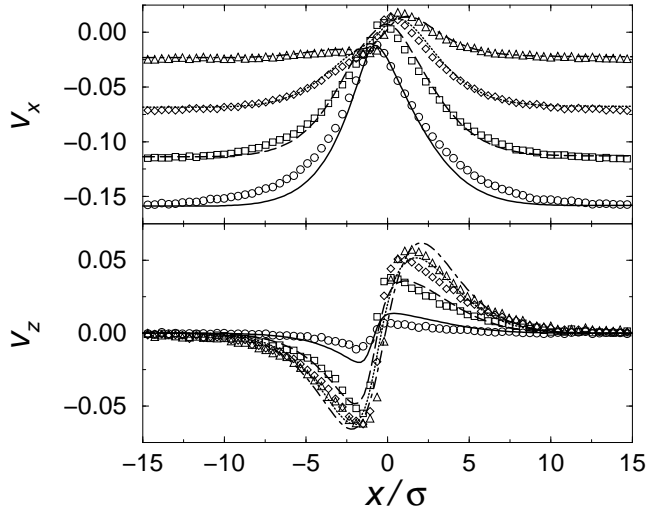,height=7.0cm}}
\caption{Comparison of the MD (symbols) and continuum (lines) 
velocity profiles ($v_x(x)$ and $v_z(x)$ at different $z$ levels)
for a symmetric case of immiscible Couette flow
($V=0.25(\epsilon/m)^{1/2}$ and $H=6.8\sigma$).
The profiles are symmetric about the center plane $z=H/2$,
hence only the lower half is shown at
$z=0.425\sigma$ (circles and solid lines), 
$1.275\sigma$ (squares and dashed lines),
$2.125\sigma$ (diamonds and dotted line), 
and $2.975\sigma$ (triangles and dot-dashed lines).
The MD velocity profiles were measured by dividing the fluid space 
into $8$ layers along $z$, each of thickness $H/8=0.85\sigma$.
}\label{4symflow}
\end{figure}

It is generally believed that continuum hydrodynamic predictions tend to
deviate more from the ``true'' MD results as the channel is further narrowed
\cite{subcontinuum}. This tendency has indeed been observed but the deviation 
is not serious for $H$ as small as $6.8\sigma$, as shown in Fig. \ref{4symflow}. 
This deviation is presumably due to the short-range molecular layering induced 
by the rigid wall \cite{israelachivili}. As the channel becomes narrower, 
the layered part of the fluids occupies a relatively larger space, 
thus making the MD-continuum comparison less satisfactory.

\subsection{Temperature effects}

\begin{figure}[h]
\bigskip
\centerline{\psfig{figure=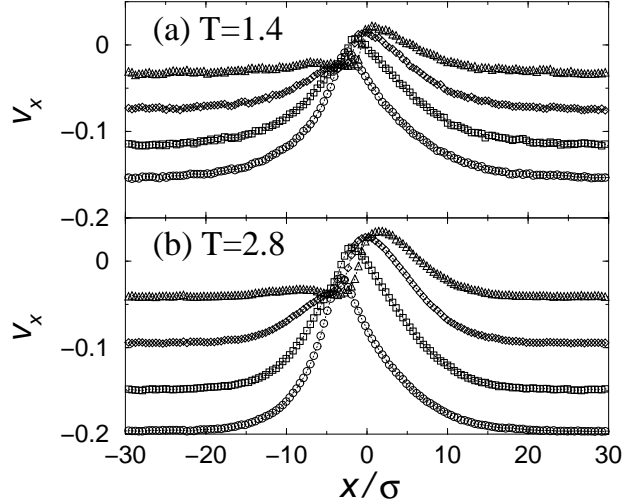,height=7.0cm}}
\caption{Comparison of the MD (symbols) and continuum (lines) 
velocity profiles ($v_x(x)$ at different $z$ levels)
for two symmetric cases of immiscible Couette flow at different temperatures.
(a) The case of $T=1.4\epsilon/k_B$, $V=0.25(\epsilon/m)^{1/2}$, and 
$H=13.6\sigma$, with a weak wall-fluid interaction 
and a high ratio of $\rho_w$ to $\rho$ (compared to case b here).
(b) The case of $T=2.8\epsilon/k_B$, $V=0.25(\epsilon/m)^{1/2}$, and 
$H=13.6\sigma$, with $\epsilon_{wf}=1.16\epsilon$, $\sigma_{wf}=1.04\sigma$, 
$\delta_{wf}=1$, and $\rho_w/\rho=2.3$.
The profiles are symmetric about the center plane $z=H/2$,
hence only the lower half is shown at
$z=0.425\sigma$ (circles), $2.125\sigma$ (squares),
$3.825\sigma$ (diamonds), and $5.525\sigma$ (triangles).
The MD velocity profiles were measured by dividing the fluid space 
into $16$ layers along $z$, each of thickness $H/16=0.85\sigma$.
The slip length for (a) is larger than that for (b). Therefore,
far away from the contact line, the slip amount in (a) 
($\approx 0.1(\epsilon/m)^{1/2}$) is larger than that in (b)
($\approx 0.05(\epsilon/m)^{1/2}$).
}\label{temperature}
\end{figure}

Most of our MD results have been obtained by setting the temperature at 
$2.8\epsilon/k_B$, above the liquid-gas coexistence region. Such a high
temperature was used to reduce the fluid layering at the solid wall 
\cite{israelachivili}. Similar two-fluid simulations have also been performed 
for temperatures ranging from $1.2\epsilon/k_B$ to $3.0\epsilon/k_B$.
We find that the MD results can always be reproduced by our continuum model,
with material parameters (e.g. viscosity, interfacial tension, and 
slip length) varying with the temperature. In Fig. \ref{temperature} 
we show the MD velocity profiles obtained at the temperatures 
$1.4\epsilon/k_B$ and $2.8\epsilon/k_B$. 
It can be seen that they are qualitatively very close to each other. 
The quantitative difference is due to the different material parameters
at different temperatures. 

Finally we list in Table \ref{table-parameters} the parameter values 
in the continuum hydrodynamic model, used for the MD-continuum comparison 
at $T=2.8\epsilon/k_B$.

\begin{table}
\centering
\caption{Parameter values used in the continuum hydrodynamic calculations
for the MD-continuum comparison at $T=2.8\epsilon/k_B$.}
\label{table-parameters}
\begin{tabular}{lll}
\hline\noalign{\smallskip}
&&\\
\noalign{\smallskip}\hline\noalign{\smallskip}
$ \rho_m\approx 0.81m/\sigma^3 $ & 
$ \eta\approx 1.95\sqrt{\epsilon m}/\sigma^2$ &  \\
$ l_{s1}=\eta/\beta_1\approx 1.3\sigma $ & 
$ l_{s2}=\eta/\beta_2\approx 1.3\sigma\;\;{\rm or}\;\;3.3\sigma$ & \\
$ \xi\approx 0.33\sigma$ &
$ \gamma\approx 5.5\epsilon/\sigma^2$ &
$ |\phi_\pm|=1$ \\
$M\approx 0.023\sigma^4/\sqrt{m\epsilon}$ &
$\Gamma\approx 0.66\sigma/\sqrt{m\epsilon}$ &
$\cos\theta_s^{surf}=0\;\;{\rm or}\;\;\approx 0.38$\\
\noalign{\smallskip}\hline
\end{tabular}
\end{table}

\section{Molecular Dynamics II}\label{md2}

We have formulated a continuum hydrodynamic model based on the CH free energy
and the GNBC. The solutions of the model equations agree with the MD results 
remarkably well. This indicates that our model captures the right physics, and hence more MD evidences can be obtained to support the continuum GNBC
(Eq. (\ref{he3})) which takes into account the uncompensated Young stress.
This necessarily requires a reliable measurement of fluid stress near 
the solid surface, plus a decomposition of the tangential stress into
two components, one being viscous and the other interfacial, as expressed by
Eq. (\ref{continuum-hydro-tangential-stress}).

\subsection{Measurement of fluid stress}

Irving and Kirkwood \cite{irving-kirkwood} have shown that in 
the hydrodynamic equation of momentum transport, the stress tensor 
(flux of momentum) may be expressed in terms of the molecular variables as
$$
{\mbox{\boldmath$\sigma$}}({\bf r},t)=
{\mbox{\boldmath$\sigma$}}_K({\bf r},t)+
{\mbox{\boldmath$\sigma$}}_U({\bf r},t),
$$
where ${\mbox{\boldmath$\sigma$}}_K$ is the kinetic contribution to 
the stress tensor, given by
$$
{\mbox{\boldmath$\sigma$}}_K({\bf r},t)=-\left\langle\sum_i m_i 
\left[\displaystyle\frac{{\bf p}_i}{m_i}-{\bf V}({\bf r},t)\right]
\left[\displaystyle\frac{{\bf p}_i}{m_i}-{\bf V}({\bf r},t)\right]
\delta({\bf x}_i-{\bf r})\right\rangle,
$$
and ${\mbox{\boldmath$\sigma$}}_U$ is the contribution of 
intermolecular forces to the stress tensor, given by
$$
{\mbox{\boldmath$\sigma$}}_U({\bf r},t)=-\displaystyle\frac{1}{2}
\left\langle\sum_i\sum_{j\ne i}({\bf x}_i-{\bf x}_j){\bf F}_{ij}
\delta({\bf x}_i-{\bf r})\right\rangle.
$$
Here $m_i$, ${\bf p}_i$, and ${\bf x}_i$ are respectively
the mass, momentum, and position of molecule $i$,
${\bf V}({\bf r},t)$ is the local average velocity, ${\bf F}_{ij}$
is the force on molecule $i$ due to molecule $j$, and 
$\langle \cdot\cdot\cdot \rangle$ means taking the average according to 
a normalized phase-space probability distribution function.

The Irving-Kirkwood expression has been widely used for 
stress measurement in MD simulations. However, as pointed out by 
the authors themselves \cite{irving-kirkwood}, the above expression for 
${\mbox{\boldmath$\sigma$}}_U$ represents only the leading term 
in an asymptotic expansion, accurate when the interaction range 
is small compared to the range of hydrodynamic variation.
As a consequence, this leading-order expression for
${\mbox{\boldmath$\sigma$}}_U$ is not accurate enough near a fluid-fluid or 
a fluid-solid interface. Unfortunately, this point has not been taken 
seriously.
For the MCL problem, a knowledge of the stress distributions at both 
the fluid-fluid and the fluid-solid interfaces is of fundamental importance 
to a correct understanding of the underlying physical mechanism. 
Therefore, a reliable stress measurement method is imperative. 

\begin{figure}[h]
\bigskip
\centerline{\psfig{figure=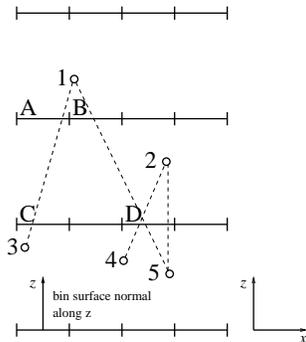,height=4.5cm}}
\caption{Schematic illustration for the measurement of 
the $zx$ component of ${\mbox{\boldmath$\sigma$}}_U$. 
The horizontal solid lines (separated by short vertical lines) 
represent bin surfaces with surface normal along the $z$ direction. 
Circles denote fluid molecules. The dashed lines connect pairs of 
interacting molecules. Here the bin surfaces and the molecules are 
projected onto the $xz$ plane. 
Molecules that appear to be close to each other may not be in 
the interaction range if their distance along $y$ is too large.
A pair of interacting molecules may act across more than one
bin surface. Here the (1,3) pair acts across the surfaces A and C
while the (1,5) pair acts across the surfaces B and D.
At each bin surface the stress measurement must run over all the
pairs acting across that surface. For surface D, there are 
three pairs of interacting molecules (1,5), (2,4), and (2,5)
that contribute to the $zx$ component of ${\mbox{\boldmath$\sigma$}}_U$.
}\label{stress-measure}
\end{figure}

To have spatial resolution along the $x$ and $z$ directions, the sampling
region was evenly divided into bins, each $\Delta x=0.425\sigma$ by 
$\Delta z =0.85\sigma$ in size. The stress components $\sigma_{xx}$ and
$\sigma_{zx}$ were obtained from the time averages of the kinetic momentum 
transfer plus the fluid-fluid interaction forces across the fixed-$x$ 
and $z$ bin surfaces. 
More precisely, we have directly measured the $x$ component of 
the fluid-fluid interaction forces acting across the $x(z)$ bin surfaces, 
in order to obtain the $xx(zx)$ component of ${\mbox{\boldmath$\sigma$}}_U$. 
For example, in measuring $\sigma_{Uzx}$ at a designated $z$-oriented
bin surface, we recorded all the pairs of fluid molecules interacting across 
that surface. Here ``acting/interacting across'' means that the line 
connecting a pair of molecules intersects the bin surface 
(the so-called Irving-Kirkwood convention \cite{irving-kirkwood}). 
For those pairs, we then computed $\sigma_{Uzx}$ at the given bin surface from
$$
\sigma_{Uzx}=\displaystyle\frac{1}{\delta s_z}\sum_{(i,j)}F_{ijx},
$$
where $\delta s_z$ is the area of $z$-oriented bin surface,
$(i,j)$ indicate all available pairs of fluid molecules interacting across 
the bin surface, with molecule $i$ being ``inside of $\hat{\bf z}\delta s_z$''
and molecule $j$ being ``outside of $\hat{\bf z}\delta s_z$''
(molecule $i$ is below molecule $j$), and $F_{ijx}$ is the $x$ component 
of the force on molecule $i$ due to molecule $j$.
For a schematic illustration see Fig. \ref{stress-measure}.

\subsection{Boundary-layer tangential force balance}
\label{tangential-force-balance}

From the data of stress measurement, we now present the MD evidence
for the BL tangential force balance, first introduced 
in Sec. \ref{tangential-fluid-force} for obtaining Eq. (\ref{gfslip})
from Eq. (\ref{gwslip}).

\subsubsection{Static tangential force balance}\label{molecular-YE}

\begin{figure}[ht]
\bigskip
\centerline{\psfig{figure=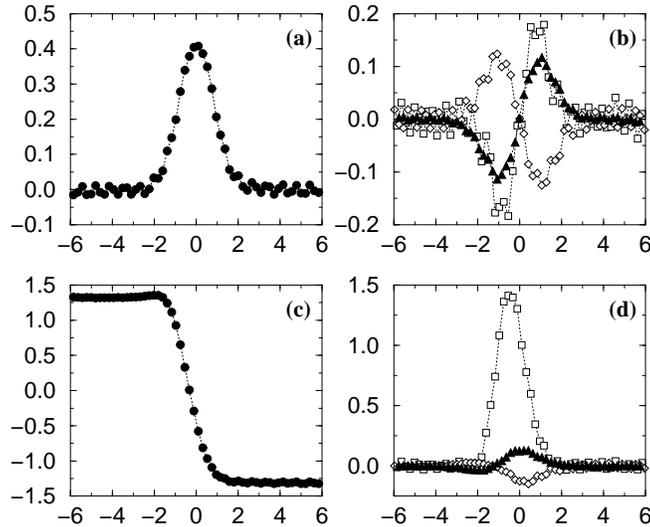,height=7.0cm}}
\caption{
Profiles of $\int_0^{z_0}dz{\sigma}^0_{xx}$, ${\sigma}_{zx}^0(z_0)$, 
${G}_x^{f0}$, and ${G}_x^{w0}$ for the lower BL. 
The horizontal axes are $x/\sigma$.
We show $\int_0^{z_0}dz{\sigma}^0_{xx}$ ($\epsilon/\sigma^2$) in (a) 
for the symmetric case and in (c) for the asymmetric case.
For clarity, ${\sigma}^0_{xx}$ has been vertically displaced such that
in the symmetric case, $\sigma_{xx}^0=0$ far from the interface, and 
in the asymmetric case, $\sigma_{xx}^0=0$ at the center of the interface.
The profiles of ${\sigma}_{zx}^0(z_0)$, ${G}_x^{f0}$, and ${G}_x^{w0}$ 
($\epsilon/\sigma^3$) are plotted in (b) for the symmetric case 
and in (d) for the asymmetric case.
The squares denote ${\sigma}_{zx}^0(z_0)$, the diamonds denote ${G}_x^{f0}$,
and the solid triangles denote ${G}_x^{w0}$.
Both (b) and (d) show $G_x^{w0}(x)=-{G}_x^{f0}(x)$.
}\label{static-balance}
\end{figure}

We start from the tangential force balance in the static configuration
($V=0$). As first pointed out in Sec. \ref{tangential-wall-force}, 
the static tangential wall force $G_x^{w0}$ shows molecular-scale features 
in the contact-line region,
due to the microscopic organization of fluid molecules there.
Then according to the local force balance, static fluid stress must
vary in such a way that the total force density vanishes. 
An integrated form of the static tangential force balance is given by
$G_x^{w0}(x)+{G}_x^{f0}(x)=0$, where 
$G_x^{w0}(x)=\int_0^{z_0}dzg_x^{w0}(x,z)$ and the static tangential
fluid force ${G}_x^{f0}(x)$ is of the form
\begin{equation}\label{static-force-expression}
{G}_x^{f0}(x)=\int_0^{z_0}dz
[\partial_x{\sigma}_{xx}^0(x,z)+\partial_z{\sigma}_{zx}^0(x,z)]
={\sigma}_{zx}^0(x,z_0)+\partial_x\int_0^{z_0}dz{\sigma}^0_{xx}(x,z).
\end{equation}
Here ${\sigma}_{xx}^0$ and ${\sigma}_{zx}^0$ are the $xx$ and $zx$
components of fluid stress in the static configuration, both measured
as reference quantities. In Fig. \ref{static-balance} we show
$\int_0^{z_0}dz{\sigma}^0_{xx}$, ${\sigma}_{zx}^0(z_0)$,
${G}_x^{f0}$, and ${G}_x^{w0}$ (which is the same as in Fig. \ref{wallforce}b).
In the symmetric case, 
$\int_{int}dx{\sigma}_{zx}^0(x,z_0)$ 
$\int_{int}dx\partial_x\int_0^{z_0}dz{\sigma}^0_{xx}(x,z)$, and
$\int_{int}dx\int_0^{z_0}dzg_x^{w0}(x,z)$ all vanish because 
$\theta_s^{surf}=90^\circ$. For the asymmetric case, 
$G_x^{w0}(x)+{G}_x^{f0}(x)=0$ means
\begin{equation}\label{micro-YE}
{\sigma}_{zx}^0(x,z_0)+\partial_x\int_0^{z_0}dz{\sigma}^0_{xx}(x,z)+
\int_0^{z_0}dzg_x^{w0}(x,z)=0,
\end{equation}
which corresponds to the continuum equation (Eq. (\ref{YE2}))
$\left[\sigma_{zx}^{Y}+\partial_x\gamma_{wf}\right]_{\phi_{eq}}=0$
at the solid surface.
Here ${\sigma}_{zx}^0(x,z_0)$ corresponds to the continuum
$\sigma_{zx}^{Y}(\phi_{eq})$ at the solid surface while
$\partial_x\int_0^{z_0}dz{\sigma}^0_{xx}+\int_0^{z_0}dzg_x^{w0}$ corresponds 
to the continuum $\partial_x\gamma_{wf}(\phi_{eq})$ at the solid surface. 
The Young's equation (Eq. (\ref{YE})) 
$\gamma\cos\theta_s^{surf}+\Delta\gamma_{wf}=0$ 
is then obtained through integration, using 
\begin{equation}\label{YE_def1}
\int_{int}dx{\sigma}_{zx}^0(x,z_0)=\int_{int}dx{\sigma}_{zx}^Y=
\gamma\cos\theta_s^{surf}
\end{equation}
and 
\begin{equation}\label{YE_def2}
\int_{int}dx\partial_x\int_0^{z_0}dz{\sigma}^0_{xx}+
\int_{int}dx\int_0^{z_0}dzg_x^{w0}=
\int_{int}dx\partial_x\gamma_{wf}=\Delta\gamma_{wf}.
\end{equation}
Here $\gamma\cos\theta_s^{surf}$ and $\Delta\gamma_{wf}$ are the two 
tangential forces per unit length along the contact line (along $y$), 
the former due to the tilt of the fluid-fluid interface ($\theta_s^{surf}
\ne 90^\circ$) while the latter due to the different wall-fluid interactions
for the two fluid species. 
In fact, equations (\ref{YE_def1}) and (\ref{YE_def2}) are the microscopic 
definitions for the two continuum quantities $\gamma\cos\theta_s^{surf}$ 
and $\Delta\gamma_{wf}$ in the Young's equation, whose validity is based on
the microscopic tangential force balance expressed in Eq. (\ref{micro-YE}).

\subsubsection{Dynamic tangential force balance}

\begin{figure}
\bigskip
\centerline{\psfig{figure=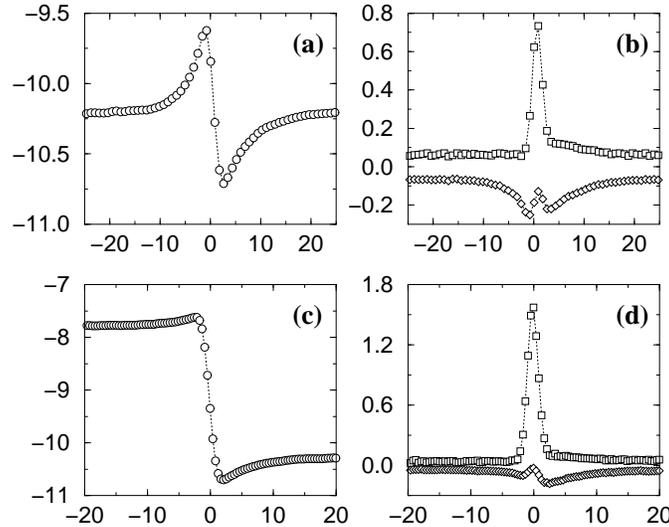,height=7.0cm}}
\caption{
Profiles of $\int_0^{z_0}dz{\sigma}_{xx}$, ${\sigma}_{zx}(z_0)$,
and ${G}_x^{w}$ for the lower BL. The horizontal axes are $x/\sigma$.
We show $\int_0^{z_0}dz{\sigma}_{xx}$ ($\epsilon/\sigma^2$) in (a) 
for the symmetric case ($V=0.25\sqrt{\epsilon/m}$ and $H=13.6\sigma$)
and in (c) for the asymmetric case 
($V=0.2\sqrt{\epsilon/m}$ and $H=13.6\sigma$).
The profiles of ${\sigma}_{zx}(z_0)$ and ${G}_x^{w}$ 
($\epsilon/\sigma^3$) are plotted in (b) for the symmetric case 
and in (d) for the asymmetric case.
The squares denote ${\sigma}_{zx}(z_0)$ and the diamonds denote ${G}_x^{w}$.
Note that in the vicinity of the contact line, ${G}_x^{w}+ 
{\sigma}_{zx}(z_0)\ne 0$. The importance of the $x$-gradient of 
the $z$-integrated normal stress $\partial_x\int_0^{z_0}dz{\sigma}_{xx}$ 
is therefore evident. 
}\label{dyn-balance}
\end{figure}

\begin{figure}
\bigskip
\centerline{\psfig{figure=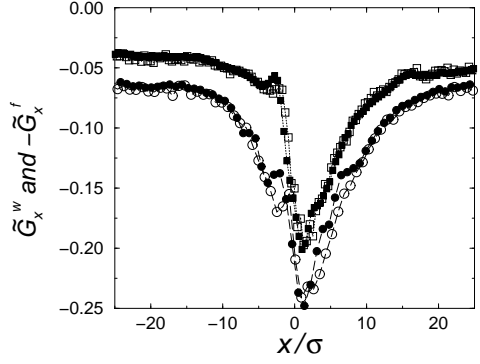,height=5.0cm}}
\caption{Hydrodynamic force balance for the lower BL. 
The circles represent the symmetric case 
($V=0.25\sqrt{\epsilon/m}$ and $H=13.6\sigma$); 
the squares represent the asymmetric case
($V=0.2\sqrt{\epsilon/m}$ and $H=13.6\sigma$).
The empty symbols denote $\tilde{G}_x^{w}$; the solid symbols denote
$-\tilde{G}_x^{f}$. It is seen that $\tilde{G}_x^{w}(x)=-\tilde{G}_x^{f}(x)$.
}\label{hydro-balance}
\end{figure}

An integrated form of the dynamic tangential force balance is given by
$G_x^{w}(x)+{G}_x^{f}(x)=0$, where 
$G_x^{w}(x)=\int_0^{z_0}dzg_x^{w}(x,z)$ and the dynamic tangential
fluid force ${G}_x^{f}(x)$ is of the form
\begin{equation}\label{dynamic-force-expression}
{G}_x^{f}(x)=\int_0^{z_0}dz
[\partial_x{\sigma}_{xx}(x,z)+\partial_z{\sigma}_{zx}(x,z)]
={\sigma}_{zx}(x,z_0)+\partial_x\int_0^{z_0}dz{\sigma}_{xx}(x,z).
\end{equation}
Here ${\sigma}_{xx}$ and ${\sigma}_{zx}$ are the $xx$ and $zx$
components of fluid stress in the dynamic configuration.
In Fig. \ref{dyn-balance} we show
$\int_0^{z_0}dz{\sigma}_{xx}$, ${\sigma}_{zx}(z_0)$, and ${G}_x^{w}$.
From a comparison between Figs. \ref{static-balance} and \ref{dyn-balance},
we can see that the dynamic quantities
$\int_0^{z_0}dz{\sigma}_{xx}$, ${\sigma}_{zx}(z_0)$, and ${G}_x^{w}$
indeed show features seen from the static quantities
$\int_0^{z_0}dz{\sigma}^0_{xx}$, ${\sigma}_{zx}^0(z_0)$, and ${G}_x^{w0}$,
respectively. This is particularly evident in the asymmetric case
where the static quantities vary more appreciably.
The reason to take the static quantities as reference quantities
is now clear: a hydrodynamic quantity must be obtained from 
the corresponding dynamic quantity by subtracting its static part, 
formally expressed as
$$\tilde{Q}=\left[Q\right]_{dynamic}-\left[Q\right]_{static},$$
where the over tilde denotes the hydrodynamic quantity
(first introduced for $\tilde{G}_x^{w}$ in Sec. \ref{tangential-wall-force}).
In Fig. \ref{hydro-balance} we show the MD evidence for 
the BL hydrodynamic tangential force balance, which is expressed as
$\tilde{G}_x^{w}(x)+\tilde{G}_x^{f}(x)=0$. This equation is necessary for
Eqs. (\ref{gwslip}) and (\ref{gfslip}) to hold simultaneously.

In summary, to verify the static/dynamic tangential force balance, 
we need to (1) identify the BL where the tangential wall force $G_x^{w(0)}$
is distributed; (2) measure the normal and tangential components
of stress $\sigma_{xx}^{(0)}$ and $\sigma_{zx}^{(0)}$ according to the
original definition of stress; (3) calculate the tangential fluid force 
$G_x^{f(0)}$ according to 
Eq. (\ref{static-force-expression})/(\ref{dynamic-force-expression}).

\subsection{Tangential Young stress}\label{stress-decomposition}

\begin{figure}[h]
\bigskip
\centerline{\psfig{figure=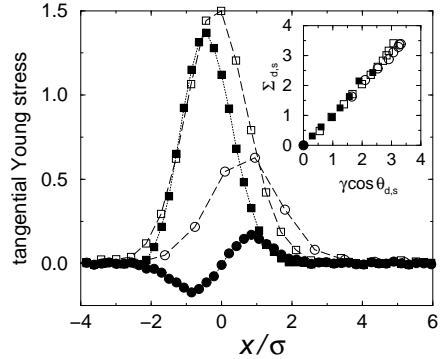,height=5.0cm}}
\caption{Dynamic and static Young stresses at $z=z_0$.
The solid circles denote $\sigma_{zx}^0$ in the static symmetric case; 
the empty circles denote $\sigma_{zx}^Y$ in the dynamic symmetric case
($V=0.25\sqrt{\epsilon/m}$ and $H=13.6\sigma$);
the solid squares denote $\sigma_{zx}^0$ in the static asymmetric case; 
the empty squares denote $\sigma_{zx}^Y$ in the dynamic asymmetric case
($V=0.2\sqrt{\epsilon/m}$ and $H=13.6\sigma$). 
Here $\sigma_{zx}^Y$ was obtained by subtracting the viscous component
$\eta(\partial_z v_x+\partial_x v_z)$ from the total tangential stress
$\sigma_{zx}$. Inset: $\Sigma_{d,s}$ plotted as a function of 
$\gamma\cos\theta_{d,s}$ at different $z$ levels. 
Here $\theta_{d,s}$ was measured from the time-averaged
interfacial profiles. The symbols have the same correspondence as 
in the main figure. The data indicate $\Sigma_{d,s}=\gamma\cos\theta_{d,s}$.
}\label{figyoung}
\end{figure}

Here we present the MD evidence for the decomposition of the tangential stress.
In a dynamic configuration, away from the interfacial region the tangential 
viscous stress $\sigma_{zx}^v=\eta(\partial_z v_x+\partial_x v_z)$ is the only 
component in the (single-fluid) tangential stress $\sigma_{zx}$. 
But in the (two-fluid) interfacial region, the tangential stress $\sigma_{zx}$
can be decomposed into a viscous component $\sigma_{zx}^v$ and 
a non-viscous component $\sigma_{zx}^Y$: 
\begin{equation}\label{decomposition}
\sigma_{zx}=\sigma_{zx}^v+\sigma_{zx}^Y,
\end{equation}
where $\sigma_{zx}^v$ is still $\eta(\partial_z v_x+\partial_x v_z)$ and
$\sigma_{zx}^Y$ is the tangential Young stress, satisfying 
\begin{equation}\label{dynamic-young-stress}
\Sigma_d\equiv\int_{int} dx \sigma_{zx}^Y(x,z)=\gamma\cos\theta_d(z).
\end{equation}
Here $\theta_d(z)$ is the dynamic interfacial angle at level $z$.
Equation (\ref{decomposition}) is essential to obtaining 
Eq. (\ref{continuum-hydro-tangential-stress}) for $\tilde{\sigma}_{zx}(x,0)$.
In a static configuration, the viscous stress $\sigma_{zx}^v$ vanishes
and $\sigma_{zx}$ becomes $\sigma_{zx}^0$, satisfying
\begin{equation}\label{static-young-stress}
\Sigma_s\equiv\int_{int} dx \sigma_{zx}^0(x,z)=\gamma\cos\theta_s(z),
\end{equation}
where $\theta_s(z)$ is the static interfacial angle at level $z$.
Figure \ref{figyoung} shows that both $\sigma_{zx}^Y$ and $\sigma_{zx}^0$
are nonzero in the interfacial region only. The inset to Fig. \ref{figyoung}
shows the evidence for Eqs. (\ref{dynamic-young-stress}) and
(\ref{static-young-stress}), which identify $\sigma_{zx}^Y$ and $\sigma_{zx}^0$
as the dynamic and static Young stresses.

\begin{figure}[ht]
\bigskip
\centerline{\psfig{figure=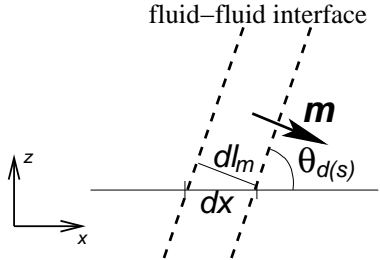,height=3.50cm}}
\caption{Schematic illustration for the origin of the tangential
Young stress $\sigma_{zx}^{Y(0)}$ as an off-diagonal component of 
the microscopic stress tensor.
}\label{figYS}
\end{figure}

Equations (\ref{dynamic-young-stress}) and (\ref{static-young-stress})
can be derived from the mechanical definition for the interfacial tension
$\gamma$ \cite{interfacial-tension}:
$$\gamma=\int_{int} dl_m\left[P_\bot(l_m)-P_\|(l_m)\right],$$ 
i.e., $\gamma$ is the integral (along the interface normal 
across the interface) of the difference between the normal and parallel 
components of the pressure, where $l_m$ is the coordinate along 
the interface normal $\bf m$, and $P_\bot$ and $P_\|$ are the pressure-tensor 
components normal and parallel to the interface, respectively.
(Note that far away from the interface the pressure is isotropic 
and $P_\bot=P_\|$). When the interfacial angle $\theta_d$ 
(or $\theta_s$) is $90^\circ$, the interface normal $\bf m=
\bf x$ and the non-viscous stress tensor in the interfacial region 
is diagonal in the $xyz$ coordinate system:
$${\mbox{\boldmath$\sigma$}}_{non-viscous}=\left[\begin{array}{ccc} 
-P_\bot & 0 & 0 \\ 0 & -P_\| & 0 \\ 0 & 0 & -P_\| \end{array}\right]=
-P_\bot{\bf I}+(P_\bot-P_\|)({\bf I}-{\bf m}{\bf m}),$$
where ${\bf I}$ is the identity matrix. According to this expression,
when the interfacial angle $\theta_d$ (or $\theta_s$) deviates from 
$90^\circ$ (see Fig. \ref{figYS}), the Young stress $\sigma_{zx}^Y$ 
(or $\sigma_{zx}^0$) arises from the interfacial stress anisotropy 
as the off-diagonal $zx$ component of the microscopic stress tensor:
$$\begin{array}{ll}
\sigma_{zx}^{Y(0)} &
={\bf z}\cdot{\mbox{\boldmath$\sigma$}}_{non-viscous}\cdot{\bf x}
=(P_\bot-P_\|)\left[{\bf z}\cdot({\bf I}-{\bf m}{\bf m})\cdot{\bf x}\right]\\
& =-(P_\bot-P_\|)m_zm_x=(P_\bot-P_\|)\cos\theta_{d(s)}\sin\theta_{d(s)},
\end{array}$$
where $m_z=-\cos\theta_{d(s)}$ and $m_x=\sin\theta_{d(s)}$.
It follows that
$$\begin{array}{ll}
\Sigma_{d(s)}\equiv\int_{int} dx \sigma_{zx}^{Y(0)}(x,z)\
&=\int_{int} dx (P_\bot-P_\|)\cos\theta_{d(s)}\sin\theta_{d(s)}\\
&=\int_{int} dl_m (P_\bot-P_\|)\cos\theta_{d(s)}\\
&=\gamma\cos\theta_{d(s)},
\end{array}$$
where $\theta_{d(s)}$ is treated as a constant along $x$
and $dx\sin\theta_{d(s)}=dl_m$.

\section{The Generalized Navier Boundary Condition}\label{gnbc-derivation}

We want to ``derive'' the continuum GNBC using the MD results in 
Secs. \ref{md1} and \ref{md2}. 
For this purpose we first need to establish the correspondence between 
the stress components measured in MD and those defined in the continuum
hydrodynamics. This correspondence is essential to obtaining 
the microscopic dynamic contact angle $\theta_d^{surf}$, which is defined in 
the continuum hydrodynamics (see Eq. (\ref{unYS1}) and (\ref{unYS2})) 
but not directly measurable in MD simulations (because of the diffuse BL). 

\subsection{MD-continuum correspondence}

It has been verified that for a BL of finite thickness, the GNBC is given by
\begin{equation}\label{gnbc-BL}
\begin{array}{ll}
\beta v^{slip}_x(x) & =G_x^f(x)-G_x^{f0}(x)\\ &
=\displaystyle\frac{\partial}{\partial x}
\int_0^{z_0}dz\left[\sigma_{xx}(x,z)-\sigma_{xx}^0(x,z)\right]+
\left[\sigma_{zx}(x,z_0)-\sigma_{zx}^0(x,z_0)\right],
\end{array}
\end{equation}
in which only MD measurable quantities are involved. 
Now we interpret these MD-measured quantities in terms of
the various continuum variables in the hydrodynamic model.
In so doing it is essential to note the following.\\
(1) $\sigma_{xx}$ can be decomposed into a molecular component
and a hydrodynamic component: $\sigma_{xx}=T_{xx}+\sigma_{xx}^{HD}$.
Meanwhile, $\sigma_{xx}^0$ can be composed into the same molecular component
and a hydrostatic component: $\sigma_{xx}^0=T_{xx}+\sigma_{xx}^{HS}$.
Physically, $T_{xx}$ is the normal stress $\sigma_{xx}^0$ for the case of
a {\it flat}, static fluid-fluid interface. Such an interface exists in the
symmetric case ($\theta_s^{surf}=90^\circ$) for any value of $H$. It
also exists in the asymmetric case ($\theta_s^{surf}\ne 90^\circ$) 
for $H\rightarrow \infty$ (with vanishing curvature $\sim 1/H$). 
In either case, the interface has zero curvature 
and the hydrostatic stress $\sigma_{xx}^{HS}$ vanishes according to 
the Laplace's equation: $\sigma_{xx}^0\rightarrow T_{xx}$
as $\sigma_{xx}^{HS}\rightarrow 0$.
(To be more precise, $\gamma\cos\theta_s^{surf}$ and $\Delta\gamma_{wf}$ 
should be defined in Eqs. (\ref{YE_def1}) and (\ref{YE_def2}) when 
the fluid-fluid interface has zero curvature. 
However, in the asymmetric case where $\Delta\gamma_{wf}$ 
$(=-\gamma\cos\theta_s^{surf})$ is nonzero, there is a static interfacial
curvature $\approx 2\cos\theta_s^{surf}/H$. This results in a hydrostatic 
$\sigma_{xx}^{HS}$, which should be subtracted from $\sigma_{xx}^0$
in the left-hand side of Eq. (\ref{YE_def2}). That is, $\Delta\gamma_{wf}$ 
should be obtained with $\sigma_{xx}^0$ replaced by $T_{xx}$. Meanwhile,
due to the static interfacial curvature, the static interfacial angle
$\theta_s(z_0)$ determined by $\cos\theta_s(z_0)=
\gamma^{-1}\int_{int}dx{\sigma}_{zx}^0(x,z_0)$ is a bit different from 
the true $\theta_s^{surf}$. In fact, $\cos\theta_s(z_0)$ deviates from 
$\cos\theta_s^{surf}$ by the BL-integrated curvature $\approx
2z_0\cos\theta_s^{surf}/H$.)
The molecular component $T_{xx}$ exists even if there is no
hydrodynamic fluid motion or fluid-fluid interfacial curvature.
On the contrary, the hydrodynamic component $\sigma_{xx}^{HD}$ arises from 
the hydrodynamic fluid motion and interfacial curvature.
In the static configuration, $\sigma_{xx}^{HD}$ becomes $\sigma_{xx}^{HS}$,
which comes from the interfacial curvature.\\
(2) $\sigma_{zx}(x,z_0)$ can be decomposed into a viscous component 
plus a Young component :
$\sigma_{zx}(x,z_0)=\sigma_{zx}^v(x,z_0)+\sigma_{zx}^Y(x,z_0)$
with $\sigma_{zx}^v=\eta(\partial_zv_x+\partial_xv_z)$
and $\int_{int}dx\sigma_{zx}^Y(x,z_0)=\gamma\cos\theta_d(z_0)$
(Eqs. (\ref{decomposition}) and (\ref{dynamic-young-stress})).\\
(3) $\sigma_{zx}^0(x,z_0)$ is the static Young stress: i.e.,
$\int_{int}dx\sigma_{zx}^0(x,z_0)=\gamma\cos\theta_s(z_0)$
(Eq. (\ref{static-young-stress})).

Using the above relations, we integrate Eq. (\ref{gnbc-BL}) along $x$
across the fluid-fluid interface and obtain
\begin{equation}\label{gnbc-BL-int0}
\begin{array}{ll}
\int_{int}dx\beta v^{slip}_x(x) = &
\Delta\left[\int_0^{z_0}dz\sigma_{xx}^{HD}(x,z)\right]
+\int_{int}dx\sigma_{zx}^v(x,z_0)+\gamma\cos\theta_d(z_0)\\ &
-\Delta\left[\int_0^{z_0}dz\sigma_{xx}^{HS}(x,z)\right]
-\gamma\cos\theta_s(z_0),
\end{array}
\end{equation}
where $\Delta\left[\int_0^{z_0}dz\sigma_{xx}^{HD(HS)}\right]$
is the change of the $z$-integrated $\sigma_{xx}^{HD(HS)}$ across 
the interface: $$\Delta\left[\int_0^{z_0}dz\sigma_{xx}^{HD(HS)}\right]
\equiv\int_{int}dx \partial x \int_0^{z_0}dz\sigma_{xx}^{HD(HS)}.$$ 
According to the Laplace's equation, the hydrostatic
stress is directly related to the static curvature $\kappa_s$:
$$-\Delta\sigma_{xx}^{HS}=\gamma\kappa_s,$$
and the $z$-integrated curvature $\int_0^{z_0}dz \kappa_s$ equals to
$\cos\theta_s(z_0)-\cos\theta_s^{surf}$. Hence, 
\begin{equation}\label{extrap-static}
-\Delta\int_0^{z_0}dz\sigma_{xx}^{HS}
=\gamma\int_0^{z_0}dz \kappa_s
=\gamma\left[\cos\theta_s(z_0)-\cos\theta_s^{surf}\right].
\end{equation}
Substituting Eq. (\ref{extrap-static}) into Eq. (\ref{gnbc-BL-int0}) yields
\begin{equation}\label{gnbc-BL-int1}
\int_{int}dx\beta v^{slip}_x(x)\\
=\Delta\int_0^{z_0}dz\sigma_{xx}^{HD}(x,z)+
\int_{int}dx\sigma_{zx}^v(x,z_0)
+\gamma\cos\theta_d(z_0)-\gamma\cos\theta_s^{surf}.
\end{equation}
In order to interpret Eq. (\ref{gnbc-BL-int1}) in the continuum hydrodynamic 
formulation with a sharp BL, it is essential to note the following.\\
(1) The sum of the first three terms on the right-hand side of 
Eq. (\ref{gnbc-BL-int1}) is the net fluid force along $x$ exerted on 
the three fluid sides of a BL fluid element in the interfacial region.\\
(2) The last term in the right-hand side of Eq. (\ref{gnbc-BL-int1}), 
$-\gamma\cos\theta_s^{surf}$, is the net wall force along $x$,
$\Delta\gamma_{wf}=\int_{int}dx\partial_x\gamma_{wf}$, which
arises from the wall-fluid interfacial free energy jump
across the fluid-fluid interface, in accordance with the Young's equation
$\Delta\gamma_{wf}+\gamma\cos\theta_s^{surf}=0$.

\subsection{Extrapolated dynamic contact angle}\label{angle-extrapolation}

Now we take the sharp boundary limit to relate the net fluid force 
$$\Delta\int_0^{z_0}dz\sigma_{xx}^{HD}(x,z)+
\int_{int}dx\sigma_{zx}^v(x,z_0)+\gamma\cos\theta_d(z_0)$$
in Eq. (\ref{gnbc-BL-int1}) to the tangential stresses 
(viscous and non-viscous) at the solid surface. 
The purpose of doing so is to obtain the surface contact angle 
$\theta_d^{surf}$ through extrapolation. Note that $\theta_d^{surf}$
is not directly measurable in MD simulations due to the diffuse BL.
Only the extrapolated $\theta_d^{surf}$ can be compared to the contact
angle in continuum calculations.

In Sec. \ref{sharp-boundary-limit}, we take the sharp boundary limit
by assuming a tangential wall force concentrated at $z=0$:
$\tilde{g}_x^w(x,z)=\tilde{G}_x^w(x)\delta(z)$.
While $\tilde{g}_x^w$ becomes a $\delta$ function, $\tilde{G}_x^w(x)$ 
per unit area remains the same. 
Using the equation of local force balance 
$\partial_x\tilde{\sigma}_{xx}+\partial_z\tilde{\sigma}_{zx}=0$ above $z=0^+$, 
we obtain $\tilde{\sigma}_{zx}(x,0^+)=\tilde{G}_x^f(x)$ as 
the tangential stress at the solid surface (Eq. (\ref{gfslipsharp})).
The extrapolation here follows this spirit.
We turn to the Stokes equation in the BL:
\begin{equation}\label{stokes-eq}
-\partial_xp+\partial_x\sigma_{xx}^v+\partial_z\sigma_{zx}^v
+\mu\partial_x\phi=0,
\end{equation}
obtained from the $x$-component of Eq. (\ref{he1}) by dropping the
inertial term and the external force term.
Integrating Eq. (\ref{stokes-eq}) along $z$ across the BL and then
along $x$ across the fluid-fluid interface, we obtain
\begin{equation}\label{stokes-eq-int}
\begin{array}{ll}
& \Delta\left\{\int_0^{z_0}dz\left[-p(x,z)+\sigma_{xx}^v(x,z)\right]\right\}
+\int_{int}dx\sigma_{zx}^v(x,z_0)+\gamma\cos\theta_d(z_0)\\
= & \int_{int}dx\sigma_{zx}^v(x,0)+\gamma\cos\theta_d^{surf}.
\end{array}
\end{equation}
Two relations have been used in obtaining Eq. (\ref{stokes-eq-int}):\\
(1) The capillary force density in the sharp interface limit 
\cite{sharp-interface-limit} is given by
$$\mu\partial_x\phi\simeq\gamma\kappa\delta(x-x_{int}),$$ 
where $\kappa$ is the interfacial curvature and $x_{int}$ the
location of the interface along $x$ (see Sec. \ref{two-phase-force}).\\ 
(2) The $z$-integrated curvature gives 
$$\int_0^{z_0}dz\kappa=\cos\theta_d(z_0)-\cos\theta_d^{surf}.$$
The local force balance along $x$ is expressed by Eq. (\ref{stokes-eq}). 
Accordingly, the tangential force balance for the BL fluids 
in the integration region ($\int_{int}dx\int_0^{z_0}dz$)
is expressed by Eq. (\ref{stokes-eq-int}), where 
$\Delta\left\{\int_0^{z_0}dz\left[-p(x,z)+\sigma_{xx}^v(x,z)\right]\right\}$
is the net fluid force on the left and right ($\mp x$-oriented) surfaces,
$\int_{int}dx\sigma_{zx}^v(x,z_0)+\gamma\cos\theta_d(z_0)$
is the fluid force on the $z=z_0$ surface, and
$\int_{int}dx\sigma_{zx}^v(x,0)+\gamma\cos\theta_d^{surf}$ is 
the tangential fluid force at the solid surface.

\begin{figure}[h]
\bigskip
\centerline{\psfig{figure=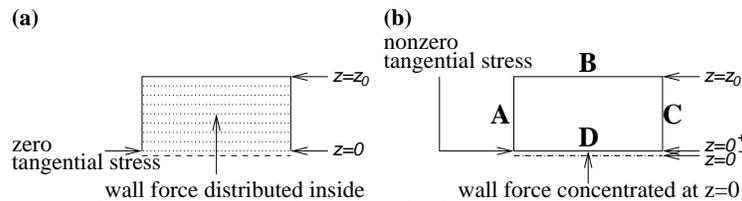,height=2.50cm}}
\caption{(a) For the real tangential wall force continuously distributed
between $z=0$ and $z=z_0$, the tangential stress in the fluid is continuous, 
vanishing at $z=0$. (b) In the sharp boundary limit, the tangential wall force 
is considered to be concentrated at $z=0$. Accordingly, the net fluid force 
on the fluids bound by A, B, C, and D (at $z=0^+$) is considered to be 
zero. In other words, the net fluid force on the three surfaces A, B, and C 
is fully transmitted to the tangential stress at the surface D at $z=0^+$ 
(Eq. (\ref{stokes-eq-int})). So there arises an abrupt change of 
the tangential stress from $0$ at $z=0^-$ to $\tilde{G}_x^f$ at $z=0^+$, 
by which the concentrated wall force is balanced.   
}\label{sharpBL}
\end{figure}

Substituting Eq. (\ref{stokes-eq-int}) into Eq. (\ref{gnbc-BL-int1}) and 
identifying the normal stress $-p+\sigma_{xx}^v$ with $\sigma_{xx}^{HD}$:
$$-p+\sigma_{xx}^v=\sigma_{xx}^{HD},$$ we obtain 
\begin{equation}\label{gnbc-BL-int2}
\int_{int}dx\beta v^{slip}_x(x)=
\int_{int}dx\sigma_{zx}^v(x,0)+\gamma\cos\theta_d^{surf}
-\gamma\cos\theta_s^{surf},
\end{equation}
which is identical to the integral of the continuum GNBC along $x$ 
across the fluid-fluid interface (Eqs. (\ref{he3}), (\ref{unYS0}), 
(\ref{unYS1}), and (\ref{unYS2})).
By doing so, we have assumed the tangential wall force is concentrated 
at $z=0$. (This leads to Eqs. (\ref{stokes-eq}) and (\ref{stokes-eq-int}) 
between $z=0^+$ and $z=z_0$.)
In essence, the above extrapolation is to obtain the tangential stresses 
(viscous and non-viscous) at the solid surface when the limit of
$\tilde{g}_x^w(x,z)=\tilde{G}_x^w(x)\delta(z)$ is taken, as illustrated
in Fig. \ref{sharpBL}.
For $\tilde{G}_x^w(x)$ distributed in the diffuse BL, there is actually
no tangential stress at the solid surface. Only in the sharp boundary limit
does a nonzero tangential stress appear at $z=0^+$, equal to the net
fluid force accumulated from $z=0^-$ to $z=z_0$ in the diffuse BL. 

It is worth emphasizing that the right-hand side of Eq. (\ref{gnbc-BL-int2})
is $\int_{int}dx\tilde{\sigma}_{zx}(x,0^+)$, where 
$\tilde{\sigma}_{zx}(x,0^+)$ is the extrapolated tangential stress
in Eq. (\ref{gfslipsharp}). Its continuum expression is given by
Eq. (\ref{continuum-hydro-tangential-stress}).
Equation (\ref{gnbc-BL-int2}) concludes our ``derivation'' of 
(an integrated form of) the continuum GNBC from the MD results
presented in Secs. \ref{md1} and \ref{md2}. 

\begin{figure}[h]
\bigskip
\centerline{\psfig{figure=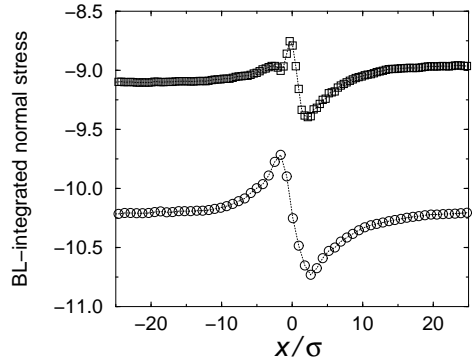,height=5.0cm}}
\caption{$\int_0^{z_0}dz\tilde{\sigma}_{xx}(x,z)=
\int_0^{z_0}dz\left[\sigma_{xx}(x,z)-\sigma_{xx}^0(x,z)\right]$ 
plotted as a function of $x$. The circles denote the symmetric case
($V=0.25\sqrt{\epsilon/m}$ and $H=13.6\sigma$)
and the squares denote the asymmetric case
($V=0.2\sqrt{\epsilon/m}$ and $H=13.6\sigma$). 
For clarity, $\sigma_{xx}^0$ was vertically displaced such that
$\sigma_{xx}^0=0$ far from the interface in the symmetric case, and 
for the asymmetric case, $\sigma_{xx}^0=0$ at the center of the interface
(same as in Fig. \ref{static-balance}).
}\label{normal-stress}
\end{figure}

\begin{figure}[ht]
\bigskip
\centerline{\psfig{figure=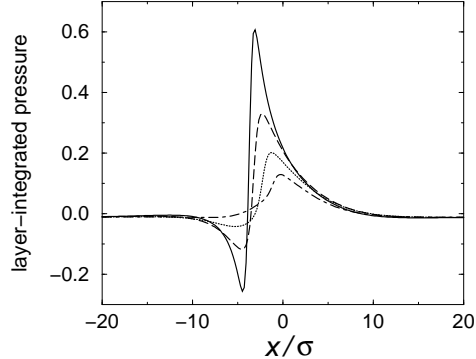,height=5.0cm}}
\caption{$\int_{z-z_0/2}^{z+z_0/2}dzp(x,z)$ obtained for 
the symmetric case of $V=0.25\sqrt{\epsilon/m}$ and $H=13.6\sigma$,
plotted as a function of $x$.
The profiles are symmetric about the center plane $z=H/2$,
hence only the lower half is shown at $z=0.425\sigma$ (solid line), 
$2.125\sigma$ (dashed line), $3.825\sigma$ (dotted line), 
and $5.525\sigma$ (dot-dashed lines).
The solid line denotes the BL-integrated pressure, 
in semi-quantitative agreement with the
$\int_0^{z_0}dz\tilde{\sigma}_{xx}(x,z)$ profile (circles) 
shown in Fig. \ref{normal-stress}. (Note that in Fig. \ref{normal-stress}
the MCL is placed at $x\approx 0$ while here it is shifted to 
$x\approx -3.9\sigma$,
same as in Figs. \ref{8symflow}, \ref{interface8}, and \ref{temperature}b.)
}\label{pressure}
\end{figure}

Equation (\ref{stokes-eq-int}) constitutes the basis for obtaining
the dynamic contact angle $\theta_d^{surf}$ from MD data.
The dominant behavior of $\int_0^{z_0}dz\left(-p+\sigma_{xx}^v\right)=
\int_0^{z_0}dz\sigma_{xx}^{HD}$ is a sharp drop across 
the fluid-fluid interface. This stress drop implies a large curvature 
in the BL, which pulls the extrapolated $\theta_d^{surf}$
closer to $\theta_s^{surf}$. We show the BL-integrated normal stress 
$\int_0^{z_0}dz\tilde{\sigma}_{xx}=
\int_0^{z_0}dz\left[\sigma_{xx}-\sigma_{xx}^0\right]$ in Fig. 
\ref{normal-stress}, where the large stress drop across the fluid-fluid 
interface is clearly seen.
(In the asymmetric case, due to the small difference between
$\sigma_{xx}^0$ and $T_{xx}$, $\tilde{\sigma}_{xx}$ 
is not precisely the hydrodynamic stress $\sigma_{xx}^{HD}$. In fact, 
$\tilde{\sigma}_{xx}=\sigma_{xx}^{HD}-\sigma_{xx}^{HS}$. 
Nevertheless, $\Delta\int_0^{z_0}dz\sigma_{xx}^{HS}$ is on the order of 
$2\gamma z_0\cos\theta_s^{surf}/H$, much smaller 
than the magnitude of the stress drop shown in Fig. \ref{normal-stress}.)
In the partial-slip region at the vicinity of the MCL, 
$\int_0^{z_0}dz\tilde{\sigma}_{xx}$ shows a fast variation along $x$ as well.
This means that the BL tangential force balance cannot be established
unless the gradient of the BL-integrated normal stress is taken into account
(see Eqs. (\ref{gf-expression}), (\ref{static-force-expression}), 
and (\ref{dynamic-force-expression})).
It is worth pointing out that the stress variation depicted in 
Fig. \ref{normal-stress} is also produced by the continuum hydrodynamic 
calculations, in semi-quantitative agreement with the MD data.
In Fig. \ref{pressure} we show the continuum profiles of 
layer-integrated pressure, obtained for a symmetric case of Couette flow.
It is readily seen that the magnitude of the pressure change across 
the fluid-fluid interface decays away from the solid surface quickly.
This conforms to the interface profile shown in Fig. \ref{interface8}:
the interfacial curvature quickly decreases with the increasing distance 
from the solid surface.

\subsection{The importance of the uncompensated Young stress}
\label{two-component}

\begin{figure}[ht]
\bigskip
\centerline{\psfig{figure=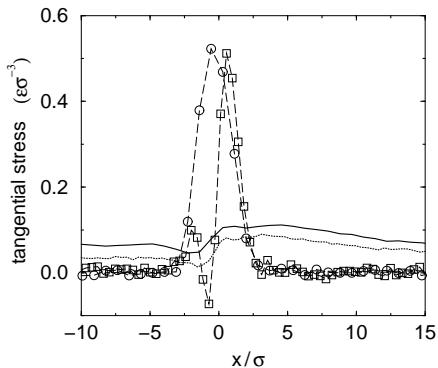,height=5.0cm}}
\caption{Two components of the hydrodynamic tangential stress at $z=z_0$,
plotted as a function of $x$.
The symbols connected by dashed lines denote $\tilde{\sigma}_{zx}^Y$;
the solid and dotted lines represent the viscous component.
Here the symmetric case is represented by circles and solid line;
the asymmetric case represented by squares and dotted line. In the contact line 
region the non-viscous component is almost one order of magnitude 
larger than the viscous component. The difference between the two components,
however, diminishes towards the boundary, $z=0$, due to the large
interfacial pressure drop (implying a large curvature) in the BL,
thereby pulling $\theta_d^{surf}$ closer to $\theta_s^{surf}$.  
}\label{shear-young}
\end{figure}

According to Eq. (\ref{decomposition}), the hydrodynamic tangential stress 
$\tilde{\sigma}_{zx}(x,z_0)$ can be decomposed into the viscous component
${\sigma}_{zx}^v$ and the non-viscous component $\tilde{\sigma}_{zx}^Y$:
$$\tilde{\sigma}_{zx}(x,z_0)=
{\sigma}_{zx}^v(z)+\tilde{\sigma}_{zx}^Y(x,z_0).$$
In Fig. \ref{shear-young} we show that away from the interfacial region
the tangential viscous stress ${\sigma}_{zx}^v(x,z_0)=
\eta(\partial_zv_x+\partial_xv_z)(x,z_0)$ is the only nonzero component, 
but in the interfacial region
$\tilde{\sigma}_{zx}^Y=\sigma_{zx}-\sigma_{zx}^v-\sigma_{zx}^0
=\sigma_{zx}^Y-\sigma_{zx}^0$ is dominant, thereby accounting for the
failure of the Navier boundary condition to describe the contact line motion.
Therefore away from the MCL region the Navier boundary condition is valid, 
but in the interfacial region 
it clearly fails to describe the contact line motion.

We have measured the $z$-integrated 
$\tilde{\sigma}_{xx}={\sigma}_{xx}-{\sigma}_{xx}^0$ in the BL.
The dominant behavior is a sharp drop across the interface, as
shown in Fig. \ref{normal-stress} for both the symmetric and asymmetric cases.
As already pointed out in Sec. \ref{angle-extrapolation}, this stress drop
means a large curvature in the BL, which pulls the extrapolated 
$\theta_d^{surf}$ closer to $\theta_s^{surf}$. 
The value of $\theta_d^{surf}$ obtained through extrapolation is 
$88\pm 0.5^\circ$ for the symmetric case and $63\pm 0.5^\circ$ for 
the asymmetric case at the lower boundary, and $64.5\pm 0.5^\circ$ 
at the upper boundary. These values are noted to be very close to 
$\theta_s^{surf}$. Yet the small difference between
the dynamic and static contact angles is essential
in accounting for the near-complete slip at the MCL.
                                                                                                                                                               
In essence, our results show that in the vicinity of the MCL,
the tangential viscous stress $\sigma_{zx}^v$ as postulated 
by the Navier boundary condition
can not give rise to the near-complete MCL slip without
taking into account the tangential Young stress $\sigma_{zx}^Y$
in combination with the gradient of the (BL-integrated) normal stress
$\sigma_{xx}$. For the static configuration, the Young stress is balanced 
by the normal stress gradient, leading to the Young's equation. 
It is only for a MCL that there is a component of the Young stress
which is no longer balanced by the normal stress gradient,
and this uncompensated Young stress is precisely the additional component
captured by the GNBC but missed by the traditional Navier boundary condition.

\section{Concluding Remarks}\label{remarks}

In summary, we have found the slip boundary condition, i.e., the GNBC, 
for the contact-line motion. Based on this finding, we have formulated 
a hydrodynamic model that is capable of reproducing the MD results of
slip profile, including the near-complete slip at the MCL.
It should be noted, however, that the present continuum formulation 
can not calculate fluctuation effects that are important in MD simulations. 

Based on the results and experiences obtained in the present study of the MCL,
we have been working on the following.

(1) Large scale MD simulations on two-phase immiscible flows 
show that associated with the MCL, there is a very large $1/x$ 
partial-slip region, where $x$ denotes the distance from the contact line. 
This power-law partial-slip region has been reproduced in large-scale 
adaptive continuum calculations \cite{ren-wang,power-law} based on 
the local, continuum hydrodynamic formulation presented in this paper.
MD simulations and continuum solutions both indicate the existence of 
a universal slip profile in the Stokes-flow regime, 
well described by $v^{slip}(x)/V=1/(1+{x}/{al_s})$, where $v^{slip}$
is the slip velocity, $V$ the speed of moving wall, $l_s$ the 
slip length, and $a$ is a numerical constant $\sim 1$ \cite{power-law}.
A large $1/x$ partial slip region is significant, 
because the outer cutoff length scale directly determines 
the integrated effects, such as the total steady-state dissipation.  
While in the past the $1/x$ stress variation away from the MCL 
has been known \cite{hua-scriven}, to our knowledge the fact 
that the partial slip also exhibits the same spatial dependence 
has not been previously seen,
even though the validity of the Navier boundary condition at 
high shear stress has been verified \cite{Th-Rob,nbc,barrat,ckb}.

(2) By explicitly taking into account the long-range wall-fluid
interactions, our hydrodynamic model for two-phase immiscible flow 
at the solid surface can be used to investigate the dry spreading
of a pure, nonvolatile liquid, attracted towards the solid by 
long-range van der Waals forces \cite{degennes}. The precursor film, 
driven by the gradient of disjoining pressure due to the long-range force
\cite{derjaguin}, was observed decades ago \cite{hardy,ghiradella,mckinley}. 
Nevertheless, the theoretical analysis remains to be difficult,
because of a wide separation of different length scales involved
\cite{degennes,hervet,pismen,glasner}. 
Application of our model to this multi-scale problem is currently underway.

(3) Decades ago it became well known that the driven cavity flow is
incompatible with the no-slip boundary condition, since the latter would
lead to stress singularity and infinite dissipation 
(known as corner-flow singularity) \cite{batchelor,moffatt-corner}.
While MD studies \cite{koplik-corner} have clearly demonstrated 
relative fluid-wall slipping near the corner intersection,
the exact rule that governs this relative slip has been left unresolved.
Based on the simulation technique developed in the present study, 
we have verified the validity of the Navier boundary condition in 
governing the fluid slipping in driven cavity flows.
We have used this discovery to formulate a continuum hydrodynamics 
whose predictions are in remarkable quantitative agreement 
with the MD simulation results at the molecular level \cite{qian-wang}.

\section*{Acknowledgments}
The authors would like to thank Weinan E and Weiqing Ren for helpful discussions.
This work was partially supported by the grants DAG03/04.SC21 and RGC-CERG 604803.



\begin{references}

\bibitem{batchelor} G.K. Batchelor, {\it An introduction to fluid dynamics},
Cambridge University Press, Cambridge, 1991.

\bibitem{Th-Rob}
P. A. Thompson and M. O. Robbins,
{\it Shear flow near solids: epitaxial order and flow boundary conditions},
Phys. Rev. A {\bf 41} (1990), pp. 6830.

\bibitem{nbc} P. A. Thompson and S. M. Troian,
{\it A general boundary condition for liquid flow at solid surfaces},
Nature {\bf 389} (1997), pp. 360.

\bibitem{barrat}
J-L. Barrat and L. Bocquet, 
{\it Large slip effect at a nonwetting fluid-solid interface},
Phys. Rev. Lett. {\bf 82} (1999), pp. 4671;
{\it Influence of wetting properties on hydrodynamic boundary conditions 
at a fluid/solid interface},
Faraday Discuss. {\bf 112} (1999), pp. 119.

\bibitem{ckb} 
M. Cieplak, J. Koplik, and J. R. Banavar,
{\it Boundary conditions at a fluid-solid interface},
Phys. Rev. Lett. {\bf 86} (2001), pp. 803.

\bibitem{navier1823}
C. L. M. H. Navier, {\it Memoire sur les lois du mouvement des fluides},
Memoires de l'Academie Royale des Sciences de l'Institut de France 
{\bf 6} (1823), pp. 389-440.

\bibitem{hua-scriven}
C. Hua and L. E. Scriven,
{\it Hydrodynamic model of steady movement of a solid/liquid/fluid
contact line,}
J. Colloid and Interface Sci. {\bf 35} (1971), pp. 85.

\bibitem{dussan} E. B. Dussan, V.,
{\it On the spreading of liquids on solid surfaces: 
static and dynamic contact lines},
Ann. Rev. Fluid Mech. {\bf 11} (1979), pp. 371.

\bibitem{degennes} P. G. de Gennes,
{\it Wetting: statics and dynamics},
Rev. Mod. Phys. {\bf 57} (1985), pp. 827.

\bibitem{blake}
T. D. Blake and J. M. Haynes, 
{\it Kinetics of liquid/liquid displacement},
J. Colloid and Interface Sci. {\bf 30} (1969), pp. 421;
T. D. Blake, {\it Dynamic contact angles and wetting kinetics},
in Wettability, edited by J. C. Berg (Marcel Dekker, Inc., 1993) pp. 251.

\bibitem{hocking}
L. M. Hocking, 
{\it A moving fluid interface. 
Part 2. The removal of the force singularity by a slip flow},
J. Fluid Mech. {\bf 79} (1977), pp. 209.

\bibitem{hua-mason} C. Huh and S. G. Mason,
{\it The steady movement of a liquid meniscus in a capillary tube},
J. Fluid Mech. {\bf 81} (1977), pp. 401.

\bibitem{sheng} 
M. Y. Zhou and P. Sheng,
{\it Dynamics of immiscible-fluid displacement in a capillary tube},
Phys. Rev. Lett. {\bf 64} (1990), pp. 882;
Ping Sheng and M. Y. Zhou, 
{\it Immiscible-fluid displacement: Contact-line dynamics and 
the velocity-dependent capillary pressure},
Phys. Rev. A {\bf 45} (1992), pp. 5694.

\bibitem{seppecher}
P. Seppecher, 
{\it Moving contact lines in the Cahn-Hilliard theory},
Int. J. Engng. Sci. {\bf 34} (1996), pp. 977.

\bibitem{jacqmin} 
D. Jacqmin,
{\it Calculation of two-phase Navier-Stokes flows using phase-field modeling},
J. Comput. Phys. {\bf 155} (1999), pp. 96;
D. Jacqmin,
{\it Contact-line dynamics of a diffuse fluid interface},
J. Fluid. Mech. {\bf 402} (2000), pp. 57.

\bibitem{vinal} 
H. Y. Chen, D. Jasnow, and J. Vinals,
{\it Interface and contact line motion in a two phase fluid under shear flow},
Phys. Rev. Lett. {\bf 85} (2000), pp. 1686.

\bibitem{pismen-pomeau}
L. M. Pismen and Y. Pomeau,
{\it Disjoining potential and spreading of thin liquid layers 
in the diffuse-interface model coupled to hydrodynamics}, 
Phys. Rev. E {\bf 62} (2000), pp. 2480; 
L. M. Pismen, 
{\it Mesoscopic hydrodynamics of contact line motion},
Colloids and Surfaces A {\bf 206} (2002), pp. 11.

\bibitem{koplik} J. Koplik, J. R. Banavar, and J. F. Willemsen,
{\it Molecular dynamics of Poiseuille flow and moving contact lines},
Phys. Rev. Lett. {\bf 60} (1988), pp. 1282;
J. Koplik, J. R. Banavar, and J. F. Willemsen,
{\it Molecular dynamics of fluid flow at solid surfaces},
Phys. Fluids A {\bf 1} (1989), pp. 781.

\bibitem{robbins} P. A. Thompson and M. O. Robbins,
{\it Simulations of contact-line motion: slip and the dynamic contact angle},
Phys. Rev. Lett. {\bf 63} (1989), pp. 766;
P. A. Thompson, W. B. Brinckerhoff, and M. O. Robbins,
{\it Microscopic studies of static and dynamic contact angles},
J. Adhesion Sci. Tech. {\bf 7} (1993), pp. 535.

\bibitem{hadji1}
N. G. Hadjiconstantinou, 
{\it Combining atomistic and continuum simulations of contact-line motion},
Phys. Rev. E {\bf 59} (1999), pp. 2475.

\bibitem{hadji2}
N. G. Hadjiconstantinou, 
{\it Hybrid atomistic-continuum formulations and the moving contact-line 
problem}, J. Comput. Phys. {\bf 154} (1999), pp. 245.

\bibitem{weiqing}
Weiqing Ren and Weinan E, 
{\it Heterogeneous multiscale method for the modeling of complex fluids 
and micro-fluidics}, J. Comput. Phys. {\bf 204} (2005), pp. 1.

\bibitem{qws} 
T. Z. Qian, X. P. Wang, and P. Sheng,
{\it Molecular scale contact line hydrodynamics of immiscible flows},
Phys. Rev. E {\bf 68} (2003), pp. 016306.

\bibitem{sensitivity}
E. B. Dussan V., {\it The moving contact line: the slip boundary condition},
J. Fluid Mech. {\bf 77} (1976), pp. 665.

\bibitem{md-book}
M. Allen and D. Tildesley,
{\it Computer Simulation of Liquids}, Clarendon, New York, 1987.

\bibitem{qian-wang} 
T. Z. Qian and X. P. Wang, 
{\it Driven cavity flow: from molecular dynamics to continuum hydrodynamics},
SIAM Multiscale Model. Simul. {\bf 3} (2005), pp. 749.

\bibitem{free-energy}
J. W. Cahn and J. E. Hilliard,
{\it Free energy of a nonuniform system. I. Interfacial free energy},
J. Chem. Phys. {\bf 28} (1958), pp. 258.

\bibitem{sharp-interface-limit}
R. Chella and J. Vinals, 
{\it Mixing of a two-phase fluid by cavity flow},
Phys. Rev. E {\bf 53} (1996), pp. 3832.

\bibitem{interfacial-tension}
J. G. Kirkwood and F. P. Buff, 
{\it The statistical mechanical theory of surface tension},
J. Chem. Phys. {\bf 17} (1949), pp. 338.

\bibitem{ren-wang}
W. Ren and X. P. Wang, 
{\it An iterative grid redistribution method for singular problems 
in multiple dimensions},
J. Comput. Phys. {\bf 159} (2000), pp. 246.

\bibitem{power-law} T. Z. Qian, X. P. Wang, and P. Sheng,
{\it Power-law slip profile of the moving contact line in two-phase
immiscible flows},
Phys. Rev. Lett. {\bf 93} (2004), pp. 094501.

\bibitem{subcontinuum}
K. P. Travis and K. E. Gubbins, 
{\it Poiseuille flow of Lennard-Jones fluids in narrow slit pores,}
J. Chem. Phys. {\bf 112} (2000), pp. 1984.

\bibitem{israelachivili}
J. N. Israelachivili, {\it Intermolecular and Surface Forces} (2nd ed),
Academic Press, London, 1992.

\bibitem{irving-kirkwood} 
J. H. Irving and J. G. Kirkwood, 
{\it The statistical mechanical theory of transport processes. 
IV. The equations of hydrodynamics,}
J. Chem. Phys. {\bf 18} (1950), pp. 817.
For the validity of the Irving-Kirkwood expression, 
see the paragraph following equation (5.15) in this paper. 

\bibitem{derjaguin}
B. V. Derjaguin, Zh. Fiz. Khim. {\bf 14} (1940), pp. 137.

\bibitem{hardy}
W. B. Hardy,
{\it The spreading of fluids on glass},
Philos. Mag. {\bf 38} (1919), pp. 49.

\bibitem{ghiradella}
H. Ghiradella, W. Radigan, and H. L. Frisch, 
{\it Electrical-resistivity changes in spreading liquid-films}, 
J. Colloid Interface Sci. {\bf 51} (1975), pp. 522.

\bibitem{mckinley}
H. P. Kavehpour, B. Ovryn, and G. H. McKinley,
{\it Microscopic and macroscopic structure of the precursor layer 
in spreading viscous drops},
Phys. Rev. Lett. {\bf 91} (2003), pp. 196104.

\bibitem{hervet}
H. Hervet and P. G. de Gennes,
{\it PHYSIQUE DES SURFACES ET DES INTERFACES. ---
Dynamique du mouillage: films precurseurs sur solide},
C. R. Acad. Sci. Paris II {\bf 299} (1984), pp. 499.

\bibitem{pismen}
L. M. Pismen, B. Y. Rubinstein, and I. Bazhlekov,
{\it Spreading of a wetting film under the action of van der Waals forces},
Phys. Fluid {\bf 12} (2000), pp. 480.

\bibitem{glasner}
K. B. Glasner,
{\it Spreading of droplets under the influence of intermolecular forces},
Phys. Fluid {\bf 15} (2003), pp. 1837.

\bibitem{moffatt-corner} H. K. Moffatt, 
{\it Viscous and resistive eddies near a sharp corner},
J. Fluid. Mech. {\bf 18} (1964), pp. 1. 

\bibitem{koplik-corner} J. Koplik and J. R. Banavar, 
{\it Corner flow in the sliding plate problem,} 
Phys. Fluids {\bf 7} (1995), pp. 3118.

\end{references}
\end{document}